\documentclass[12pt]{iopart}
\usepackage{graphicx}
\usepackage[utf8]{inputenc}
\usepackage[T1]{fontenc}

\expandafter\let\csname equation*\endcsname\relax
\expandafter\let\csname endequation*\endcsname\relax
\usepackage{amsmath}

\usepackage{hyperref}
\usepackage[capitalize]{cleveref}
\usepackage{xcolor}
\usepackage{soul}

\begin{document}

\setstcolor{red}

\title[ITG stability near the magnetic axis]{Ion-temperature-gradient stability near the magnetic axis of quasisymmetric stellarators}

\author{R. Jorge, M. Landreman}

\address{Institute for Research in Electronics and Applied Physics, University of Maryland, College
Park, MD 20742, USA}
\ead{rjorge@umd.edu}

\begin{abstract}
The stability of the ion-temperature gradient mode in quasisymmetric stellarators is assessed. This is performed using a set of analytical estimates together with linear gyrokinetic simulations. The peak growth rates, their corresponding real frequencies and wave-vectors are identified. A comparison is made between a first-order near-axis expansion model and eleven realistic designs obtained using numerical optimization methods. It is found that while the near-axis expansion is able to replicate the growth rates, real frequencies and perpendicular wave-vector at the inner core (both using simplified dispersion relations and first-principle gyrokinetic simulations), it leads to an overestimation of the growth rate at larger radii. An approximate analytic solution of the ITG dispersion relation for the non-resonant limit suggests growth rates could be systematically higher in quasi-axisymmetric (QA) configurations compared to quasi-helically (QH) symmetric ones. However except for very close to the axis, linear gyrokinetic simulations do not show systematic differences between QA and QH configurations.
\end{abstract}

\section{Introduction}
\label{sec:intro}

The feasibility of magnetic confinement fusion relies on a delicate balance between upholding a hot enough plasma in the core of the reactor and the inevitable transport of particles and energy to the wall.
Such transport is termed neoclassical when associated with geometrical effects in standard collisional transport theories, and termed turbulent when associated with perturbations in the plasma driven by underlying instabilities.
The reduction of neoclassical transport to acceptable levels has been, to an extent, attained with the tokamak and omnigeneous (such as quasisymmetric and quasi-isodynamic) stellarator designs.
The reduction of turbulent transport, however, has been a focus of theory and modelling of magnetic confinement plasma physics in the past few decades.
In the core, two microinstabilities are thought to be the main drive of turbulent transport: the Ion Temperature Gradient (ITG) mode and the Trapped Electron Mode (TEM) \cite{Xanthopoulos2007,Helander2015}.
In this work, we focus on the question of linear ITG stability and its parameter dependence in the core of a particular class of omnigeneous stellarator designs, namely quasisymmetric stellarators.

Quasisymmetry is a property of toroidal magnetic fields that ensures omnigenity, i.e., that collisionless particle trajectories remain confined inside the device over long periods of time.
Although there are numerous equivalent definitions of quasisymmetry,  here we define it as the condition that the magnitude of the magnetic field $B=|\mathbf B|$ varies on a surface of constant toroidal flux only through a fixed combination of the poloidal $\theta$ and toroidal $\varphi$ Boozer angles $B=B(\psi, M \theta - N \varphi)$ for some integers $M$ and $N$ \cite{Nuhrenberg1988a,Boozer1995,Garabedian1996}.
There are currently two ways of developing quasisymmetric stellarator designs: analytically, using the near-axis expansion \cite{Garren1991a,Landreman2018,Jorge2020b}
or a near-axisymmetry formulation \cite{Plunk2020};
or numerically, using one of the stellarator optimization methods available \cite{Nelson2003,Canik2007,Drevlak2019}.
In this work, the microstability of quasisymmetric designs near the magnetic axis will be assessed using both methods, and the advantages and shortcomings of using either method are highlighted.

Microinstabilities are driven by the presence of free energy associated with temperature and density gradients and are, therefore, ubiquitous in tokamaks and stellarators.
These develop locally in a plasma at the gyroradius scale, as opposed to macroinstabilities that affect the plasma globally.
Microinstabilities have regions of favorable development in the device, usually assessed by their relative "good" or "bad" curvature properties,  evaluated using the sign and magnitude of the magnetic drift frequency $\omega_d \simeq T \mathbf B \times \nabla B \cdot \mathbf k_\perp/(e B^3)$ where $T$ is the background temperature, $e$ the elementary charge and $\mathbf k_\perp$ the wave-vector of the mode perpendicular to the magnetic field $\mathbf B$.
If this frequency is positive (or, more accurately, if it has the same sign as the diamagnetic drift-frequency), trapped particle modes are likely to be unstable \cite{Helander2013a}.
While $\omega_d$ is positive in the outboard side of tokamaks, rendering unstable the region where trapped particles spend most of their time, in stellarators, the locations of bad curvature need to be evaluated numerically and are not necessarily in the outboard region.
As a result, many studies aimed at assessing the influence of magnetic field geometry on plasma stability and turbulence have been performed, particularly using gyrokinetic theory \cite{Helander2015}.
The freedom associated with the length and location of the bad curvature region has also allowed for stellarator optimization criteria based on the sign of $\omega_d$ and the non-overlap of trapped and "bad" curvature regions \cite{Mynick2011}.
While the first non-axisymmetric gyrokinetic studies were performed using the linear eigenvalue FULL code \cite{Rewoldt1982,Rewoldt1987}, only recently have nonlinear simulations been performed \cite{Xanthopoulos2007a} together with stellarator optimization efforts aimed at reducing turbulent transport \cite{Mynick2010}.

In this work, the ITG instability is analyzed first by incorporating the near-axis expansion into several known analytical estimates and then by performing linear gyrokinetic simulations for a set of eleven quasisymmetric designs using the GS2 code \cite{Dorland2000a}.
This instability can either be destabilized by the presence of an equilibrium temperature gradient which alters the temperature of a perturbation travelling at the sound speed that is in phase with a perturbed $\mathbf E \times \mathbf B$ drift, usually called the slab branch of the mode, or by the coupling of the grad-$B$ and curvature drifts with the equilibrium temperature gradient, usually called the toroidal branch of the mode \cite{Zocco2018}.
We mention that, although this work focuses on the properties of the linearized gyrokinetic equation, many non-linear studies of gyrokinetic ITG turbulence and saturation mechanisms have been carried out, showing the interplay between plasma geometry and stability \cite{Xanthopoulos2016,Hegna2018,McKinney2019,Wang2020}.

The purpose of this paper is twofold.
First, we show how quasisymmetric near-axis designs could lead to simplified yet accurate models of ITG stability.
Such models can then be used as optimization metrics when seeking for new stellarator designs with minimal levels of microinstability.
Second, we extend previous studies comparing first order near-axis expansions with the designs based on numerical optimization methods \cite{Landreman2019,Jorge2020c}.
While in Ref. \cite{Jorge2020c} a comparison was made between the geometric quantities used in linear and non-linear models of plasma dynamics, here we aim at using metrics that allow for a more quantitative comparison, such as the growth rate of underlying unstable modes.
Owing to the difficulty of simulating complex geometries in the flux-tube approximation \cite{Faber2018,Martin2018}, convergence tests are shown for the selected base case simulation scenario for every stellarator design.
The details of the implementation of non-axisymmetric geometries in the GS2 code can be found in Ref. \cite{Baumgaertel2011}.

This paper is structured as follows.
In \cref{sec:physicsgeometry}, the gyrokinetic equation and its underlying approximations used to assess microstability are presented. Also, the near-axis model is introduced together with the eleven quasisymmetric geometries being studied.
In \cref{sec:ITGmodels} various estimates aimed at assessing linear and non-linear stellarator microstability are rewritten leveraging the simplifying assumptions used in the near-axis expansion framework.
A benchmark configuration from a near-axis solution is used in \cref{sec:nabenchmark} to detail the analysis employed here when dealing with linear gyrokinetic simulations, including scans over the physical parameters at play, the resulting eigenmodes and eigenfrequencies.
The analysis performed for the benchmark case of \cref{sec:nabenchmark} is carried out in \cref{sec:numerical} for the selected eleven quasisymmetric designs.
The conclusions follow.

\section{Physical Model and Geometry}
\label{sec:physicsgeometry}

\subsection{Gyrokinetic Equation}

In this work, we employ the gyrokinetic approximation, i.e., we assume that the plasma is strongly magnetized, and that fluctuations have small amplitude and low frequency \cite{Howes2006}.
These assumptions are, in general, satisfied in magnetic confinement nuclear fusion devices.
In here, we find that the particle (electron and ion) Larmor radius, $\rho=v_{th}/\Omega$ with $v_{th}=\sqrt{2 T/m}$ the thermal velocity, $\Omega=q B/m$ the cyclotron frequency, $T$ the temperature, $m$ the mass, $q$ the charge and $B$ the magnetic field strength, is much smaller than typical length scales of the macroscopic equilibrium, $L$.
Furthermore, the frequency of fluctuations, $\omega$, is much smaller than the cyclotron frequency, $\Omega$.
This allows us to average the particle's equations of motion, effectively turning their Larmor orbit into a guiding-center one via the gyroaverage operator $\left<...\right>$.
Using the small fluctuation assumption and focusing on an electrostatic regime, we split the distribution function of each species, $f=F_0+\delta f$, and the electrostatic potential $\phi=\delta \phi$, into an equilibrium (denoted with a subscript $0$) and a fluctuating (denoted with $\delta$) part, where $\delta f \ll F_0$ and $q \phi/T\ll 1$.
Furthermore, the fluctuating distribution function is split into an adiabatic and non-adiabatic part as $\delta f = h - (q \phi/T)F_0$.
For the equilibrium, we take $F_0$ to be a Maxwell-Boltzmann distribution with thermal velocity $v_{th}$ and density $n$, and both the background density $n=n(\psi)$ and temperature $T=T(\psi)$ are taken to be functions of the magnetic toroidal flux $\psi$ only, with characteristic lengths $L_n$ and $L_T$, respectively.
Due to its importance in assessing regions of stability in the gyrokinetic space of parameters, we define $\eta=L_n/L_T$ as the ratio between the two characteristic lengths.
In the following, for simplicity, we denote $\left<\phi\right>$ and $\left<h\right>$ as $\phi$ and $h$, respectively, we assume that electrons are adiabatic with density $n$ and temperature $T_e$ and that a single ion species is present, with its gyroaveraged distribution function given by $h$, density $n$ and temperature $T_i = \tau_{T} T_e$.

The nonlinear gyrokinetic equation solved by GS2, in the linear electrostatic limit, can be written as \cite{Frieman1982a}
\begin{equation}
    \frac{\partial h}{\partial t}+(v_\parallel \mathbf b + \mathbf v_d) \cdot \nabla h + \frac{\mathbf b \times \nabla \phi}{B}  \cdot \nabla F_0 = \left<C\right> + \frac{q F_0}{T}\frac{\partial \phi}{\partial t}.
\label{eq:gyro}
\end{equation}
The particle's grad-B and curvature drifts are contained in
\begin{equation}
    \mathbf v_d=\frac{\mathbf b}{\Omega}\times\left(v_\parallel^2\mathbf b \cdot \nabla \mathbf b + \frac{v_\perp^2}{2}\frac{\nabla B}{B}\right).
\end{equation}
We note that, in \cref{eq:gyro}, the spatial derivatives are taken at constant energy $\mathcal{E}=m v^2/2$ and magnetic moment $\mu=m v_\perp^2/2B$ and collisions are modeled via the collision operator $C$.
The system of equations is closed via the quasineutrality equation which, in the limit of adiabatic electrons, reads
\begin{equation}
    \int d \mathbf v \left< h \right>_{\mathbf{r}} = n (1+\tau_{T})\frac{q \phi}{T},
\label{eq:quasin}
\end{equation}
where $\left<h\right>_{\mathbf{r}}$ denotes a gyroaverage performed at constant position vector $\mathbf r$ as opposed to $\left<h\right>$ which is performed at constant guiding-center locaion $\mathbf R$.
A relation between the two can be found by Fourier decomposing $h$ into a parallel and perpendicular wave-vectors $\mathbf k=k_\parallel \mathbf b + \mathbf k_\perp$ with $\mathbf b=\mathbf B/B$, allowing us to write, in Fourier space, $ \left<h\right>_{\mathbf{r}}=J_0(k_\perp v_\perp/\Omega)\left<h\right>$.

In order to perform a stability analysis, we write \cref{eq:gyro} in the field-line following limit.
Additionally, due to the high temperature of the plasma in the core, we look at the collisionless limit $\nu_e \ll \omega$ where $\nu_e$ is the electron-ion collision frequency, and set $C=0$.
In the field-line following limit, we write the perturbed quantities $h$ and $\phi$ as
\begin{equation}
    h=\hat{h}(l)e^{i(S-\omega t)},
\end{equation}
where $l$ is the distance along a magnetic field line  and $\mathbf b \cdot \nabla S=0$, leading us to define $\mathbf k_\perp\equiv\nabla S$.
%
%
In this limit, and normalizing $\phi$ by $q/T$, \cref{eq:gyro} can be written as
\begin{equation}
    i v_\parallel \mathbf b \cdot \nabla \hat{g}+(\omega-\tilde\omega_d)\hat{g}=\hat\phi F_0 (\omega - \tilde \omega_*)J_0\left(\frac{v_\perp k_\perp}{\Omega}\right),
\label{eq:lingk}
\end{equation}
with $\tilde \omega_d=\mathbf k_\perp \cdot \mathbf v_d$ the magnetic drift-frequency.
When a low $\beta$ limit is taken, $\tilde \omega_d$ can be written as $\tilde \omega_d=\omega_d (v_\parallel^2+v_\perp^2/2)/v_{th}^2$ with $\omega_d$ its velocity-independent counterpart.
The velocity-dependent diamagnetic frequency $\tilde \omega_*$ is given by
\begin{equation}
    \tilde \omega_*=\omega_*\left[1+\eta\left(\frac{v^2}{v_{th}^2}-\frac{3}{2}\right)\right],
\label{eq:tildeomega}
\end{equation}
where $\omega_*=\mathbf b \times \mathbf k_\perp \cdot \nabla F_0/(m \Omega T F_0)$ usually written as $\omega_*=(T k_\alpha/q)d \ln n/d\psi$ when a Clebsch representation for the magnetic field is used.

As a final note, we mention here the normalizations used to output GS2 quantities.
%
%
Lengths are normalized to a fixed length $a$ (a measure of the minor radius of the device), time is normalized to $a/v_{th}$ and perturbed quantities are scaled up by $a/\rho_{th}$.
In the axisymmetric case, $a$ is half of the diameter of the last closed flux surface measured at the elevation of the magnetic axis.
Here, however, we identify $a$ with the effective minor radius computed by VMEC  \cite{Hirshman1983}, named Aminor\_p in its wout*.nc file, defined as $\pi$ (Aminor\_p$)^2=({1}/{2\pi})\int_0^{2\pi}S(\phi)d\phi$ where $\phi$ is the toroidal angle associated with the cylindrical coordinate system $(R,Z,\phi)$ ($\phi$ is not to be confused with the electrostatic potential) and $S(\phi)$ is the area of the outer surface's cross-section in the $R-Z$ plane.

\subsection{Coordinates and Magnetic Geometry}

We now rewrite the differential operators in the linearized gyrokinetic equation, \cref{eq:lingk}, in terms of the straight-field-line magnetic coordinates commonly referred to as Boozer coordinates \cite{Boozer1981} and employ the Garren-Boozer near-axis expansion formalism \cite{Garren1991a,Garren1991}.
We note that such formalism uses an inverse coordinate approach, writing $\mathbf r$ as a function of Boozer coordinates, as opposed to formalisms using a direct coordinate approach, which write the toroidal flux as a function of the spatial coordinates $\mathbf r$ \cite{Mercier1964,Solovev1970,Jorge2020}.
First, we express the magnetic field $\mathbf B$ using a Clebsch representation of the form
\begin{equation}
    \mathbf B = \nabla \psi \times \nabla \alpha,
\label{eq:clebsch}
\end{equation}
where $2 \pi \psi$ is the toroidal flux and $\alpha$ is a field line label.
Taking as a spatial coordinate system the set $(\psi, \alpha, z)$ with $z$ any quantity independent of $\psi$ and $\alpha$, the eight independent geometrical quantities $\mathbf Q$ needed to solve \cref{eq:lingk} are \cite{Jorge2020c}
\begin{align}
    \mathbf Q=&\left\{B, \mathbf b \cdot \nabla z, |\nabla \psi|^2, |\nabla \alpha|^2, \nabla \psi \cdot \nabla \alpha,\right.\nonumber\\
    &\left.(\mathbf b \times \nabla B) \cdot \nabla \alpha,(\mathbf b \times \nabla B) \cdot \nabla \psi, (\mathbf b \times \boldsymbol \kappa)\cdot \nabla \alpha\right\}.
\label{eq:geomQ}
\end{align}
From here onward, we employ Boozer coordinates $(\psi, \theta, \varphi)$ with $\theta$ and $\varphi$ the poloidal and toroidal angles, respectively, and choose $z=\varphi$.
For convenience, a helical angle $\vartheta=\theta-N \varphi$ with $N$ an integer is introduced, such that the magnetic field $\mathbf B$ can be written as
\begin{align}
\mathbf{B} = \nabla\psi \times\nabla\vartheta + \iota_N \nabla\varphi \times\nabla\psi,
\label{eq:straight_field_lines_h}
\\
 = \beta \nabla\psi + I \nabla\vartheta + (G+NI) \nabla\varphi,
\label{eq:Boozer_h}
\end{align}
where $\iota_N = \iota$ with $\iota$ the rotational transform and $I, G$ and $\iota$ are constants on $\psi$ surfaces, i.e., $I=I(\psi)$, $G=G(\psi)$ and $\iota = \iota(\psi)$.
A relation between $\alpha$ and the angles $\vartheta$ and $\varphi$ can be found by comparing \cref{eq:clebsch} and \cref{eq:straight_field_lines_h}, yielding $\alpha=\vartheta-\iota_N \varphi=\theta-\iota \varphi$.

The geometry coefficients $\mathbf Q$ in \cref{eq:geomQ} are evaluated at lowest order in $\epsilon=r/\mathcal{R}$, where $r$ is defined as $2\pi|\psi|=\pi r^2 B_0$ and $\mathcal{R}$ a scale length representing the major radius of the device (e.g., the inverse of the maximum axis curvature).
Furthermore, we focus on quasisymmetric magnetic fields, i.e., we use magnetic geometries where $B=B(\psi, M \vartheta - N \varphi)$ with $M$ and $N$ integers.
As shown in Ref. \cite{Jorge2020c}, this yields the following set of geometric coefficients
\begin{align}
    B&=B_0(1+r \overline \eta \cos \vartheta),\label{eq:g1}\\
    \mathbf b \cdot \nabla z&=s_G \frac{2\pi}{L}(1+r \overline \eta \cos \vartheta),\label{eq:g2}\\
    |\nabla \psi|^2&=r^2 \frac{B_0^2}{\overline \eta^2 \kappa^2}\left[\overline \eta^4 \sin ^2\vartheta+\kappa^4 (\cos \vartheta-\sigma\sin \vartheta )^2\right],\label{eq:g3}\\
    |\nabla \alpha|^2&=\frac{1}{r^2\overline \eta^2 \kappa^2}\left[\overline \eta^4 \cos ^2\vartheta+\kappa^4 (\sigma \cos \vartheta +\sin\vartheta)^2\right],\label{eq:g4}\\
    \nabla \psi \cdot \nabla \alpha&=\frac{s_\psi B_0}{2 \overline \eta^2 \kappa^2}\left( \left[\overline \eta^4+\kappa^4 \left(\sigma^2-1\right)\right]\sin 2 \vartheta-2 \kappa^4 \sigma \cos 2 \vartheta\right),\label{eq:g5}\\
    (\mathbf b \times \nabla B)\cdot \nabla \alpha&=\frac{s_\psi}{r} B_0 \overline \eta \cos \vartheta \label{eq:g6},\\
    (\mathbf b \times \nabla B)\cdot \nabla \psi&=r B_0^2 \overline \eta \sin \vartheta, \label{eq:g7}\\
    (\mathbf b \times \boldsymbol \kappa)\cdot \nabla \alpha&=\frac{s_\psi}{r}\overline \eta \cos \vartheta \label{eq:g8}.
\end{align}
In the set of \cref{eq:g1,eq:g2,eq:g3,eq:g4,eq:g5,eq:g6,eq:g7,eq:g8}, $B_0$ is the constant magnetic field on-axis, $r=\sqrt{2\psi/B_0}$ is an effective minor radius, $\overline \eta$ is a constant that is related to the elongation of the elliptical flux-surfaces, $\kappa=\kappa(\varphi)$ is the axis curvature, $s_\psi=${sgn}$(\psi)$, $s_G=${sgn}$(G)$, and $\sigma=\sigma(\varphi)$ is the solution of
\begin{equation}
    \frac{d\sigma}{d\varphi}+(\iota_0-N)\left(\frac{\overline \eta^4}{\kappa^4}+1+\sigma^2\right)-\frac{s_G L \overline \eta^2}{\pi \kappa^2}\left(\frac{I_2}{B_0}-s_\psi\tau\right)=0,
\label{eq:sigma}
\end{equation}
where $\tau=\tau(\varphi)$ is the axis torsion, $\iota_0$ is the on-axis rotational transform, $I_2$ is the lowest order toroidal current $I\simeq I_2 r^2$ and $L$ is the total length of the magnetic axis curve $\mathbf r_0(\phi)$
\begin{equation}
    L=\int_0^{2\pi} \left|\frac{d \mathbf r_0}{d \phi}\right| d\phi.
\end{equation}
Alternatively, due to \cref{eq:g1}, $\overline \eta$ can be regarded as the constant parameter that reflects the magnitude of variation in $B$.

We now introduce dimensionless $x(\psi)$ and $y(\alpha)$ coordinates that correspond to scaled versions of $\psi$ and $\alpha$:
\begin{align}
    x&=\frac{dx}{d\psi}(\psi-\psi_0),
\end{align}
and
\begin{align}
    y&=\frac{dy}{d\alpha}(\alpha-\alpha_0),
\end{align}
with $dx/d\psi$ and $dy/d\alpha$ constant.
Here, we take $\psi_0=\alpha_0=0$ and define $dx/d\psi$ and $dy/d\alpha$ such that $x$ reduces to the usual minor radius in the cylindrical limit and that the product $x dx/d\psi$ equals the inverse of a reference magnetic field, $B{\textsubscript{ref}}$.
More specifically, we take $x=a r/r_a$ and $y=\alpha a r/r_a$ with $r_a$ the value of $r$ at the VMEC boundary and $a=$Aminor\_p the effective minor radius.
The linearized gyrokinetic equation, \cref{eq:lingk}, is then Fourier decomposed along the $x$ and $y$ directions with $\mathbf k_\perp=k_x \nabla x + k_y \nabla y$ the perpendicular wave-vector.

\begin{figure}
    \centering
    \includegraphics[width=.5\textwidth]{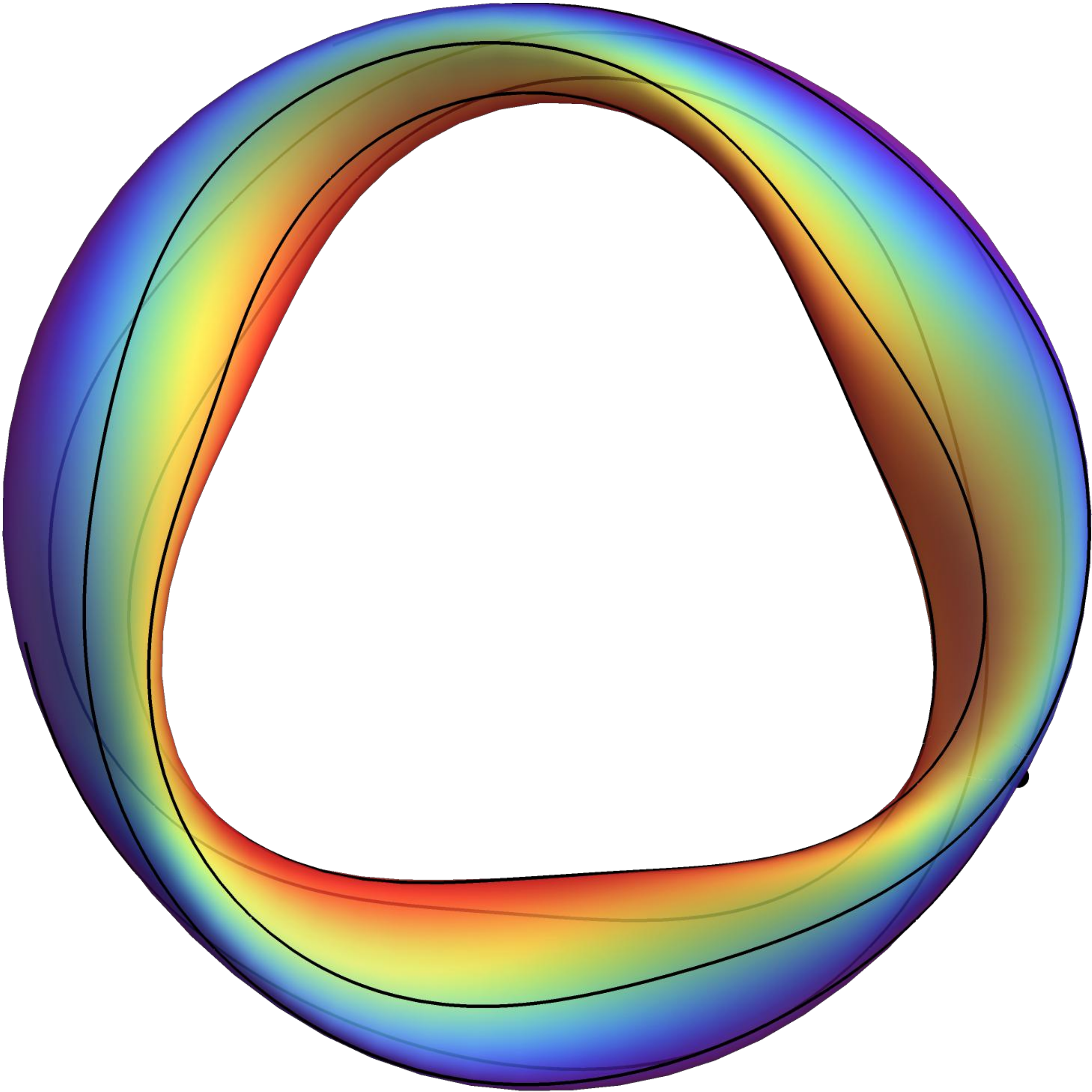}
    \includegraphics[width=.49\textwidth]{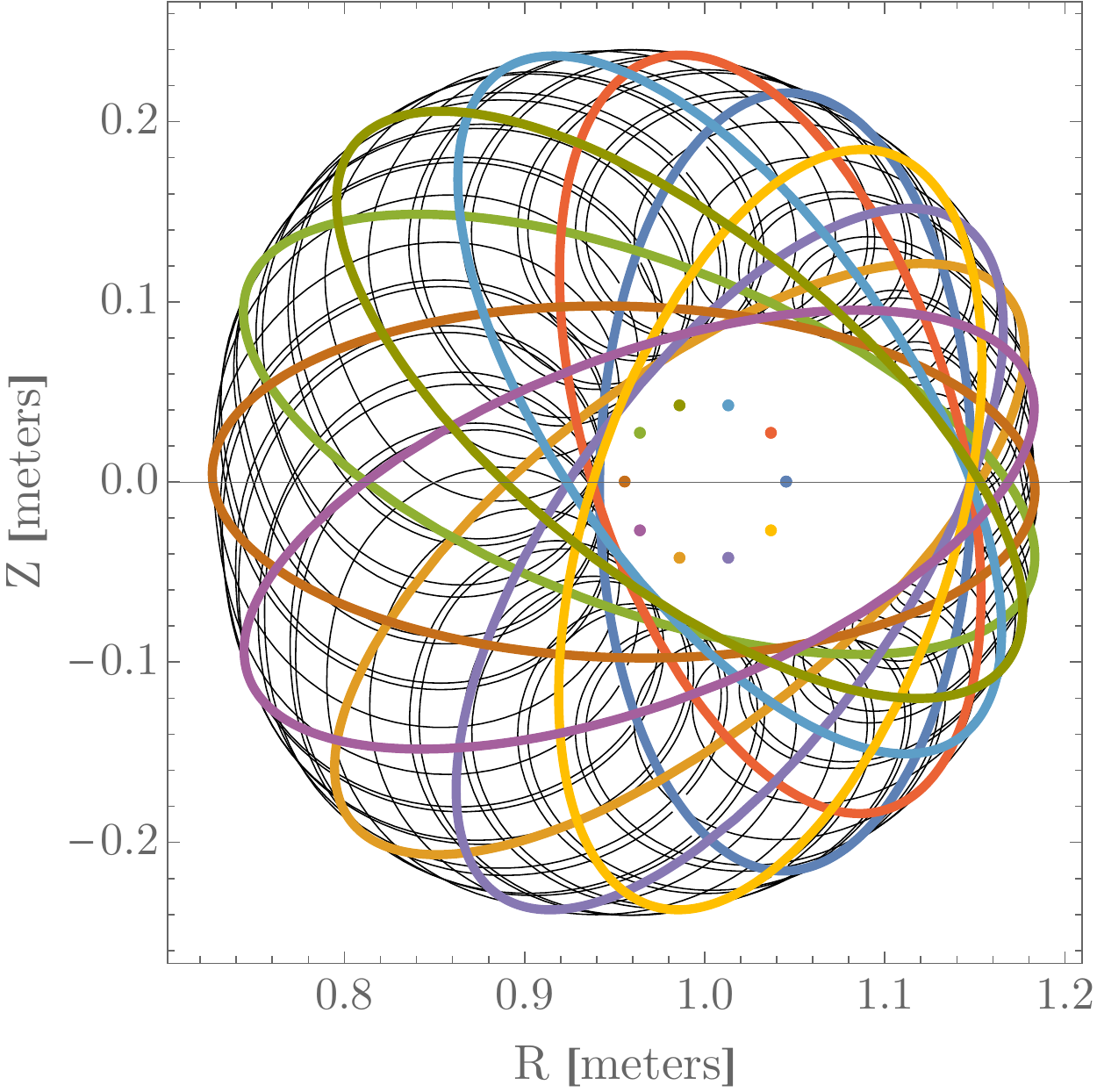}
    \includegraphics[width=.85\textwidth]{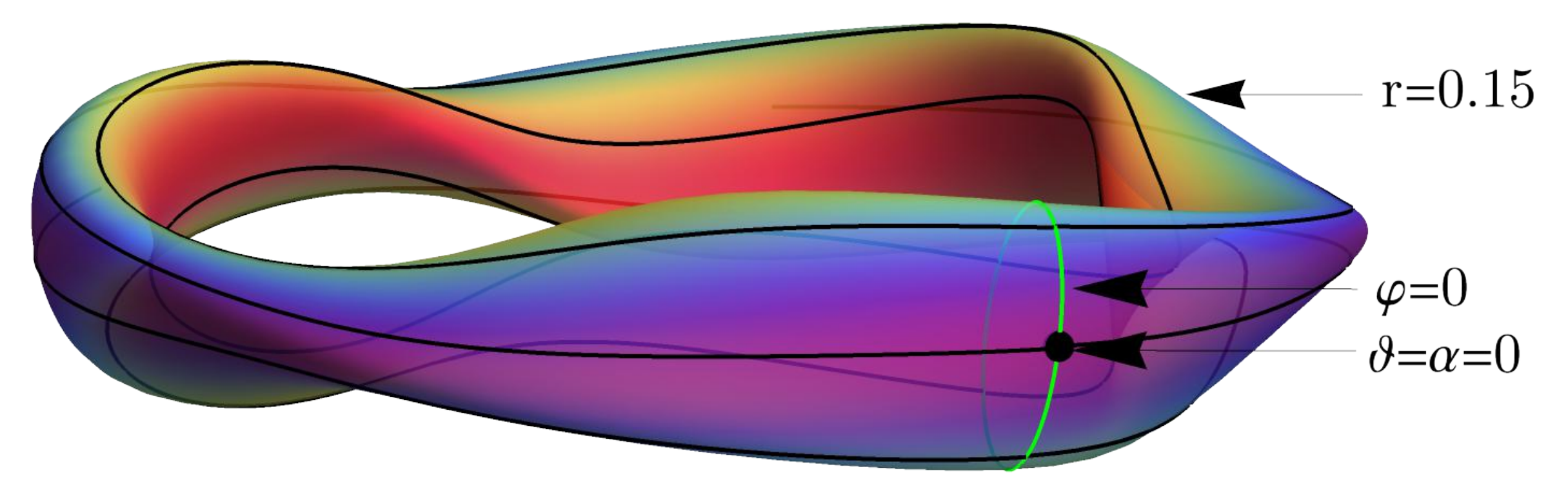}
    \caption{Top left: top view of a first-order quasi-axisymmetric surface of constant toroidal flux $\psi$ (Section 5.1 of Ref. \cite{Landreman2018}). Colour indicates $B$ on the outermost flux surface. An example field-line with $-2\pi<\vartheta<2\pi$, $\alpha=0$ and $r=0.15$ is shown in black together with the location of $\vartheta=0$. Top right: ten poloidal planes and their respective axis location, together with a field line with $-10\pi<\vartheta<10\pi$ and $\alpha=0$. Bottom: side view of the toroidal flux surface.}
    \label{fig:fluxtube}
\end{figure}

An example first-order near-axis stellarator geometry is shown in \cref{fig:fluxtube}, namely the quasi-axisymmetric design of Section 5.1 of Ref. \cite{Landreman2018}. A field line with $r=0.15$, $\alpha=0$ and $-2\pi<\vartheta<2\pi$ is shown, together with the location of $\vartheta=0$.
A poloidal projection of a longer field-line with $-10\pi<\vartheta<10\pi$ and ten poloidal cross-sections of the flux surface are shown in the top-right of \cref{fig:fluxtube}.
The flux-tube approach employed in GS2 to simulate plasma conditions in the field-line following limit can then thought of as a series of field-lines similar to the one in \cref{fig:fluxtube} with a finite (but small) extension along the radial ($\psi$) and poloidal ($\alpha$) directions.

The stellarator geometries used in this study, apart from a near-axis benchmark case, consist of eleven quasisymmetric stellarator designs.
These were developed by several
independent research teams using different optimization codes, with varying degrees of quasisymmetry.
By order of decreasing number of field periods, these are the NZ1988 design of Ref. \cite{Nuhrenberg1988}, the Drevlak design of Ref. \cite{Drevlak2017} recently developed at the Max Planck Institute for Plasma Physics in Greifswald, Germany, HSX \cite{Anderson1995}, the KuQHS48 design of Ref. \cite{Ku2011}, the WISTELL-A configuration of Ref. \cite{Bader2020} recently developed at the University of Wisconsin-Madison, USA, NCSX (configuration LI383,  \cite{Zarnstorff2001}), ARIES-CS configuration N3ARE \cite{Najmabadi2008}, the QAS2 configuration of Ref. \cite{Garabedian2008} with a vanishing on-axis current, ESTELL \cite{Drevlak2013}, CFQS \cite{Shimizu2018a} and the Henneberg design of Ref. \cite{Henneberg2019}.
The properties of each stellarator design are listed in Table 1 of Ref. \cite{Jorge2020c}, together with the $B_0$ and $\overline \eta$ constants used for each configuration and a comparison between the original and the resulting near-axis geometric coefficients.
The original coefficients $\mathbf Q$ are obtained from the VMEC file corresponding to each design using the geometry module of the \textit{stella} code \cite{Barnes2019}.

Finally, we review the main input and output parameters of the model equations when combined with the near-axis geometry described above.
For each configuration, we use its axis shape to find the axis curvature $\kappa$ and torsion $\tau$.
The parameters $B_0$ and $\overline \eta$ are found using the fitting procedure described in Ref. \cite{Jorge2020c} allowing us to find the $\sigma$ function for each configuration, hence to fully specify the geometry coefficients in \cref{eq:g1,eq:g2,eq:g3,eq:g4,eq:g5,eq:g6,eq:g7,eq:g8}.
The physical input parameters used by GS2 to solve the linear gyrokinetic equation, \cref{eq:lingk}, are the $x$ and $y$ components of the perpendicular wave-vector, $k_x$ and $k_y$, respectively, and the physical parameters related to the particle species.
We note that, at $\alpha=0$, choosing a flux tube with $k_x=0$ yields the largest growth rate \cite{Proll2013}.
Therefore, in the following, all simulations are performed at $\alpha=k_x=0$.
As electrons are considered adiabatic, the input characteristic gradient lengths consist only of the ion density, $L_n$, and temperature, $L_T$, lengths.
The atomic number is set to $Z=1$.
The temporal and spatial resolution parameters are set using convergence scans.

\section{Models of Near-Axis ITG stability}
\label{sec:ITGmodels}

In this section, we use the geometric coefficients $\mathbf Q$ in the set of \cref{eq:g1,eq:g2,eq:g3,eq:g4,eq:g5,eq:g6,eq:g7,eq:g8} in order to simplify the characteristic ITG frequencies $\omega_*$ and $\omega_d$, together with the perpendicular wave-vector $\mathbf k_\perp$.
Then, we plug these simplified expressions into known limits of the gyrokinetic equation commonly employed in ITG stability studies.
The resulting models allow for a more intuitive understanding of the properties of the ITG mode near the magnetic axis and are used in the subsequent sections to interpret the linear gyrokinetic results.

At lowest order in the distance to the axis, and noting that $k_\alpha = a r k_y/r_a$, the diamagnetic frequency $\omega_*$ can be written as
\begin{equation}
    \frac{\omega_*}{\Omega}=\frac{a}{L_n}\frac{\rho_{th}}{r_a}k_y \rho_{th},
\end{equation}
where $\rho_{thi}=v_{th}/\Omega$ and $L_n=d \ln n/dr$ (and similarly for $L_T$), while the drift-frequency reads
\begin{equation}
    \frac{\omega_d}{\Omega}=s_\psi\frac{a}{r_a}\overline \eta \rho_{th} k_y \rho_{th} \cos \vartheta.
\end{equation}
In the following, we further assume that the most unstable linear modes peak at $k_\psi=0$ and write the square of the normalized perpendicular wave-vector $b$ as
\begin{equation}
    b={k_\perp^2 \rho_{th}^2}=\frac{a^2}{r_a^2}k_y^2 \rho_{th}^2\left[\frac{\overline \eta^2}{\kappa^2}\cos^2\vartheta+\frac{\kappa^2}{\overline \eta^2}(\sigma \cos \vartheta+\sin \vartheta)^2\right].
\end{equation}
Taking $z=\vartheta$, the parallel derivative operator can be written as
\begin{equation}
    \mathbf b \cdot \nabla \phi = \nabla_\parallel \phi=s_G\frac{2 \pi}{L}\iota_N(1+r \overline \eta \cos \vartheta)\frac{\partial \phi}{\partial \vartheta}.
\end{equation}
Furthermore, by defining the normalized quantities $\tilde k_y = k_y \rho_{th}$, $\tilde L_n=L_n 2\pi/L$, $\tilde \eta = \overline \eta L/2\pi$, $\tilde \kappa = \kappa/\overline \eta$, $\tilde \omega = \omega/\omega_t$ with the transit frequency $\omega_t=2 \pi v_{th}/L$ (similarly for $\omega_d$ and $\omega_*$), we rewrite the ITG parameters above as
\begin{equation}
    \tilde \omega_*=\frac{\tilde k_y}{\tilde L_n},
\end{equation}
and
\begin{equation}
    \tilde \omega_d = s_\psi \tilde k_y \tilde \eta \cos \vartheta,
\end{equation}
and
\begin{equation}
    b=\frac{\tilde k_y^2}{\tilde \kappa^2}\left[\cos^2 \vartheta + \tilde \kappa^4(\sigma \cos \vartheta+\sin \vartheta)^2\right],
\end{equation}
where the approximation $a \simeq r_a$ was used.
This reduces the set of dimensionless parameters entering the linearized gyrokinetic equation to $\tau_{T}, \tilde k_y, \tilde L_n, \eta, \tilde \eta, \tilde \kappa$ and $\sigma$.
We note that for stellarator-symmetric designs $\sigma(\varphi=0)=0$ and that according to \cref{eq:sigma}
\begin{equation}
    \sigma'(0)=\frac{1}{(\iota_0-N)}\left.\frac{d \sigma}{d \varphi}\right|_{0}=\frac{2\tilde \kappa(0)^2[\tilde I - s_\psi \tilde \tau(0)]}{(\iota_0-N)^2[1+\tilde \kappa(0)^4]},
\end{equation}
where $\tilde I = I_2 L /(2 \pi B_0)$ is the normalized current and $\tilde \tau = \tau L/2\pi$ is the normalized torsion.

One limit in which further analytic simplification is possible is the case in which the eigenfunction is localized near $\vartheta=0$.
We can then expand the ITG parameters $\omega_d$ and $b$ around $\vartheta=0$ as
\begin{equation}
    \tilde \omega_d=\tilde \omega_{d0}\left(1-\frac{\vartheta^2}{2} \right),
\label{eq:omegadhighlylocalized}
\end{equation}
and
\begin{equation}
    b=b_0(1+s \vartheta^2),
\label{eq:bhighlylocalized}
\end{equation}
where we find $\tilde \omega_{d0}=s_\psi \tilde k_y \tilde \eta$, $b_0=\tilde k_y^2/\tilde \kappa^2$, the effective shear parameter
\begin{equation}
    s=-\left[1+\frac{\tilde \kappa''(0)}{(\iota_0-N)^2\tilde \kappa(0)}-\tilde \kappa(0)^4\left(1+\frac{\sigma'(0)}{\iota_0-N}{}\right)^2\right],
\label{eq:shighlylocalized}
\end{equation}

\subsection{Nonlocal small transit frequency approximation in the non-resonant limit}

Following Ref. \cite{Plunk2014}, we consider the limit where the transit frequency $\omega_t$ is small when compared with the mode frequency $\omega$.
Furthermore, we adopt the non-resonant limit where $\omega/\omega_*^T=\delta \ll 1$ and $\omega_d/\omega\sim\omega_t^2/\omega^2\sim k_\perp^2 \rho^2 \sim \delta$, yielding the following second order differential equation for the electrostatic potential \cite{Zocco2020}
\begin{equation}
    \frac{\omega_*^T v_{th}^2}{2 \omega^3}\nabla_\parallel^2 \phi=\left[\tau_{T}+(1-b \eta)\frac{\omega_*}{\omega}+\frac{\omega_d \omega_*^T}{\omega^2}\right]\phi,
\label{eq:ITGsimple1}
\end{equation}
where $\omega_*^T=\eta \omega_*$ and
$\nabla_\parallel^2 \phi \simeq (\iota_0-N)^2 ({4 \pi^2}/{L^2}){\partial^2 \phi}/{\partial \vartheta^2}$.
In the highly localized approximation of \cref{eq:omegadhighlylocalized,eq:bhighlylocalized}, \cref{eq:ITGsimple1} turns into a Schr\"odinger equation of the Weber type whose bound solutions have the form $\phi \sim H_n(\sqrt{\delta}\vartheta)\exp(\delta \vartheta^2/2)$ where $H_n$ is a Hermite polynomial of order $n=0,1,2,...$ and the width $\delta$ is given by
\begin{equation}
    \delta^2=-\tilde \omega  \tilde k_y\left( s_\psi \tilde \eta +\frac{2\tilde k_y s^2}{\tilde \kappa^2}\tilde \omega\right),
\end{equation}
where $\omega$ is the root of the dispersion relation
\begin{equation}
    \frac{\tau_{T} \tilde L_n}{\tilde k_y \eta} \tilde \omega^3+\left(1-\frac{\eta \tilde k_y^2}{\tilde \kappa^2}\right)\frac{\tilde \omega^2}{\eta}+s_\psi \tilde k_y \tilde \eta \tilde  \omega+\left(n+\frac{1}{2}\right)\delta^+=0,
\label{eq:analyticalDR1}
\end{equation}
for which Re$[\delta]>0$ (denoted as $\delta^+$).
We note that, in the dispersion relation above, the effective shear parameter $s$ only contributes to $\delta^2$ as an $s^2$ factor yielding eigenfunctions that are symmetric with respect to $s$.

\begin{figure}
    \centering
    \includegraphics[width=.99\textwidth]{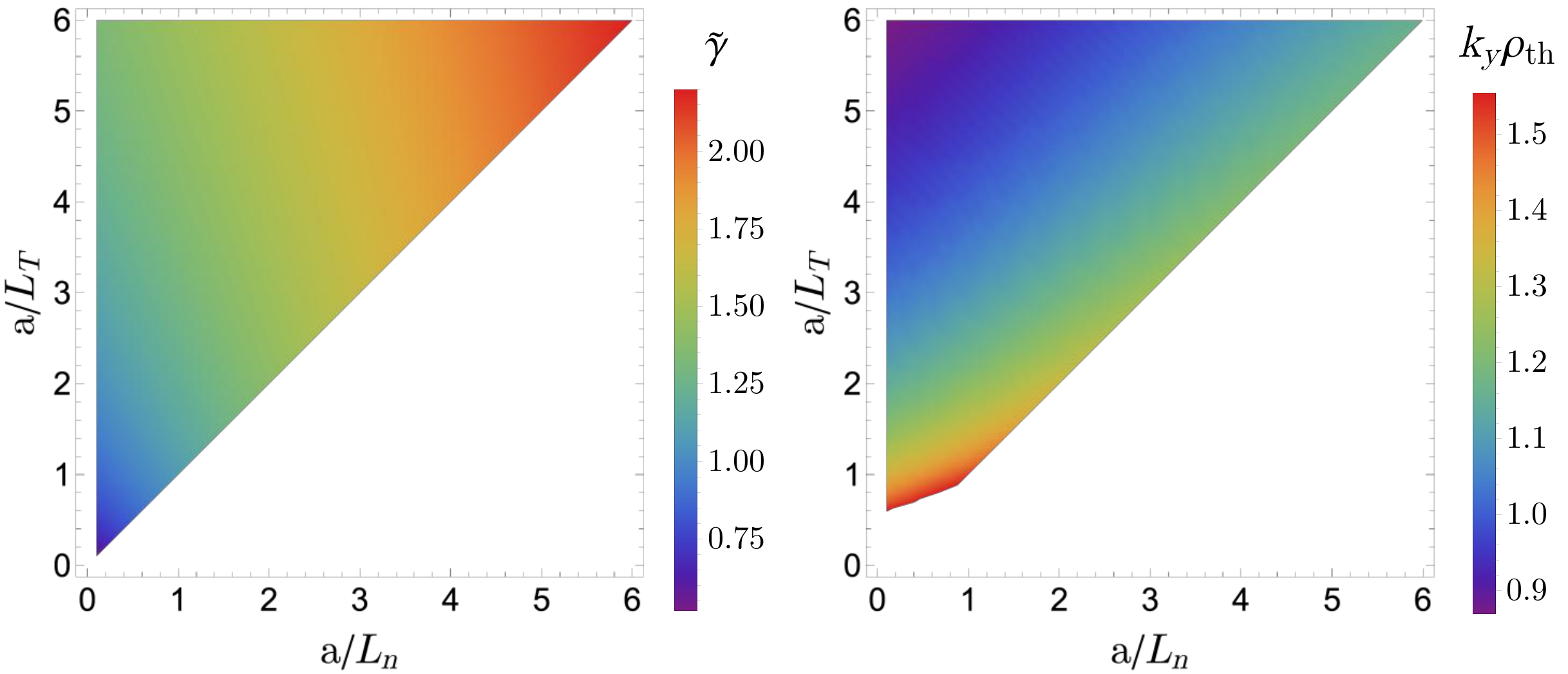}
    \includegraphics[width=.99\textwidth]{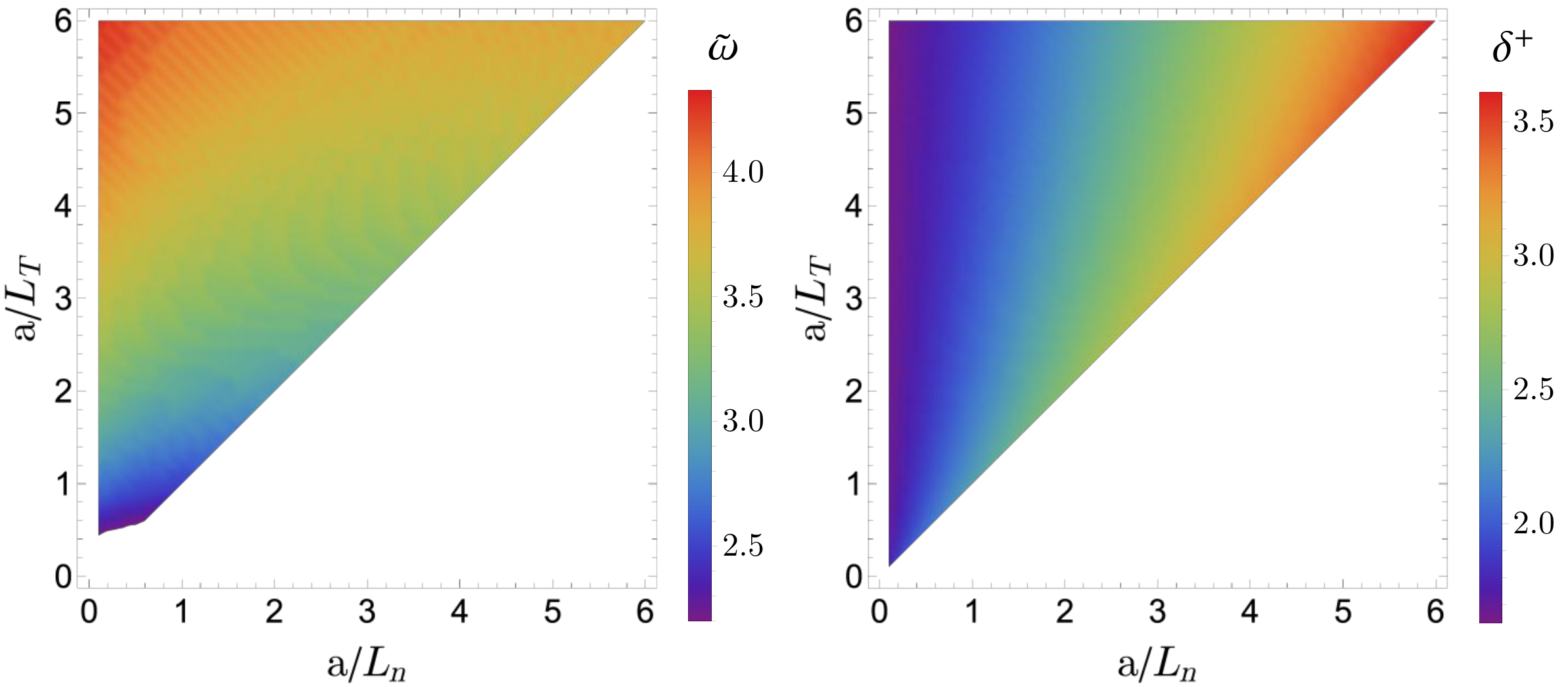}
    \caption{
    Solution of the ITG dispersion relation \cref{eq:analyticalDR1} with $\eta>1$ for the case $s=0.1, \tilde \eta=1$ and $\tilde \kappa=1$, namely the peak growth rate $\gamma$ (top left), the corresponding $k_y \rho_{th}$ (top right), real frequency $\omega$ (bottom left) and the eigenfunction width $\delta^+$ (bottom right). The frequencies $\gamma$ and $\omega$ are normalized to the transit frequency $\omega_t=2 \pi v_{th}/L$.}
    \label{fig:analyticalDR1}
\end{figure}

An example numerical solution for the dispersion relation in \cref{eq:analyticalDR1} with $s=0.1, \tilde \eta=1$ and $\tilde \kappa=1$ is shown in \cref{fig:analyticalDR1}.
There, the peak growth rate $\gamma$ and the corresponding value of $\tilde k_y$ and $\delta^+$ are shown for a range of $L_n$ and $L_T$ values satisfying $\eta=L_n/L_T>1$.
It is seen that, in general, the ITG growth rate $\gamma$ and its corresponding mode width $\delta^+$ increase with increasing $a/L_n$ and $a/L_T$ while the opposite is found for $k_y$.

\begin{figure}
    \centering
    \includegraphics[width=.99\textwidth]{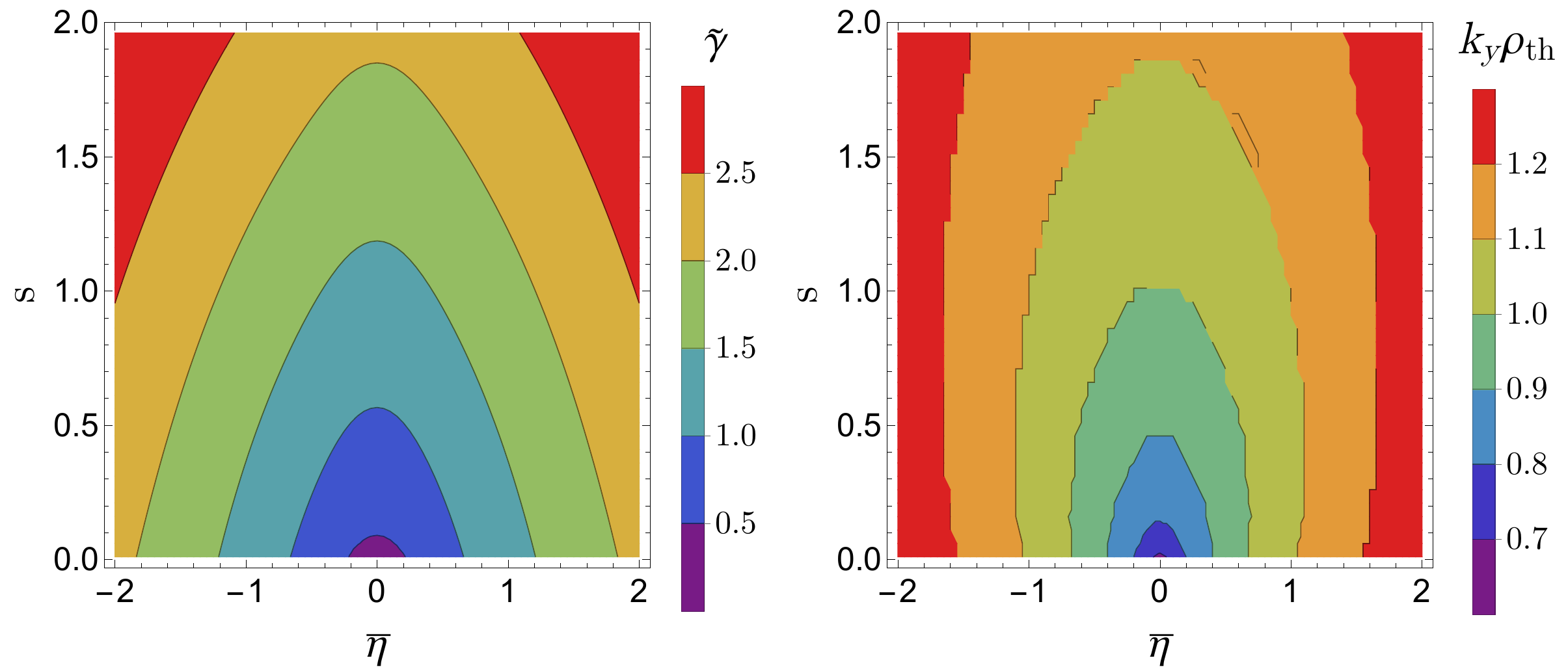}
    \includegraphics[width=.99\textwidth]{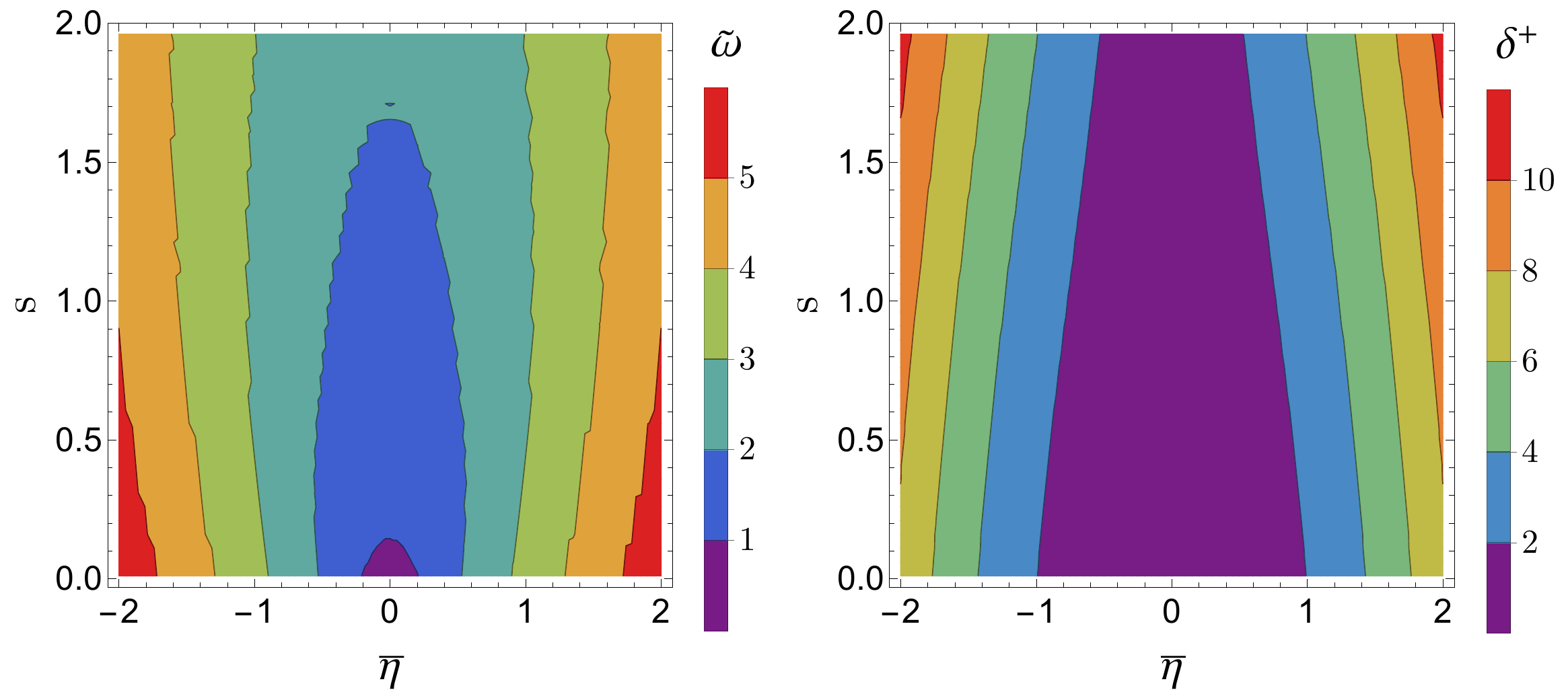}
    \caption{Solution of the ITG dispersion relation \cref{eq:analyticalDR1} for the case $\eta=3$ and $a/L_n=1$, namely the peak growth rate $\gamma$ (top left), the corresponding $k_y \rho_{th}$ (top right), real frequency $\omega$ (bottom left) and the eigenfunction width $\delta^+$ (bottom right). The frequencies $\gamma$ and $\omega$ are normalized to the transit frequency $\omega_t=2 \pi v_{th}/L$ while $\overline \eta$ is normalized to $L/2\pi$.}
    \label{fig:analyticalDR2}
\end{figure}

Besides a scan over $a/L_n$ and $a/L_T$, the dispersion relation in \cref{eq:analyticalDR1} can be used to assess the impact of $s$ and $\eta_b$ (which are quantities directly related to the near-axis expansion parameters) in the stability of the ITG mode.
In \cref{fig:analyticalDR2} a parameter scan over $s$ and $\eta_b$ is shown for a fixed value of $\eta=3$ and $a/L_n=1$.
It is seen that the results are symmetric with respect to the sign of $\overline \eta$ and that all four parameters increase as either $s$ or $\overline \eta$ are increased.

\begin{figure}
    \centering
    \includegraphics[width=.7\textwidth]{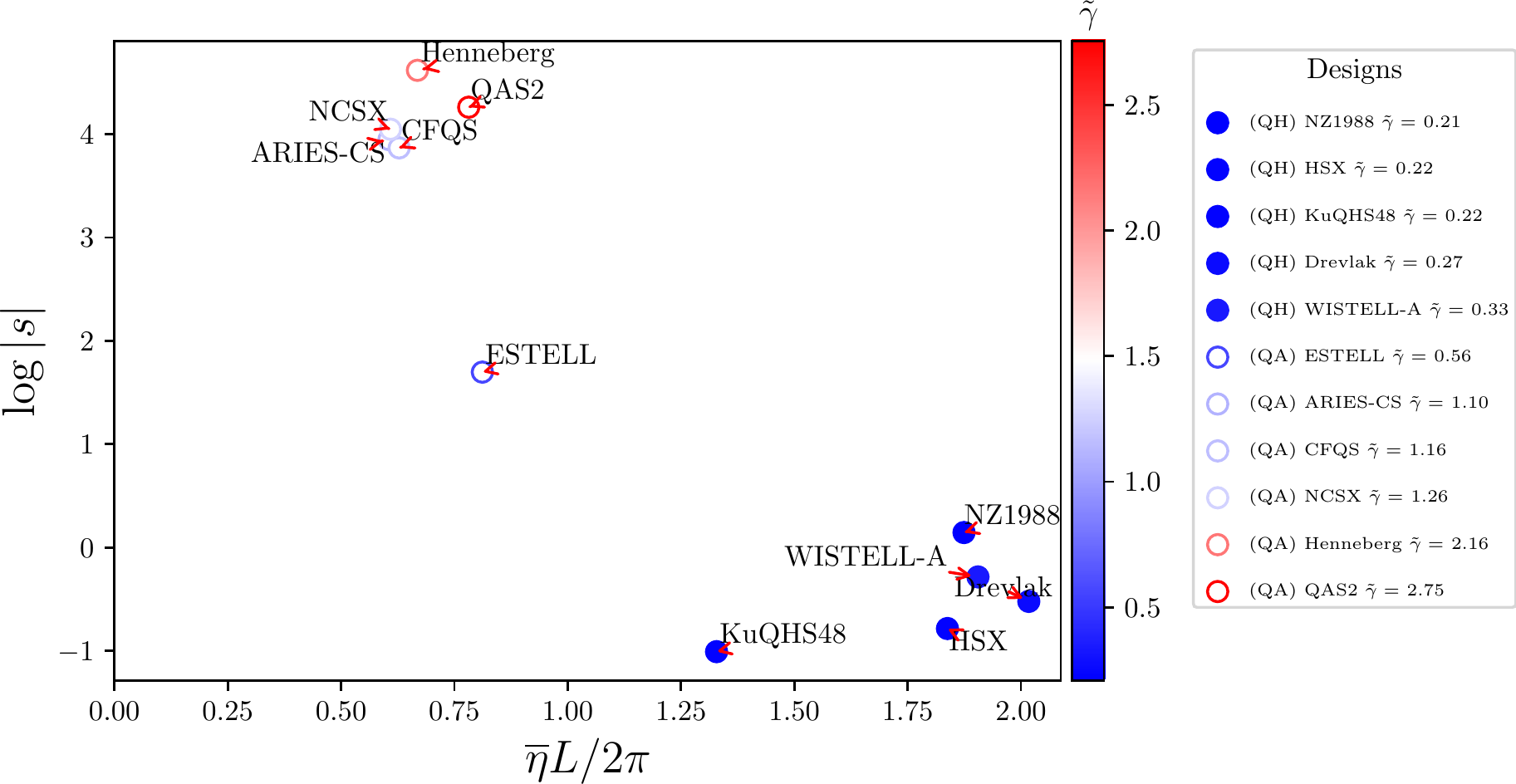}
    \caption{Effective shear parameter $s$ and $
    \overline \eta$ for each of the eleven quasisymmetric designs considered in this study computed using \cref{eq:shighlylocalized}. The designs with hollow circles are quasi-axisymmetric, while the ones with filled circles are quasi-helically symmetric. The color represents the growth rate $\gamma$ computed from the dispersion relation in \cref{eq:analyticalDR1} with $\tilde L_n=0.6$ and $\eta=2.5$.}
    \label{fig:sforQS}
\end{figure}

Finally, we apply the dispersion relation in \cref{eq:analyticalDR1} to the eleven quasisymmetric designs used in this study.
For each design, the value of $\overline \eta$ is read from Table 1 in Ref. \cite{Jorge2020c}, the axis parameters are read from each VMEC file (used to compute the curvature and torsion) and $\sigma$ is calculated using \cref{eq:sigma}, yielding the values of $\iota_0$ in Table 1 of Ref. \cite{Jorge2020c}.
The values of the effective shear parameter $s$ for each design is shown in \cref{fig:sforQS}.
It is found that quasi-helically symmetric stellarators have, in general, values of $s$ that are one to two orders of magnitude above the ones of quasi-axisymmetric stellarators.
We also show in \cref{fig:sforQS} the resulting growth rate estimate and the corresponding values of $\overline \eta L/2\pi$ and $s$.
The designs with $\overline \eta L/2\pi<1$ are quasi-axisymmetric, while the ones with $\overline \eta L/2\pi>1$ are quasi-helically symmetric.
The results in \cref{fig:sforQS} suggest that quasi-helically symmetric configurations are prone to having higher values of $\overline \eta L/2\pi$, lower values of $s$ and lower growth rates (normalized to $\omega_t$) when compared with quasi-axisymmetric designs.

\subsection{Bounce-averaged metric}

We now show how a bounce-averaged metric $Q{\textsubscript{bounce}}$, which takes into account the role of good and bad curvature regions in the particle motion, can be simplified using the near-axis expansion.
This parameter, although more focused on TEM rather than ITG, allows us to show how metrics based on bounce-averaged quantities, in the near-axis expansion formalism, can be reduced to one-dimensional elliptic integrals.
Here, we employ the metric used in Ref. \cite{Proll2015a}
\begin{equation}
    Q{\textsubscript{bounce}}=-\int_{1/B{\textsubscript{max}}}^{1/B{\textsubscript{min}}}\overline \omega_d(\lambda) d\lambda,
\label{eq:Qbounce}
\end{equation}
where $\overline \omega_d(\lambda)$ is the bounce-averaged drift-frequency
\begin{equation}
    \overline \omega_d(\lambda)=\frac{\int_{l_1}^{l_2} \tilde \omega_d {dl}/{v_\parallel}}{\int_{l_1}^{l_2} {dl}/{ v_\parallel}}.
\end{equation}
The minimization of this metric aims at distributing a large fraction of trapped particles over a favorable bounce-averaged curvature.
Although \cref{eq:Qbounce} has been used in Ref. \cite{Proll2015a} to stabilize TEMs, similar metrics are often employed in the analysis of tokamak and stellarator microstability \cite{Beeke2020}.

\begin{figure}
    \centering
    \includegraphics[width=.56\textwidth]{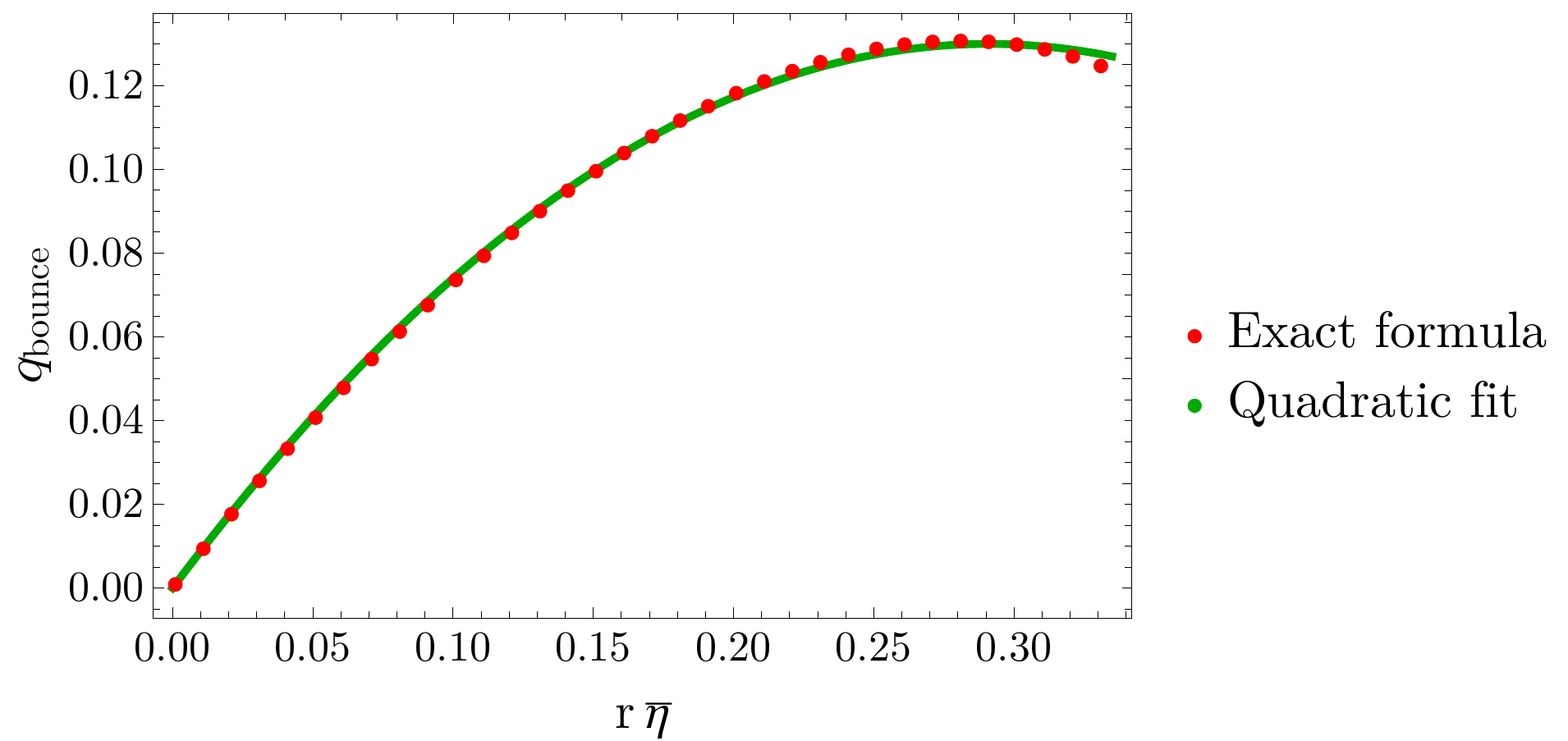}
    \includegraphics[width=.33\textwidth]{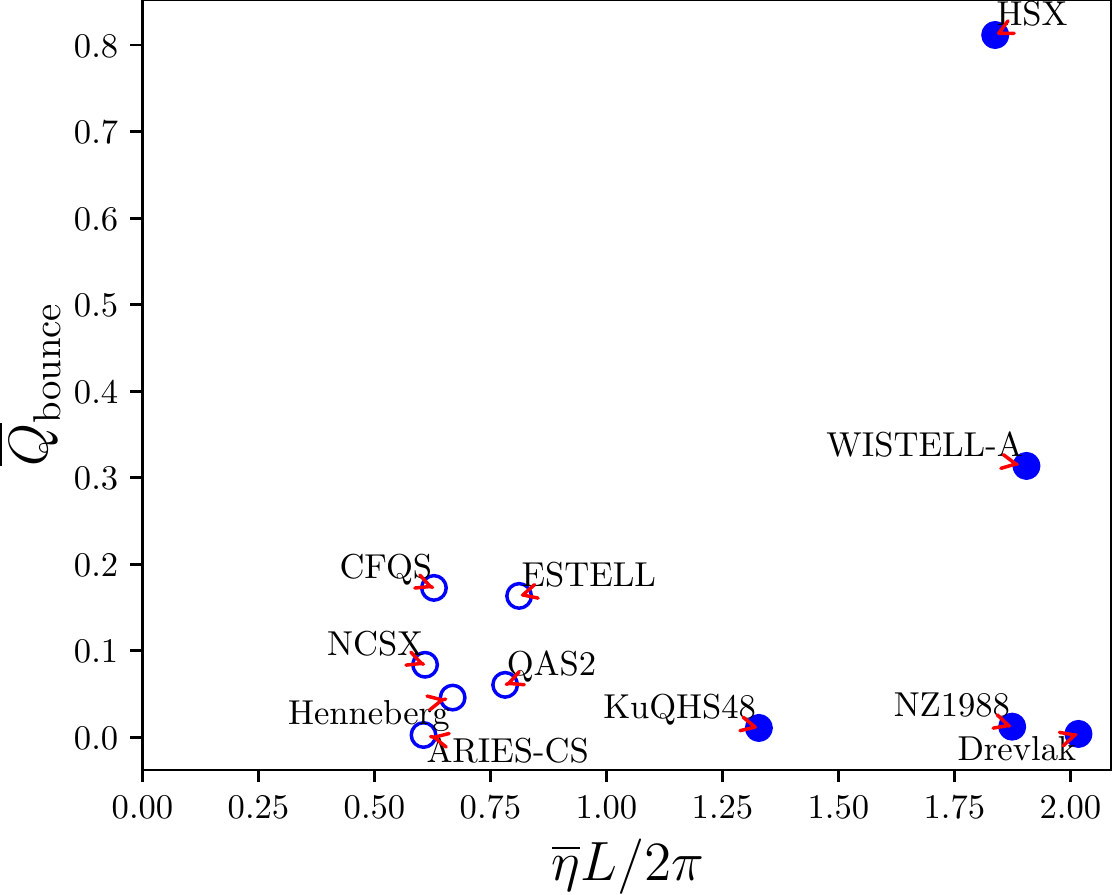}
    \caption{Left: Analytical form and the corresponding quadratic fit $0.893 x - 1.53 x^2$ for the function $q{\textsubscript{bounce}}$ entering the heat flux proxy $Q\textsubscript{bounce}$ in \cref{eq:qbounce2}. Right: $Q\textsubscript{bounce}$ normalized to $k_y \rho_{th} v^2/B_0 v_{th}$ as a function of $\overline \eta L/2\pi$ for the eleven configurations considered in this study and $r=0.1$. The designs with hollow circles are quasi-axisymmetric, while the ones with filled circles are quasi-helically symmetric.}
    \label{fig:qbounce}
\end{figure}

In the near-axis expansion formalism, at first order in $r$, the metric $Q{\textsubscript{bounce}}$ in \cref{eq:Qbounce} can be simplified by noting that $v_\parallel=\pm v \sqrt{1-\lambda B}$ with $B=B_0(1+r \overline \eta \cos \vartheta)$.
The resulting integral in $l$ is rewritten by parametrizing the field-line coordinate $l$ using $d\vartheta/dl=\mathbf b \cdot \nabla \vartheta = |\iota_N| B/(G+\iota I)\simeq (2\pi/L)|\iota_N|(1+r \overline \eta \cos \vartheta)$, yielding the bounce points $l_1$ and $l_2=-l_1$ at $\vartheta_1=-\vartheta_2=-2 \arcsin(1/a)$ with $a^2=-2 r \overline \eta \lambda B_0/([1-\lambda B_0(1+r \overline \eta)]>1$.
By performing a variable substitution $x=a \sin \vartheta/2$ and assuming $s_\psi=1$, the following expression for $Q{\textsubscript{bounce}}$ is found
\begin{equation}
    Q{\textsubscript{bounce}}=\frac{8}{\sqrt{2}}\frac{k_y \rho_{th}\overline \eta v^2}{v_{th}}\int_{[B_0(1+r \overline \eta)]^{-1}}^{[B_0(1-r \overline \eta)]^{-1}}d\lambda \sqrt{2-\lambda B_0(1+r \overline \eta)}\frac{\Pi_c(1/c,t,1/a)}{\Pi(t,1/a)},
\label{eq:analyticalQbounce}
\end{equation}
with $\Pi_c$ and $\Pi$ the elliptic integrals
\begin{equation}
    \Pi_c(1/c,t,1/a)=\int_0^1\frac{(1-2x^2/a^2)\sqrt{1-x^2/c^2}}{\sqrt{1-x^2}\sqrt{1-x^2/a^2}}\frac{dx}{1-t x^2},
\end{equation}
and
\begin{equation}
    \Pi(t,1/a)=\int_0^1\frac{1}{\sqrt{1-x^2}\sqrt{1-x^2/a^2}}\frac{dx}{1-t x^2},
\end{equation}
respectively.
In \cref{eq:analyticalQbounce}, $t=2 r \overline \eta/[(1+r \overline \eta)a^2]$ and $c^2=[2-\lambda B_0(1+r \overline \eta)]/[1-\lambda B_0(1+r \overline \eta)]$.
Noting that the integrand of \cref{eq:analyticalQbounce} is a function of $r \overline \eta$ and $\lambda B_0$ only, we rewrite $Q{\textsubscript{bounce}}$ by turning the integral over $\lambda$ to an integral over $\lambda B_0$, yielding.
\begin{equation}
     Q{\textsubscript{bounce}}=\frac{8}{\sqrt{2}}\frac{k_y \rho_{th}\overline \eta v^2}{v_{th}B_0}q{\textsubscript{bounce}}(r \overline \eta).
\label{eq:qbounce2}
\end{equation}
As shown in \cref{fig:qbounce}, the function $q{\textsubscript{bounce}}$ can be reasonably approximated by a quadratic function of $r \overline \eta$, leading to the following approximate model for $Q{\textsubscript{bounce}}$ close to the magnetic axis
\begin{equation}
     Q{\textsubscript{bounce}}\simeq 5.051\frac{k_y \rho_{th}\overline \eta v^2}{B_0v_{th}}r \overline \eta[1-1.172 (r \overline \eta)^2 ].
\label{eq:qbouncefinal}
\end{equation}
The positive sign of $Q{\textsubscript{bounce}}$ in \cref{eq:qbouncefinal} suggests that, for quasisymmetric stellarators, the majority of trapped particles experience bad curvature.
Furthermore, \cref{eq:qbouncefinal} shows that there is little freedom to optimize quasi-symmetric stellarators for trapped electron modes besides varying the value of $\overline \eta$.
Finally, an estimate for $Q\textsubscript{bounce}$ for the eleven quasisymmetric designs considered in this study using \cref{eq:qbouncefinal} is shown in \cref{fig:qbounce}.
We note that in \cref{fig:qbounce} there are are no systematic differences between quasi-axisymmetric and quasi-helically symmetric stellarators.

\subsection{Reduction of Turbulent Transport via Quasilinear Modelling}

Using a quasilinear model, in Ref. \cite{Mynick2011,Mynick2010} the following proxy function $Q{\textsubscript{prox}}$ for ITG transport was derived
\begin{equation}
    Q{\textsubscript{prox}}=-\frac{c_D}{4 \overline B} \frac{\gamma}{k_\psi^2}\frac{|\nabla \psi|^2}{\psi^2}\frac{T}{L_T},
\label{eq:qprox}
\end{equation}
with $c_D$ a constant, $\overline B$ a reference magnetic field and $\gamma$ the growth rate resulting from the following simplified ITG dispersion relation
\begin{equation}
    \gamma \simeq \omega_* L_n |\tau_{T} \kappa_1(\kappa_p-\kappa{\textsubscript{cr}})|^{1/2}H(\kappa_p-\kappa{\textsubscript{cr}})H(-\kappa_1),
\label{eq:simplegrowthrateITG}
\end{equation}
where $\kappa_1$ is the radial component of the curvature vector $\mathbf b \cdot \nabla \mathbf b$, $\kappa_p=(1+\eta)/L_n$ is the pressure gradient length and $\kappa{\textsubscript{cr}}$ a critical pressure gradient length stemming from a contribution of the ITG slab branch.
Minimization of this proxy was shown to be an effective way to reduce turbulent transport levels, raising the prospect of a new class of stellarators with improved overall confinement \cite{Mynick2010}.

Using the expression for $|\nabla\psi|^2$ in \cref{eq:g3} and choosing $\overline B=B_0$, we can write the proxy function in \cref{eq:qprox} as
\begin{equation}
 Q{\textsubscript{prox}}=Q_0\frac{\sqrt{\overline \eta \cos \vartheta}}{B_0}H(\cos \vartheta)\left[\frac{\overline \eta^2}{\kappa^2}\sin^2 \vartheta + \frac{\kappa^2}{\overline \eta^2}(\cos \vartheta-\sigma \sin \vartheta)^2\right]
\label{eq:qprox2}
\end{equation}
where
\begin{equation}
    Q_0=-\frac{c_D}{k_\psi^2}\frac{T}{L_T}\frac{k_y\rho_{th} c_s}{a}\sqrt{\frac{\tau_{T}}{L_n}}\sqrt{1+\eta+\kappa{\textsubscript{cr}}L_n}H(1+\eta+\kappa{\textsubscript{cr}}L_n).
\label{eq:Q0func}
\end{equation}
%
%
The function $Q{\textsubscript{prox}}$ in \cref{eq:qprox2} can then be used as a proxy for the reduction of ITG turbulence near the axis of quasisymmetric stellarators.

\begin{figure}
    \centering
    \includegraphics[width=.6\textwidth]{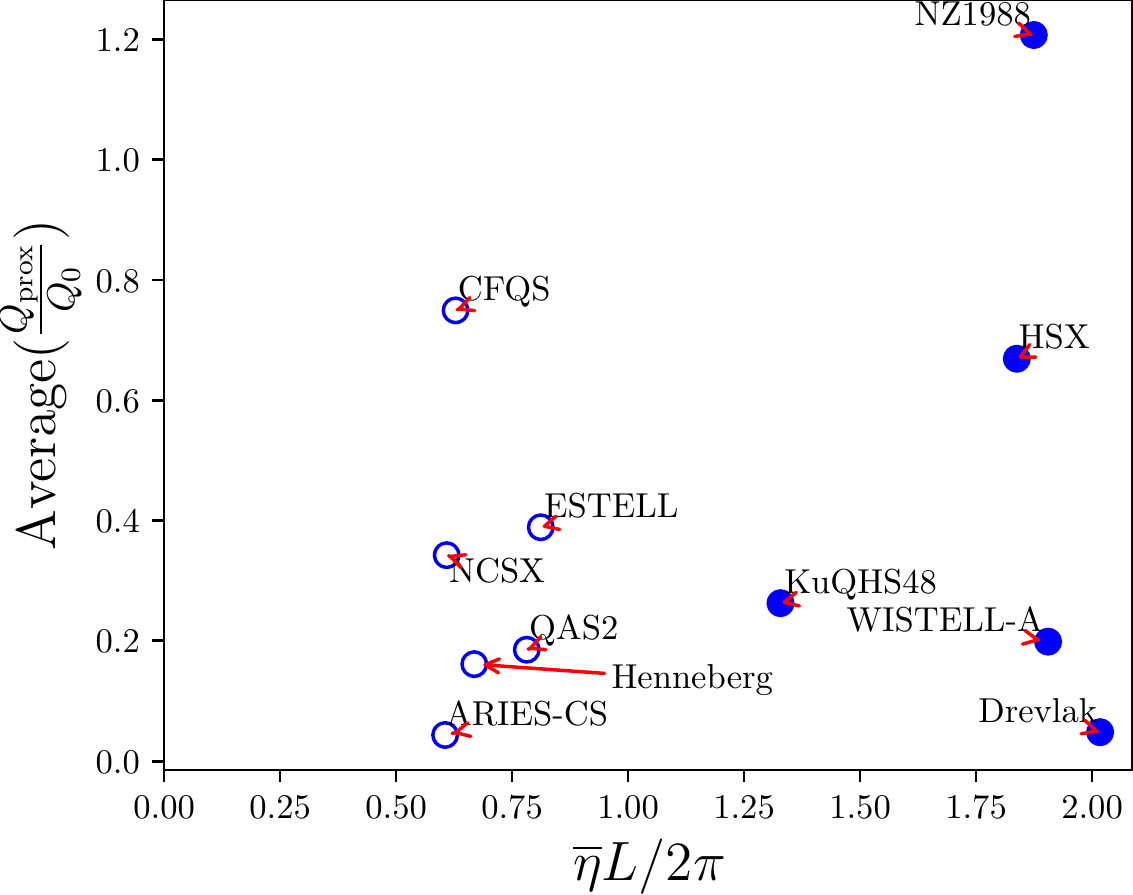}
    \caption{Normalized proxy function $Q{\textsubscript{prox}}/Q_0$ computed using \cref{eq:qprox2} for each of the eleven quasisymmetric designs considered in this study averaged along a field line with $\alpha=0$. The designs with hollow circles are quasi-axisymmetric, while the ones with filled circles are quasi-helically symmetric.}
    \label{fig:QproxMeanforQS}
\end{figure}

In \cref{fig:QproxMeanforQS}, the normalized proxy function $Q{\textsubscript{prox}}/Q_0$ is evaluated for the configurations considered in this study, averaged along a field line with $\alpha=0$.
Due to the fact that the parameter $Q_0$ in \cref{eq:Q0func} is not directly dependent on the geometry used, it is considered constant in \cref{fig:QproxMeanforQS}.
The resulting averaged proxy function in \cref{fig:QproxMeanforQS} suggests that there is not a significant difference in $Q$ between quasi-axisymmetric and quasi-helically symmetric configurations.
Indeed, the range of values of $Q\textsubscript{prox}$ in \cref{fig:QproxMeanforQS} can vary two orders of magnitude in each set of quasisymmetric configurations.
We note that this analysis assumes $Q_0$ is constant across all stellarator designs, which may not be true in practice due to possible variations of characteristic wave-vectors and critical gradients between configurations.



\section{Numerical Benchmark with a Near-Axis Geometry}
\label{sec:nabenchmark}

\begin{figure}
    \centering
    \includegraphics[width=.485\textwidth]{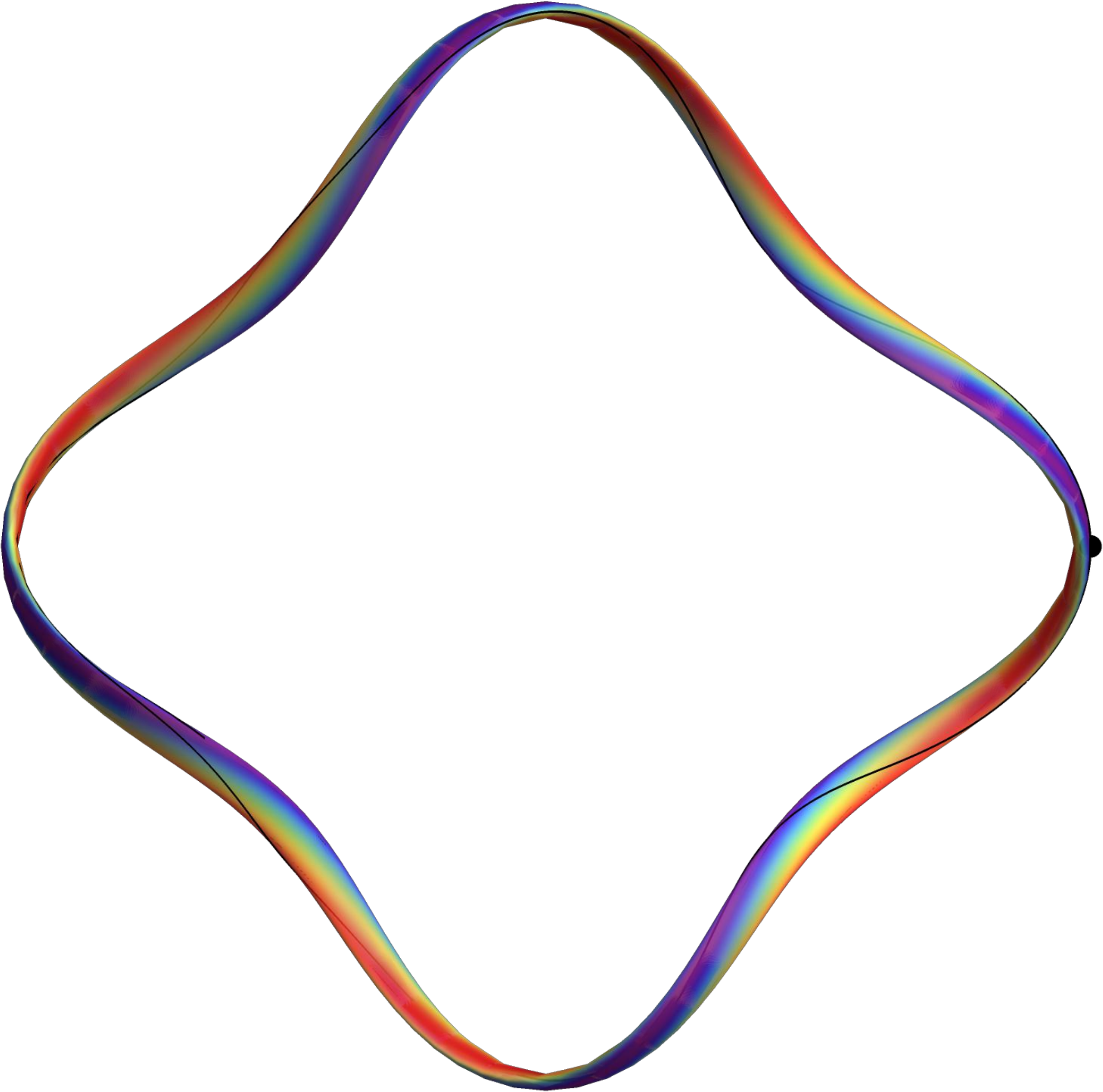}
    \includegraphics[width=.505\textwidth]{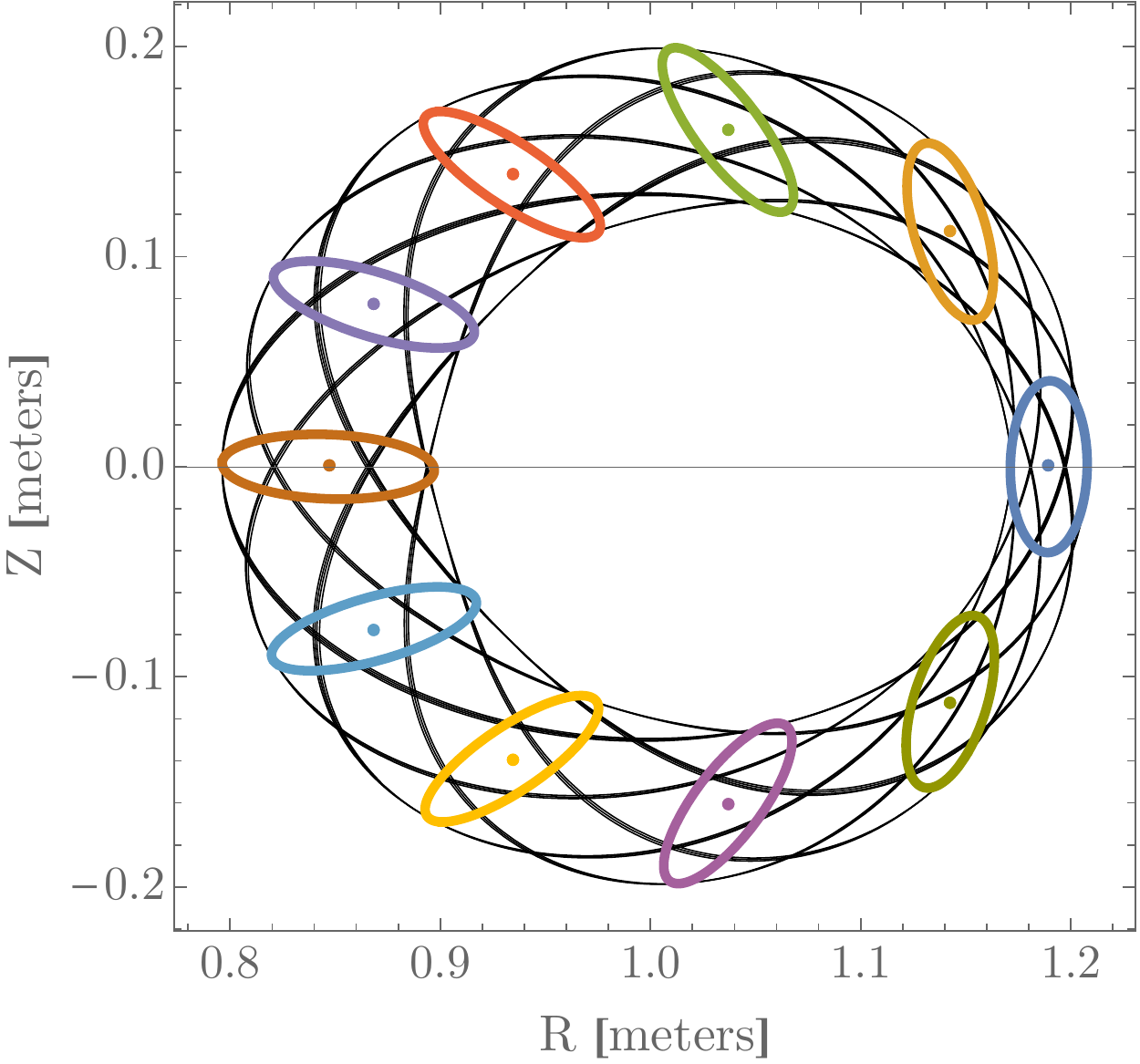}
    \includegraphics[width=.85\textwidth]{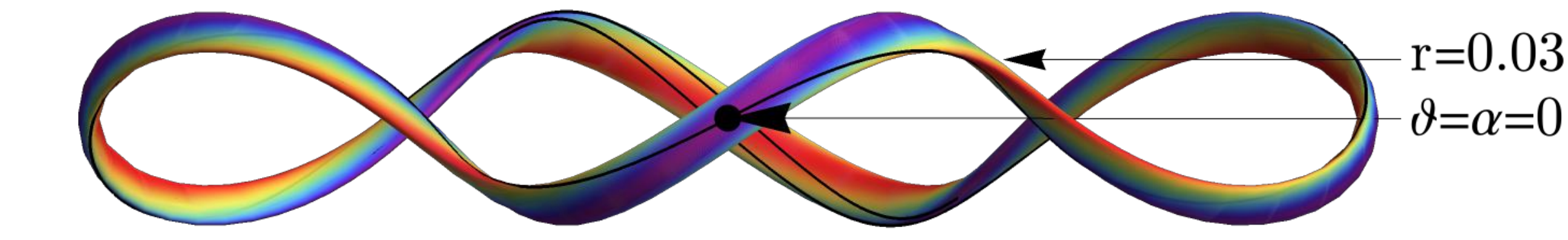}
    \caption{Top left: top view of a surface of constant toroidal flux $\psi$ for the quasi-helically symmetric NAQS stellarator (Section 5.4 of Ref. \cite{Landreman2019b}) at first order in $\epsilon$. Colour indicates $B$ on the outermost flux surface. An example field-line with $-8\pi<\vartheta<8\pi$, $\alpha=0$ and $r=0.03$ is shown in black together with the location of $\vartheta=0$. Top right: ten poloidal planes and their respective axis location, together with a field line with $-40\pi<\vartheta<40\pi$ and $\alpha=0$. Bottom: side view of the toroidal flux surface.}
    \label{fig:my_label}
\end{figure}

\begin{figure}
    \centering
    \includegraphics[width=0.9\textwidth]{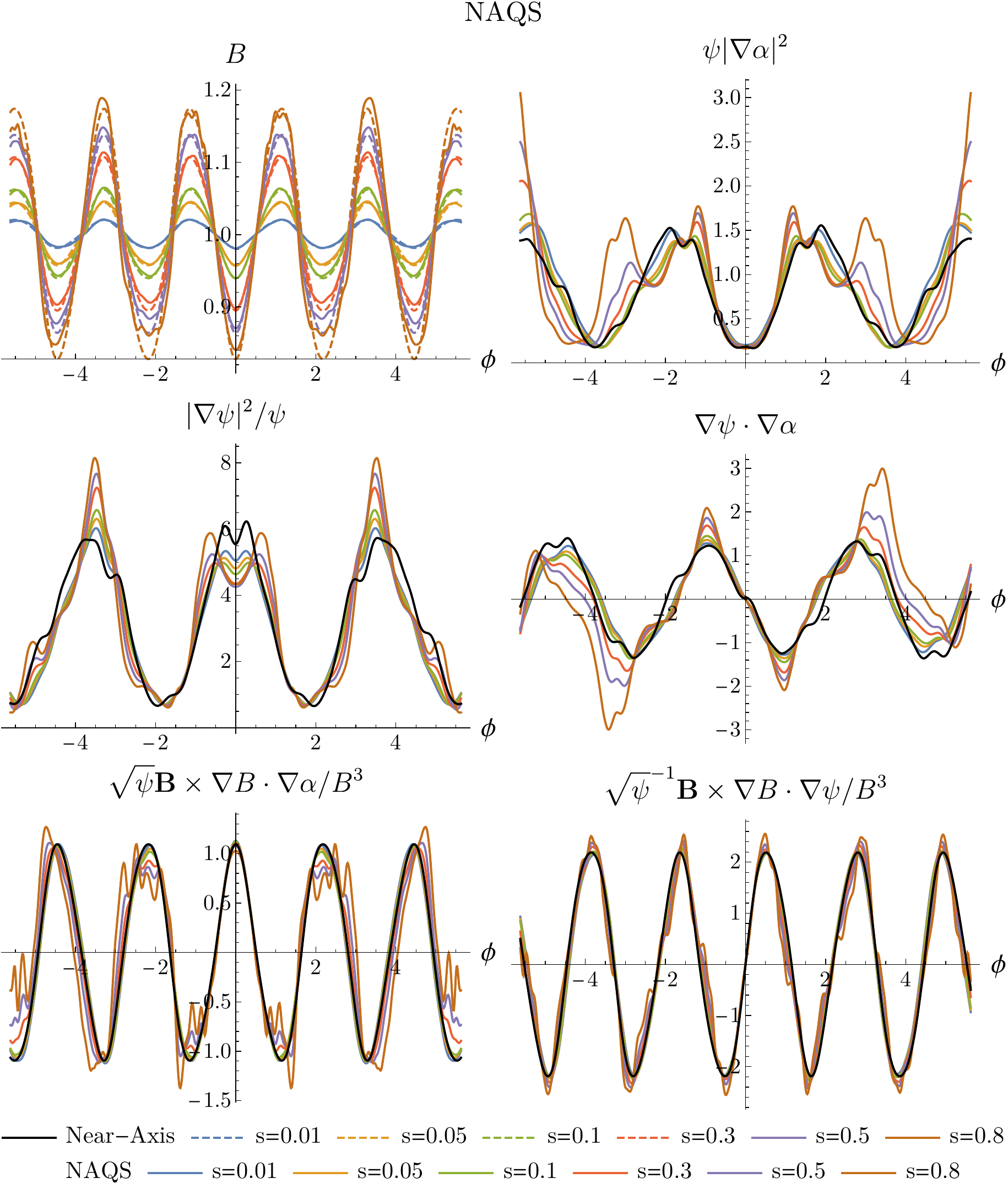}
    \caption{Comparison between the fitted and the original near-axis geometry coefficients for the NAQS configuration of Section 5.4 of Ref \cite{Landreman2019a} along the cylindrical toroidal angle $\phi$ at several radial locations $s={\psi/\psi_a}$ with $\psi_a$ the toroidal flux at the plasma boundary. All quantities are expressed in SI units.}
    \label{fig:naqs}
\end{figure}

Before assessing the linear stability of the eleven quasisymmetric designs and its comparison with the near-axis expansion, we start with a comparison between a solution of the near-axis equations and a VMEC configuration based on this solution.
This allows us to examine the details of the input parameters and respective convergence tests, the comparison between eigenfunctions, the dependence of the growth rate on the perpendicular wave-vector and the variation of the maximum growth rate with the density and temperature gradient scale lengths before presenting the main results in the next section.
For this analysis, we take the quasi-helically symmetric solution of Section 5.4 in Ref. \cite{Landreman2019a} (hereafter referred to as NAQS) and generate a corresponding VMEC output file with a plasma boundary at $\psi=0.002$ T m$^2$, $B_0=1.0$ T and $\overline \eta = 2.25$, with a resulting rotational transform on-axis of $\iota_0=1.93$ when solving \cref{eq:sigma}.
The comparison between the geometrical quantities is shown in \cref{fig:naqs} using flux surfaces with $s=\psi/\psi_a$ (with $\psi_a$ the toroidal flux at the plasma boundary) ranging from 0.01 to 0.8 and a field line with $\alpha=0$.
Only six of the eight quantities are shown as \cref{eq:g1,eq:g2} and \cref{eq:g6,eq:g8} are independent in the limit of near-axis quasisymmetric fields used here.
Although, overall, there is good agreement, some differences between the two methods arise, even at $s=0.01$.
The fact that a magnetic field solution obtained from setting the boundary value of $\mathbf B$ from a first-order near-axis solution at a particular radius $r$ is not perfectly quasisymmetric is a phenomenon that has been recently been observed in Ref. \cite{Landreman2019a} and addressed using second-order solutions in Ref. \cite{Landreman2019b}.
%
%

\begin{figure}
    \centering
    \includegraphics[trim={0 1.3cm 4.1cm 0},clip,width=0.32\textwidth]{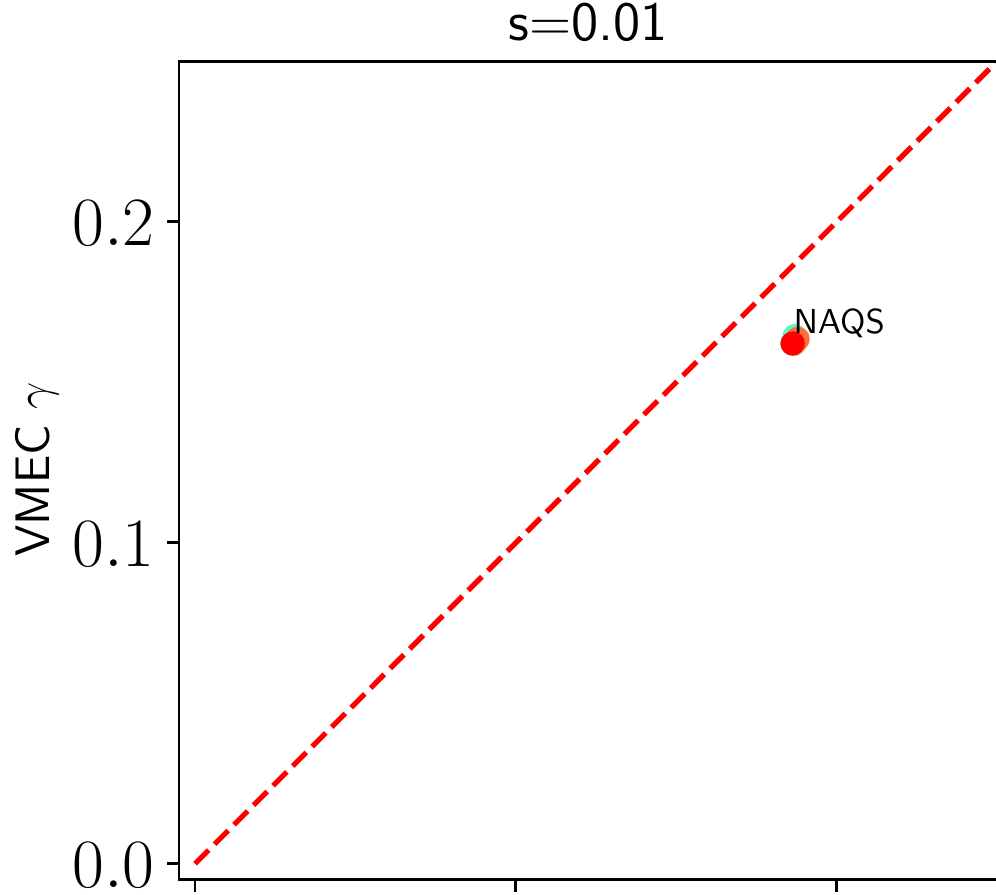}
    \includegraphics[trim={1.7cm 1.3cm 4.1cm 0},clip,width=0.266\textwidth]{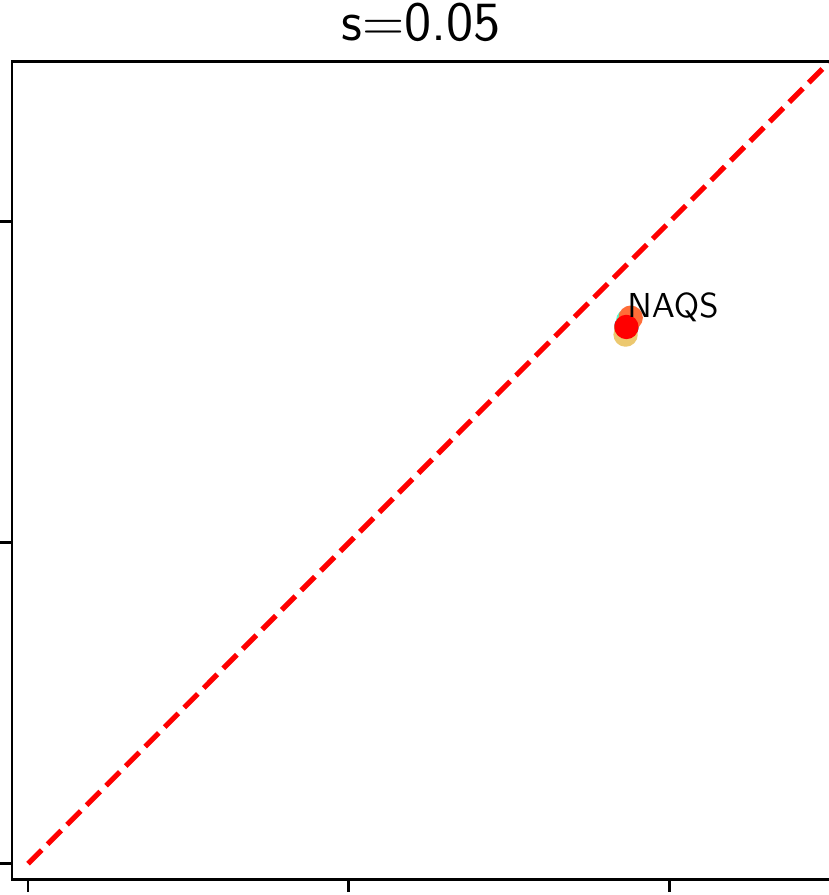}
    \includegraphics[trim={1.7cm 1.3cm 0 0},clip,width=0.397\textwidth]{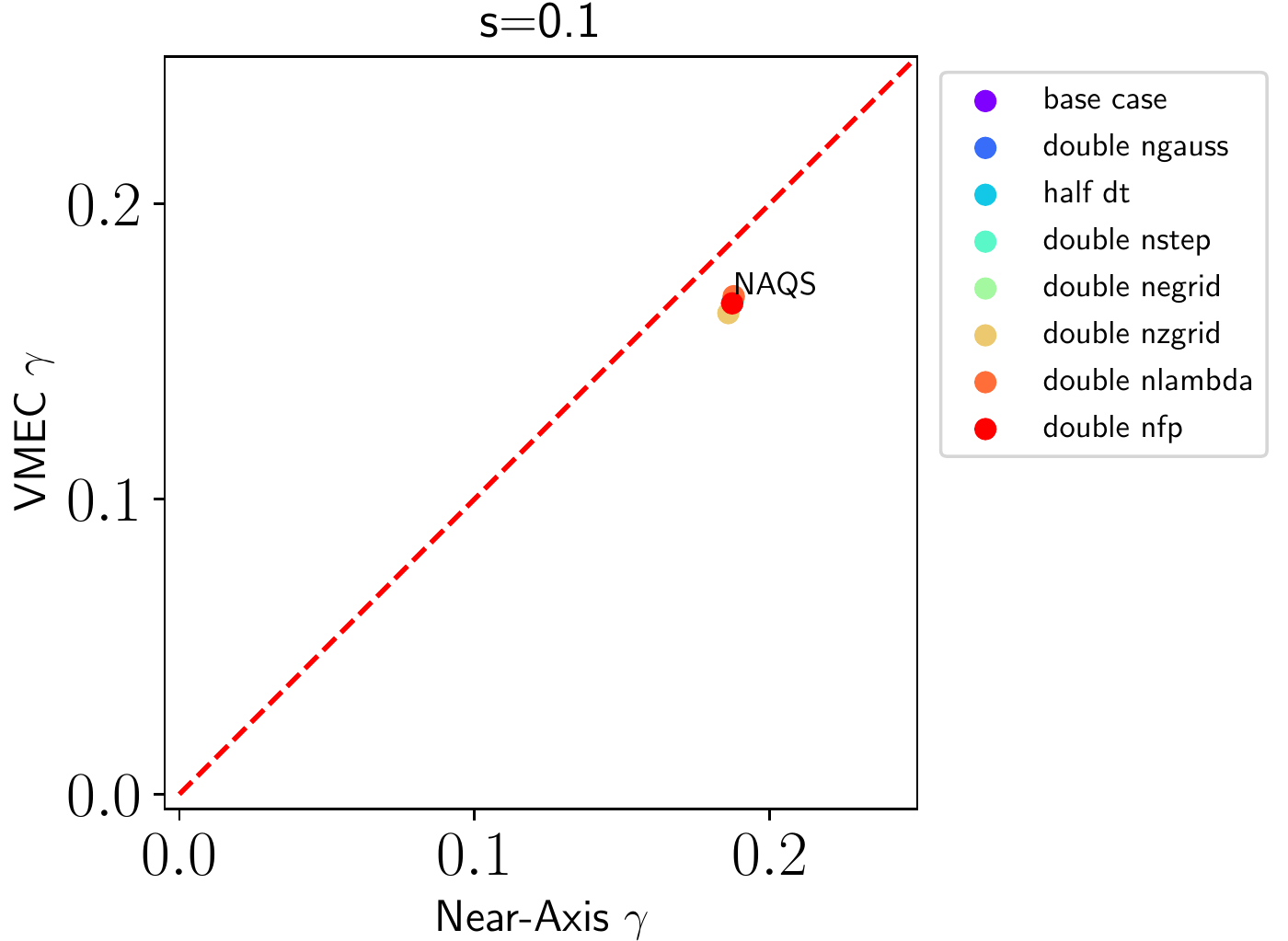}
    \includegraphics[trim={0 0 4.1cm 0},clip,width=0.32\textwidth]{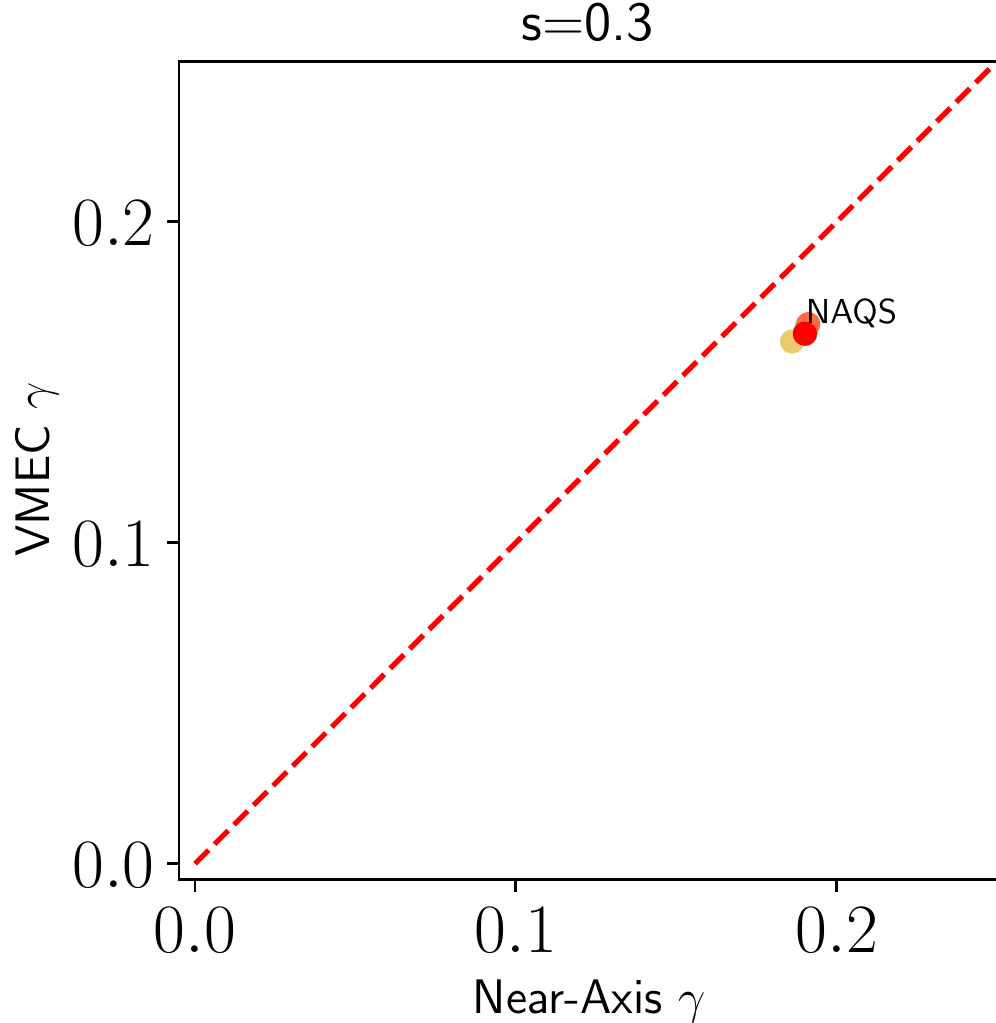}
    \includegraphics[trim={1.6cm 0 4.1cm 0},clip,width=0.266\textwidth]{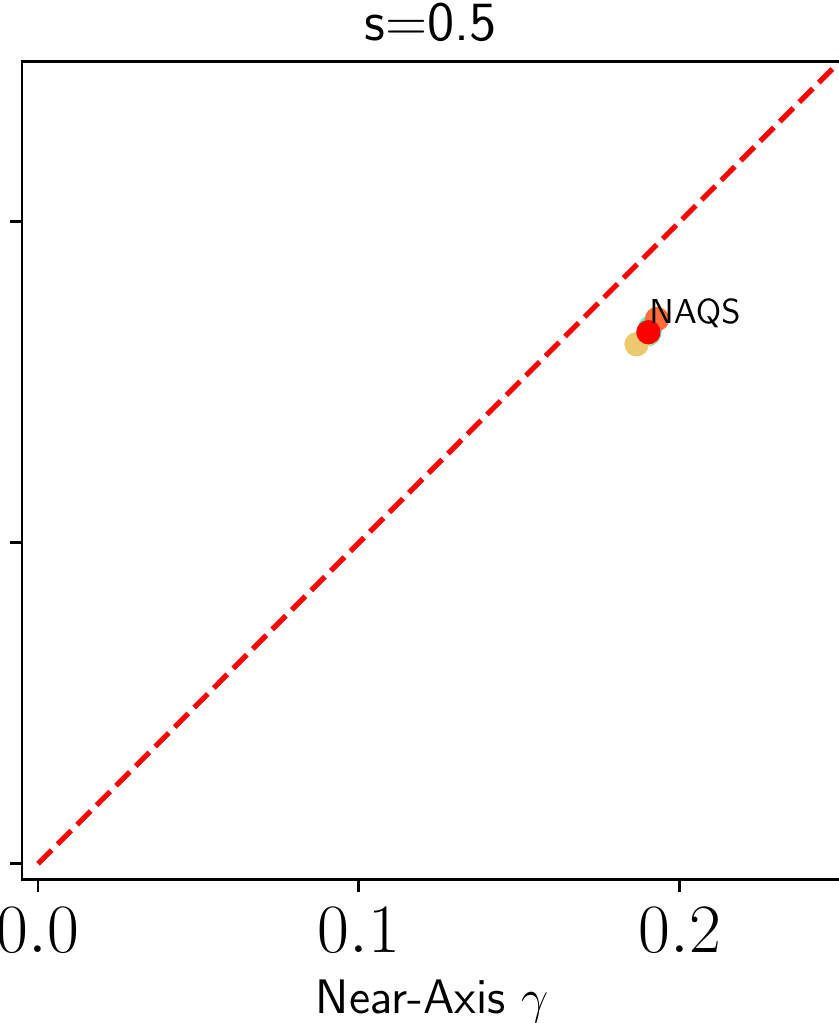}
    \includegraphics[trim={1.6cm 0 0 0},clip,width=0.397\textwidth]{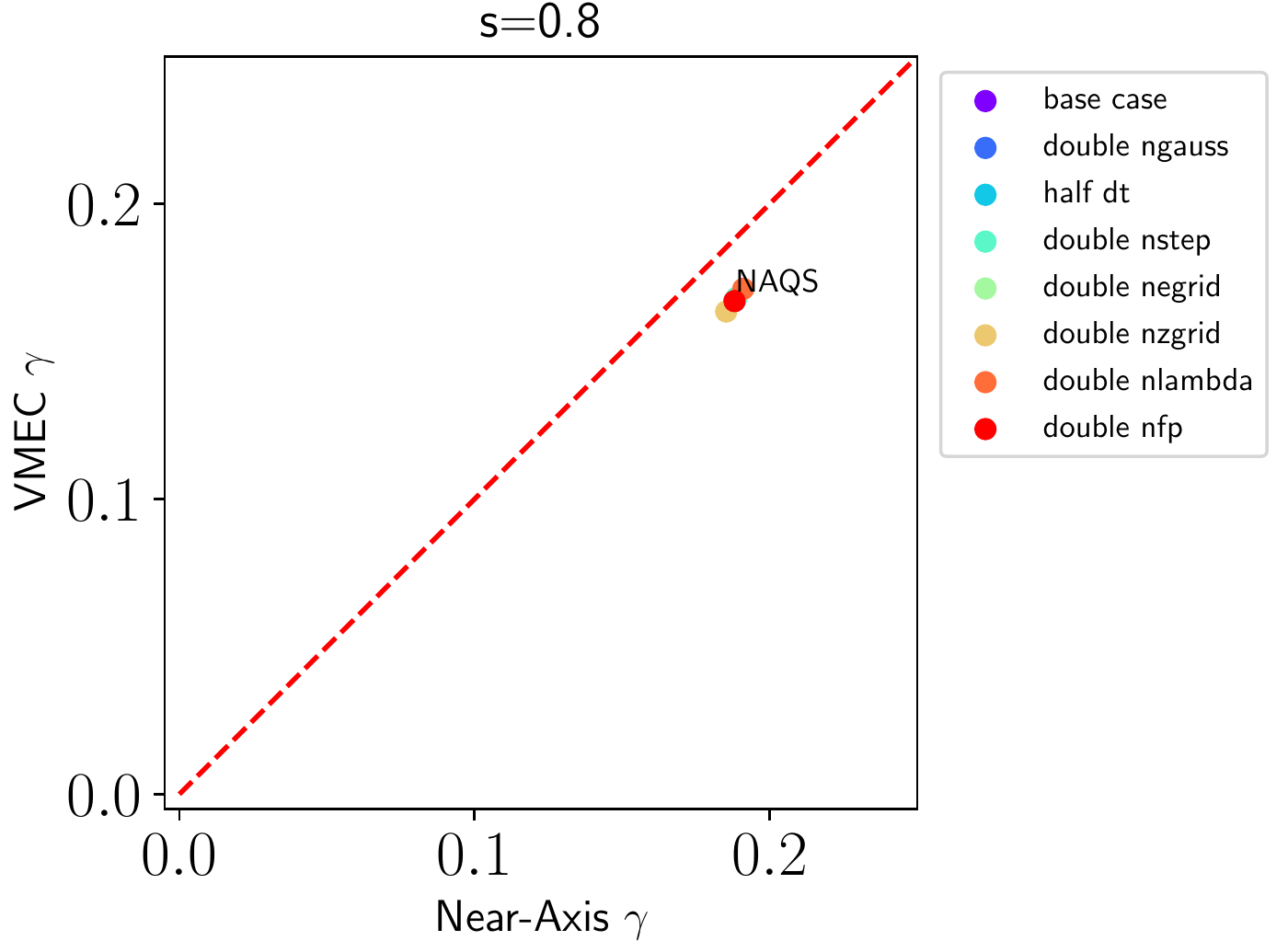}
    \caption{Growth rate convergence test of the base case scenario for the VMEC and near-axis NAQS geometries using the gyrokinetic GS2 code at different radii $s=\psi/\psi_a$ and $k_y \rho=1$, $a/L_T=3$ and $a/L_n=1$.}
    \label{fig:NAQSconv}
\end{figure}

We proceed by using the GS2 code to assess the properties of ITG modes in both the VMEC and the near-axis solution.
For each simulation, the temporal evolution of $|\hat \phi|^2$ is fitted to an exponential of the form $e^{2 \gamma t}$ with $\gamma$ the growth rate.
As a base case scenario, we use \textit{ngauss}=3 (half of the number of untrapped pitch-angles moving in one direction along field line), \textit{deltat}=0.4 (the temporal increment between time steps), \textit{nstep=150} (the total number of time steps), \textit{ngrid}=10 (the total number of energy points), \textit{nzgrid}=100 (the total number of points in $z$ along the field line), \textit{nlambda}=24 (number of trapped-pitch angles along the field line) and \textit{nfp}=4 (number of field periods, a measure of the total length of the field line).
In \cref{fig:NAQSconv}, a convergence test is performed where the base case growth rate for both the VMEC and the near-axis geometries is computed, along with the growth rate resulting from doubling the resolution along each parameter.
As physical input parameters, we use $k_y \rho = 1.0$, $a/L_T=3$ and $a/L_n=1$.
A base case scenario is considered to be converged if the scatter plot cluster of all simulations has a deviation smaller than 5\%, a criterion that is satisfied for all radii in \cref{fig:NAQSconv}.

\begin{figure}
    \centering
    \includegraphics[width=.538\textwidth]{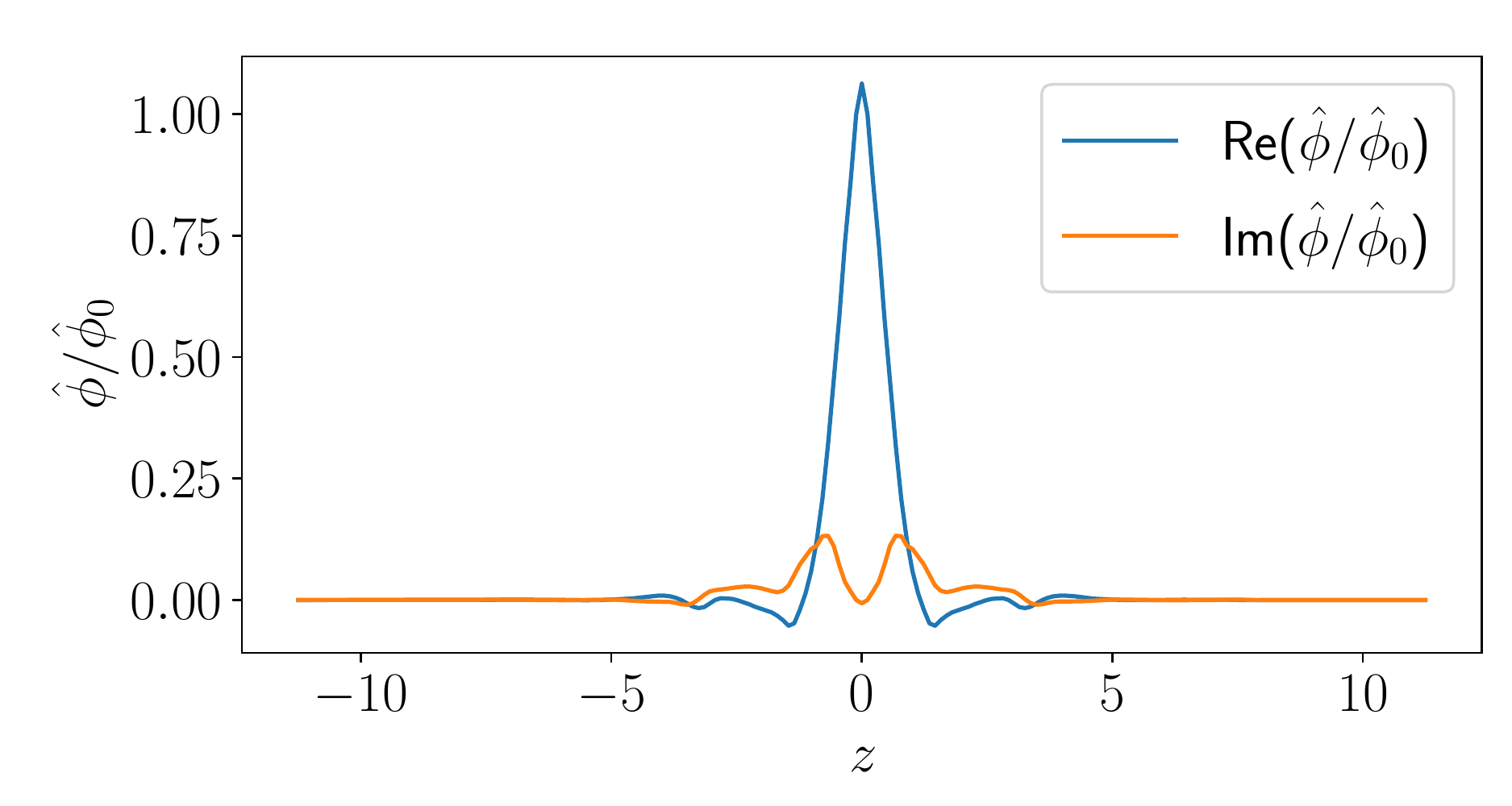}
    \includegraphics[trim={3.01cm 0 0 0},clip,width=.452\textwidth]{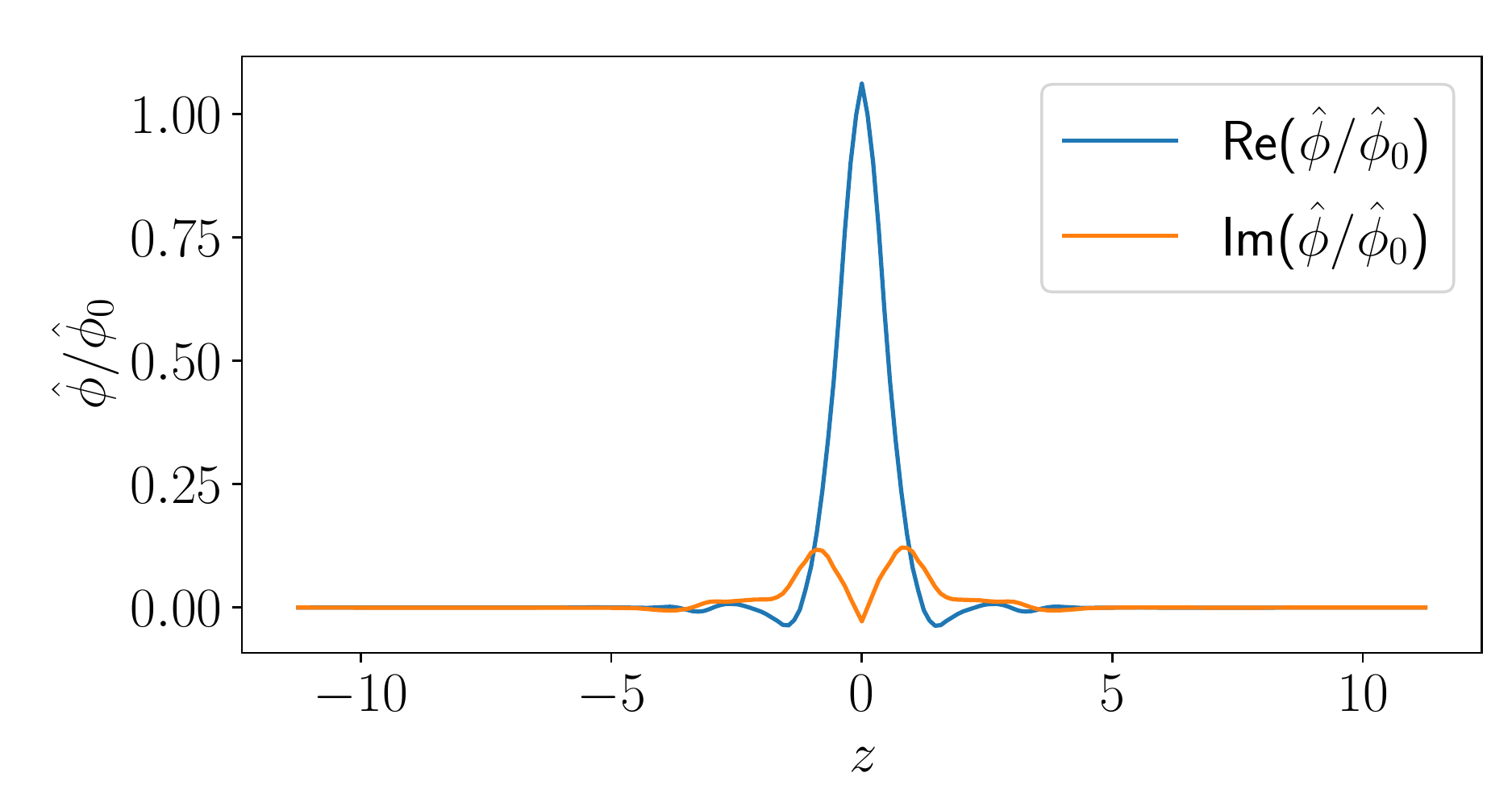}
    \caption{Real and imaginary eigenfunctions $\hat \phi$ at the end of the simulation period for both the VMEC (left) and near-axis (right) geometries at $s=0.3$. Each eigenfunction is normalized to its value at $z=0$.}
    \label{fig:NAQSeigenfuncs}
\end{figure}

For the base case scenario at $s=0.3$, we show in \cref{fig:NAQSeigenfuncs} the resulting real and imaginary electrostatic potential eigenfunctions $\hat \phi$ at the end of the simulation period for both the VMEC (left) and near-axis (right) geometries.
Overall, both eigenfunctions look practically indistinguishable from each other for every simulation of the convergence test.
Finally, we note that these are smooth functions of the parallel coordinate $z$ and reach insignificant values before the end of the simulation domain, which provides additional evidence for the identification of a converged simulation scenario.

\begin{figure}
    \centering
    \includegraphics[width=0.6\textwidth]{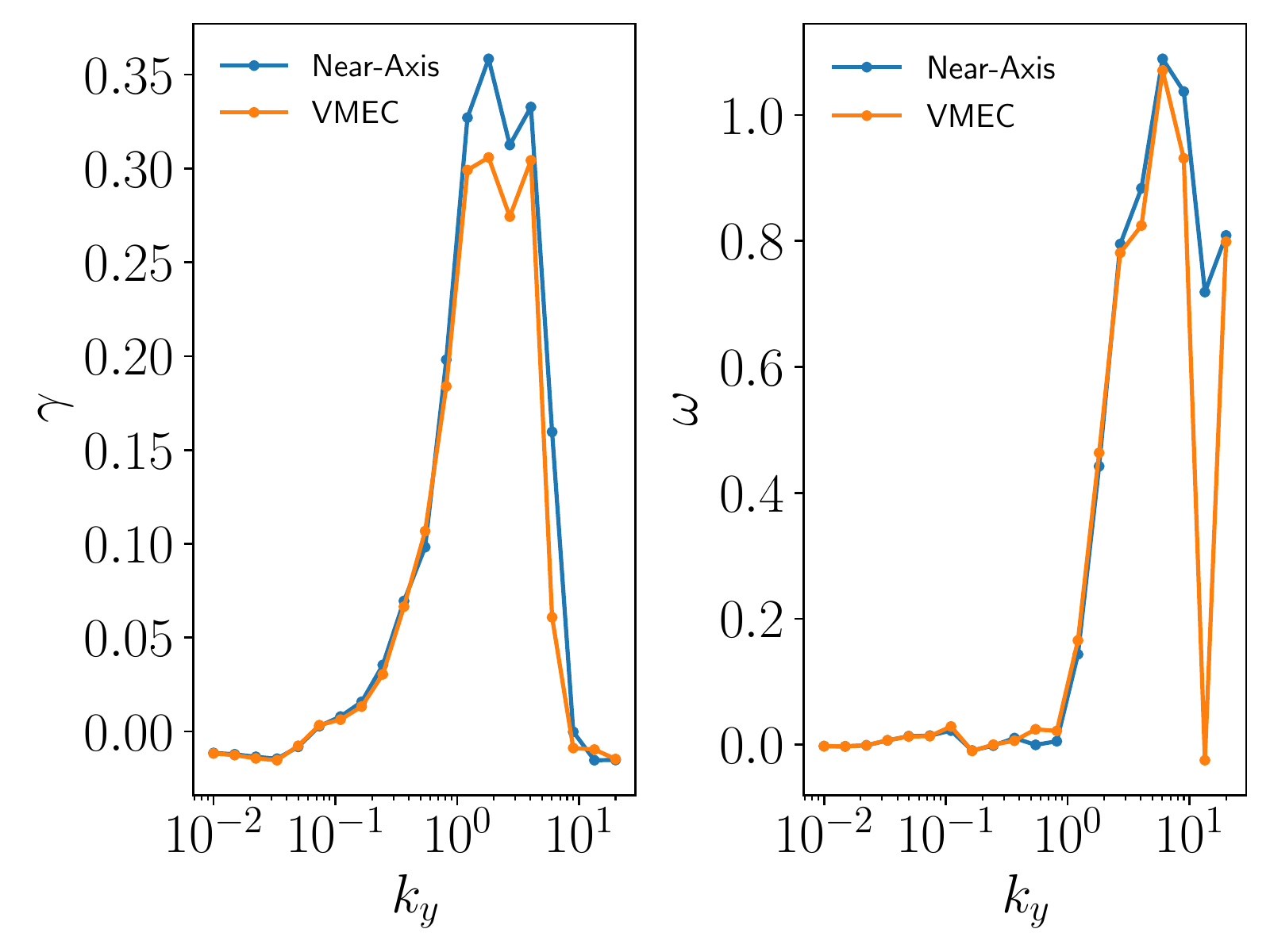}
    \caption{Dependence of $\gamma=\gamma(k_y)$ and $\omega=\omega(k_y)$ for $0\le k_y \rho\le20$ for the VMEC and near-axis NAQS configuration for $a/L_T=5$ and $a/L_n=2$ at $s=0.3$.
    }
    \label{fig:NAQSgammaomegaky}
\end{figure}

Next, we perform a scan over $k_y$ in order to determine the peak growth rate $\gamma$ for each value of $L_n$ and $L_T$.
For the values of $a/L_T=5$ and $a/L_n=2$ considered above, we show in \cref{fig:NAQSgammaomegaky} the dependence of $\gamma=\gamma(k_y)$ and $\omega=\omega(k_y)$ for $0\le k_y \rho\le20$ for both configurations at $s=0.3$.
The peak growth rate is found at $k_y \rho = 2$, a value within the expected range of the ITG instability.
Also, the near-axis expansion appears to yield a larger value of the growth rate across the majority of the spectrum, with a particular incidence at the peak values of $k_y \sim 1$.

\begin{figure}
    \centering
    \includegraphics[width=.9\textwidth]{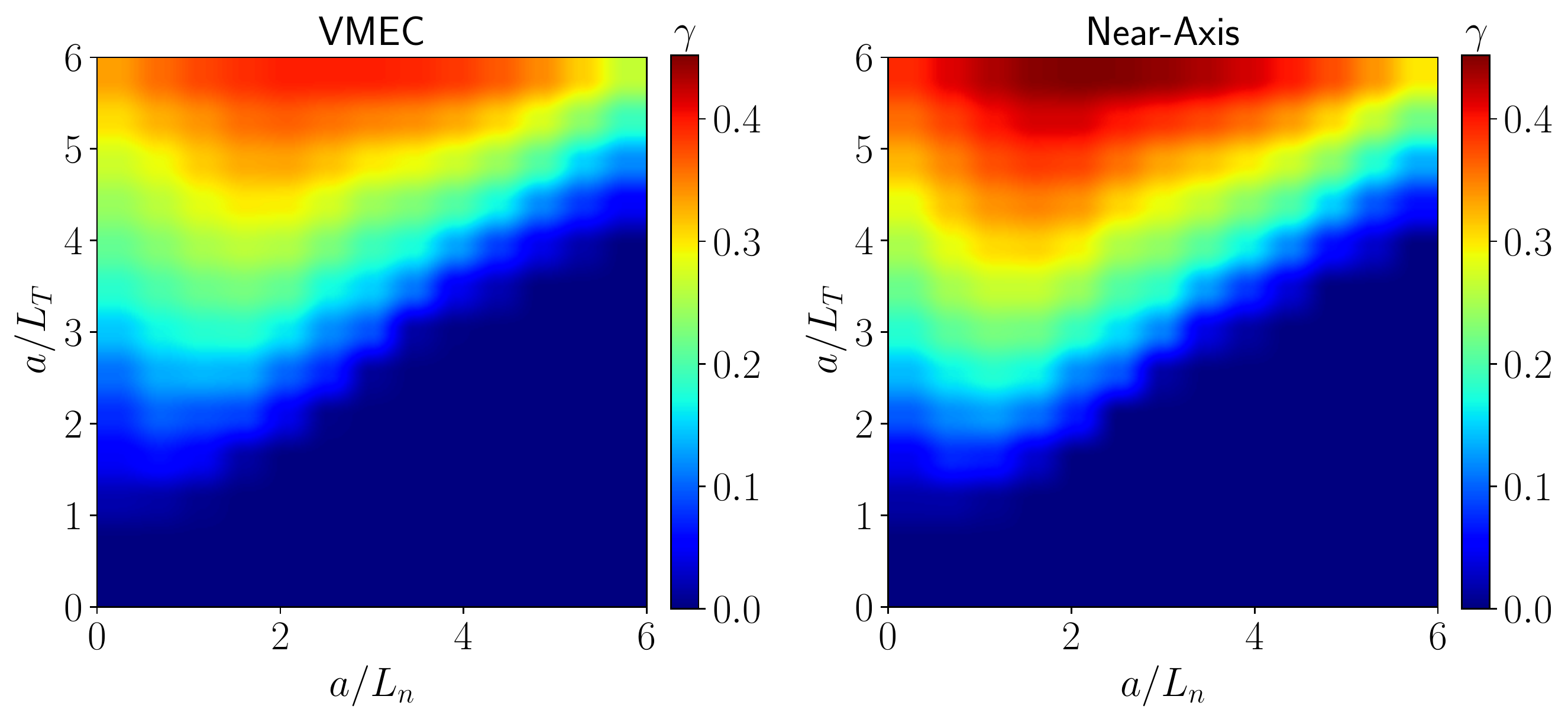}
    \includegraphics[width=.9\textwidth]{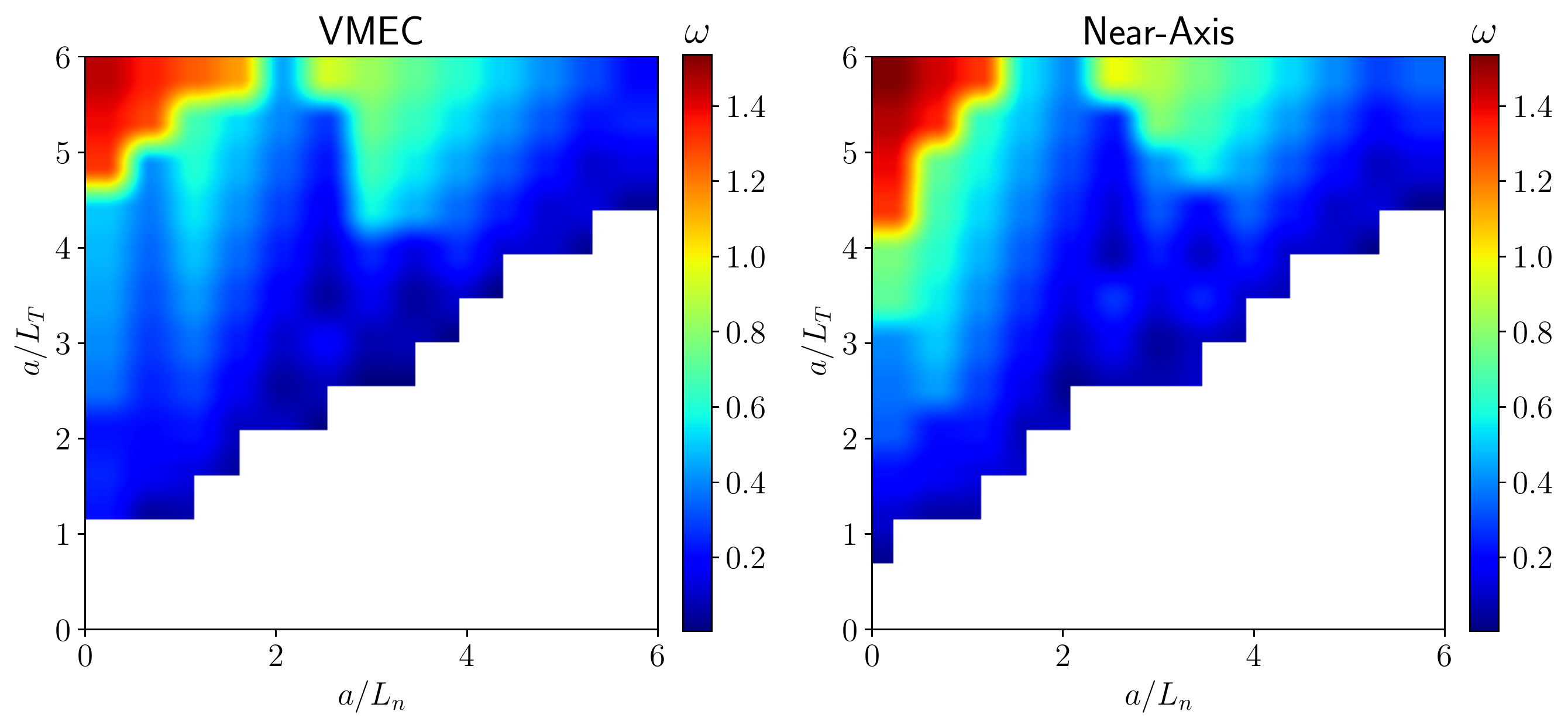}
    \includegraphics[width=.9\textwidth]{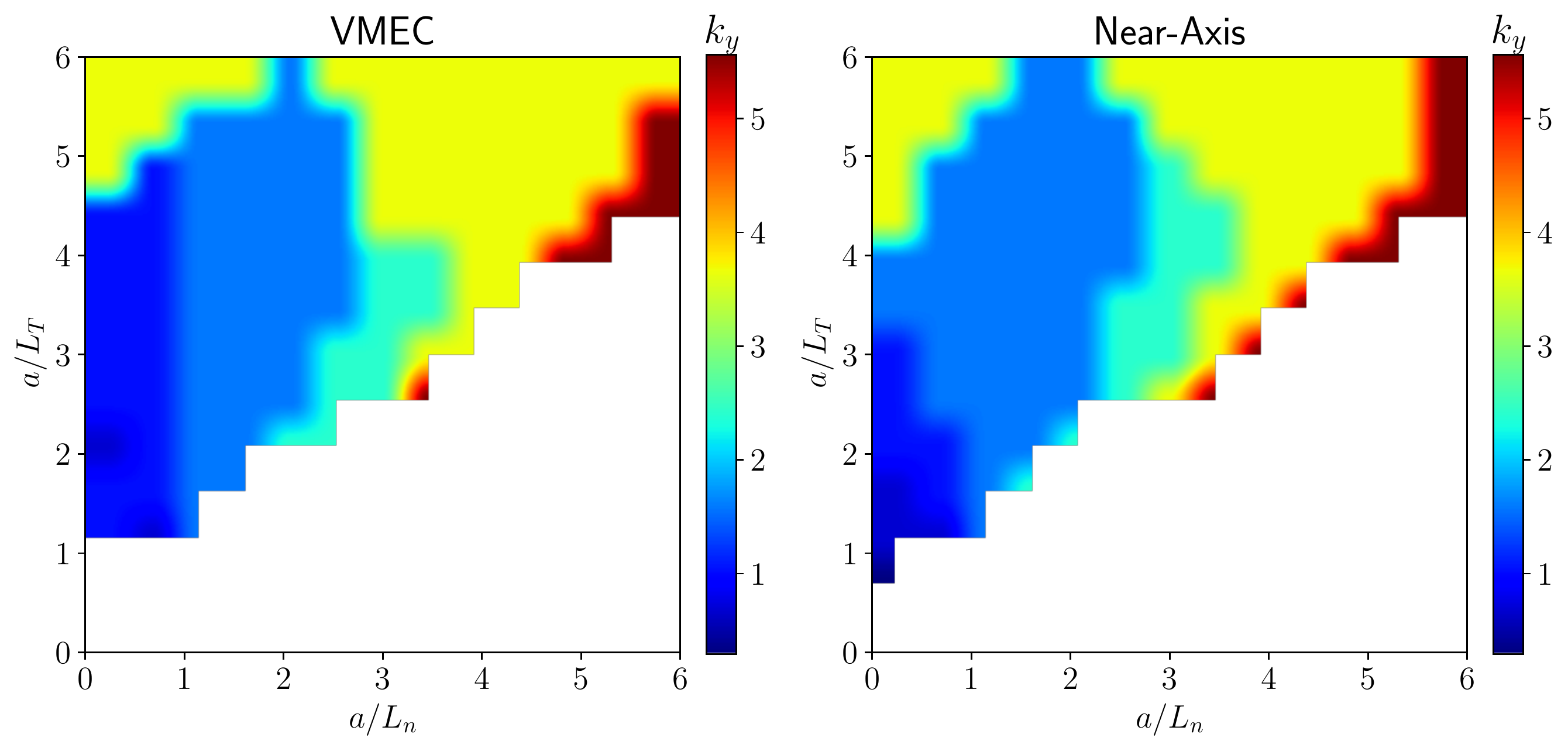}
    \caption{Comparison of the peak growth rates $\gamma$ and associated real frequencies $\omega$ and wave-numbers $k_y$  between the VMEC and near-axis geometries for the NAQS stellarator at $s=0.01$. The gradient lengths are scanned over $0\le a/L_n\le6$ and $0\le a/L_T\le6$.}
    \label{fig:NAQSmaxgammaomegaky}
\end{figure}

Finally, we vary the density and temperature gradient scale lengths, $a/L_n$ and $a/L_T$, and pick the fastest growing mode by selecting the maximum value of $\gamma(k_y)$ for each $L_n$ and $L_T$.
The resulting peak growth rates, associated $\omega$ and $k_y$ within $0\le a/L_n\le6$ and $0\le a/L_T\le6$ are shown in \cref{fig:NAQSmaxgammaomegaky}.
The locations of $\omega$ and $k_y$ where $\gamma \le 0$ are shown in white.
As is characteristic of the ITG mode, the peak growth rate is negative for a range of values where $\eta=L_n/L_T>\eta{\textsubscript{crit}}$ with $\eta_{crit} \sim 1$ \cite{Romanelli1989}, such as the condition for instability $\eta>2/3$ of the toroidal branch of the ITG mode found in \cite{Biglari1989}.
A more detailed understanding of the mechanisms that set the ITG critical gradient using first-principle approaches, such as the one in Ref. \cite{Roberg-Clark2020}, are left for future work.

\section{Numerical results}
\label{sec:numerical}

In this section, we carry out the analysis performed in \cref{sec:nabenchmark} for a single benchmark case, now for the eleven realistic quasisymmetric designs under study.
In particular, we perform a convergence scan to determine a base case scenario that satisfies the convergence criterion for all stellarator designs, perform a scan over $k_y$, $L_n$ and $L_T$ and pick the fastest growth rate in order to assess the critical gradients $\eta{\textsubscript{crit}}$ and peak growth rates $\gamma{\textsubscript{max}}$.

\begin{figure}
    \centering
    \includegraphics[trim={0 1.3cm 4.1cm 0},clip,width=0.32\textwidth]{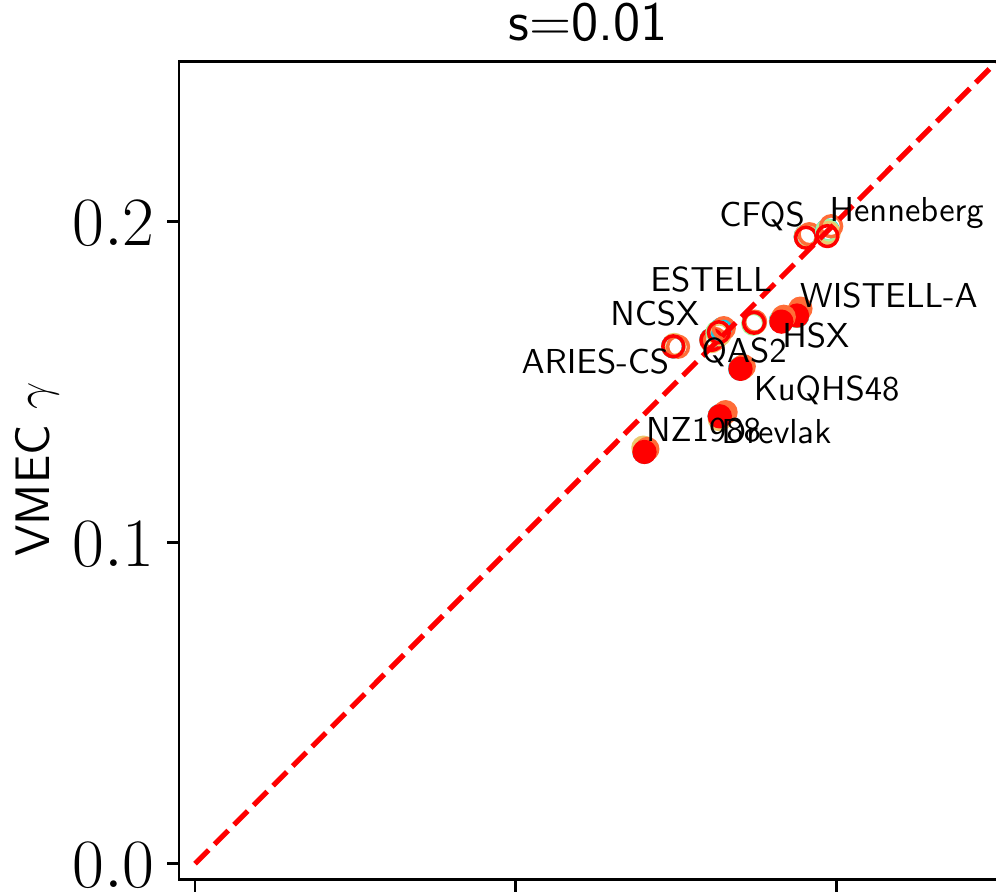}
    \includegraphics[trim={1.7cm 1.3cm 4.1cm 0},clip,width=0.266\textwidth]{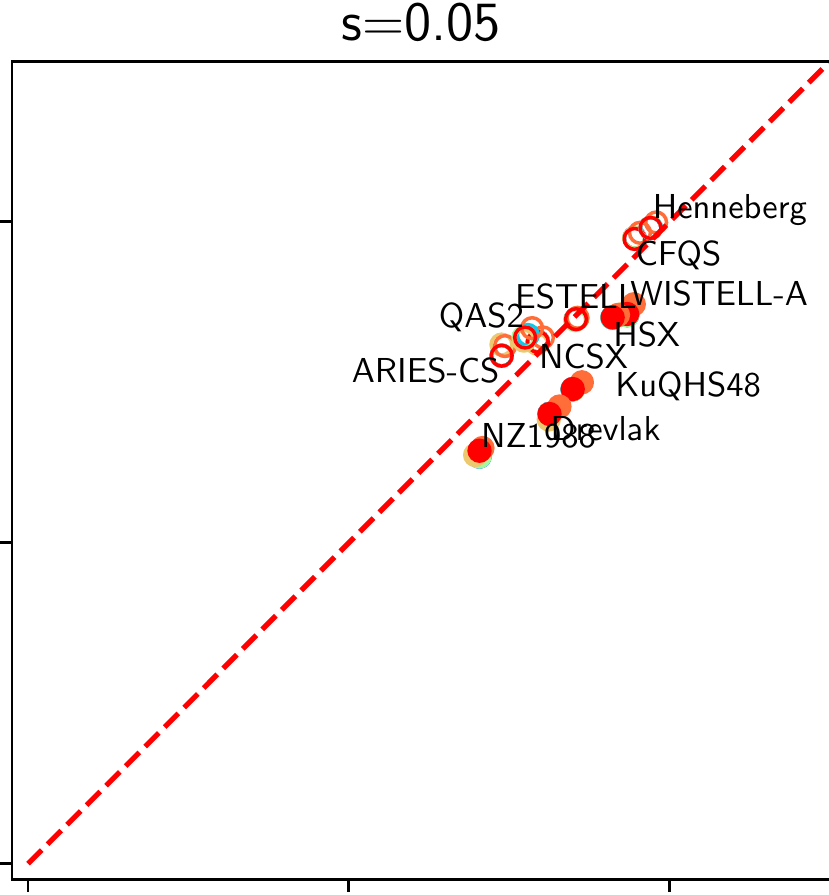}
    \includegraphics[trim={1.7cm 1.3cm 0 0},clip,width=0.397\textwidth]{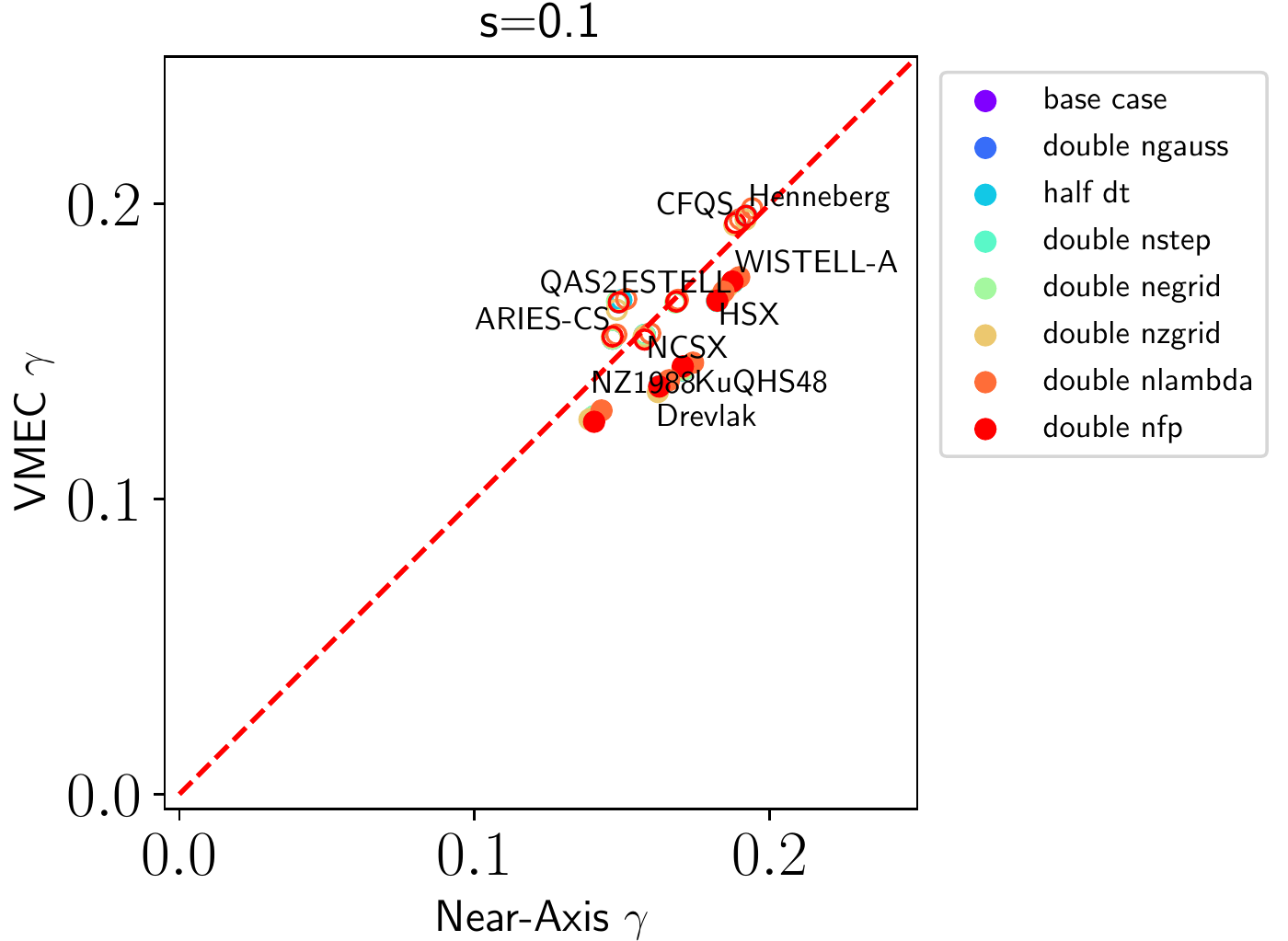}
    \includegraphics[trim={0 0 4.1cm 0},clip,width=0.32\textwidth]{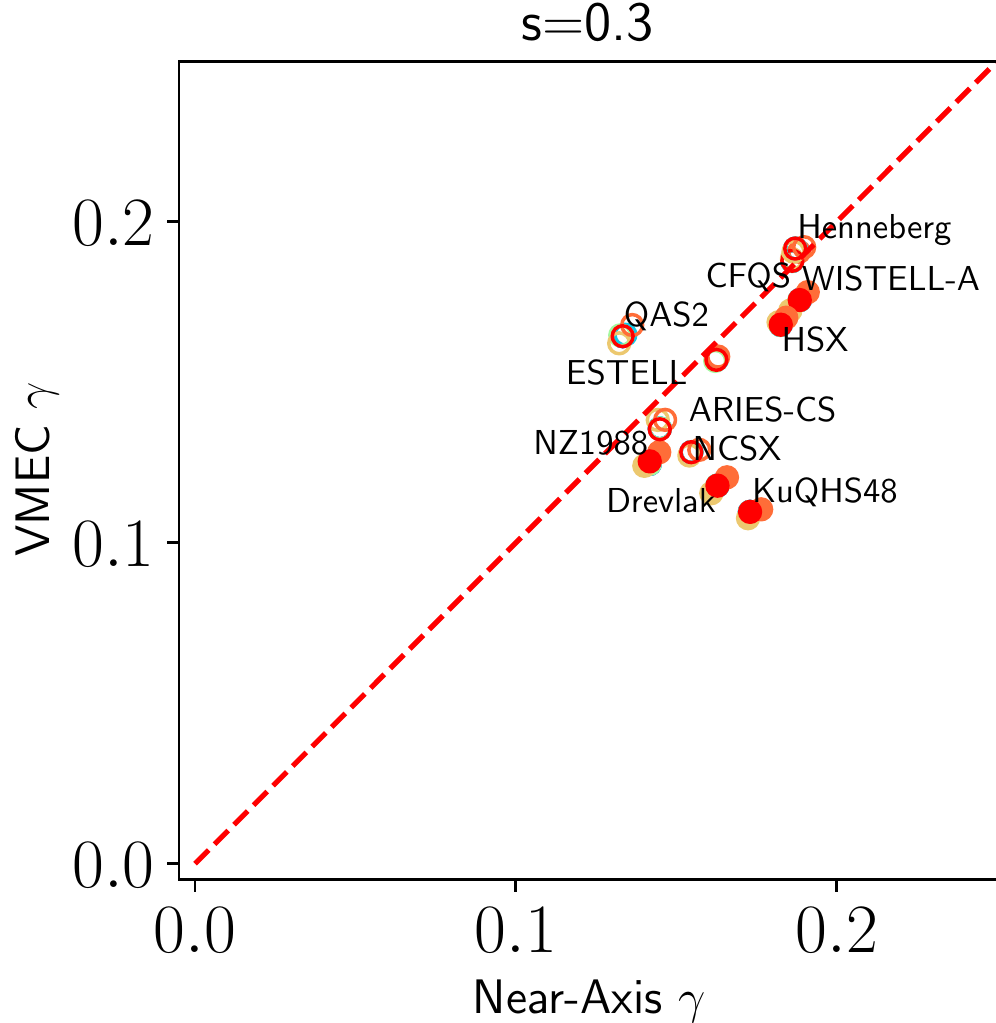}
    \includegraphics[trim={1.6cm 0 4.1cm 0},clip,width=0.266\textwidth]{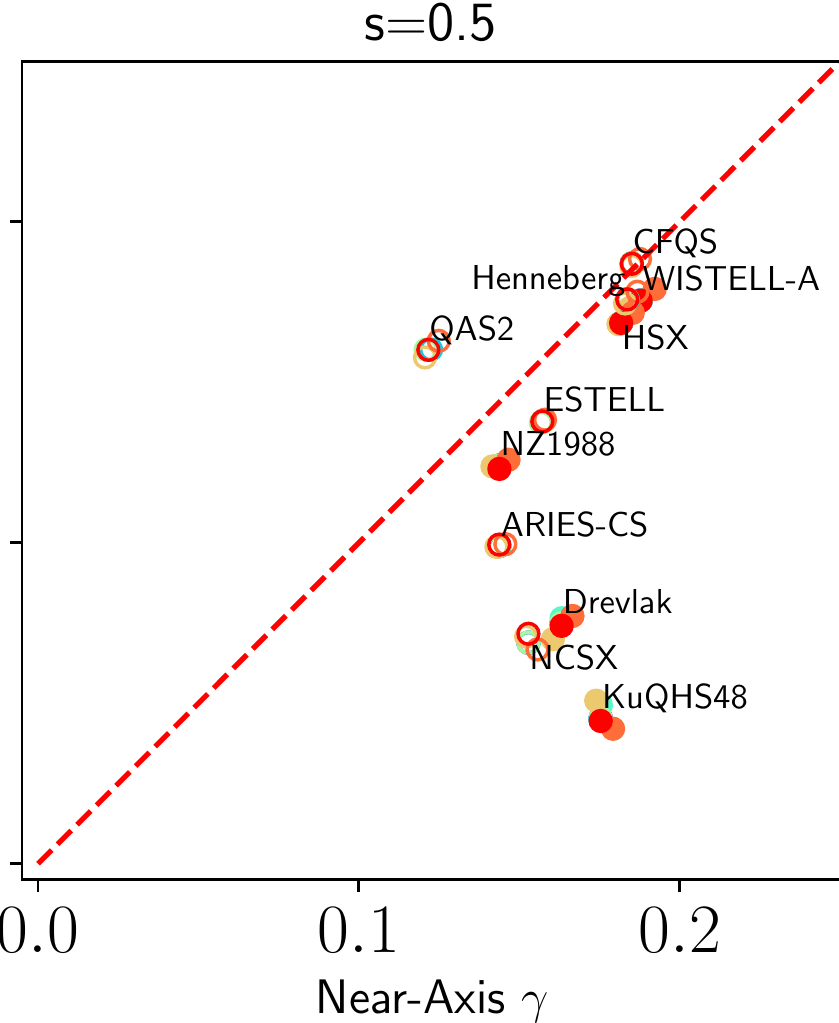}
    \includegraphics[trim={1.6cm 0 0 0},clip,width=0.397\textwidth]{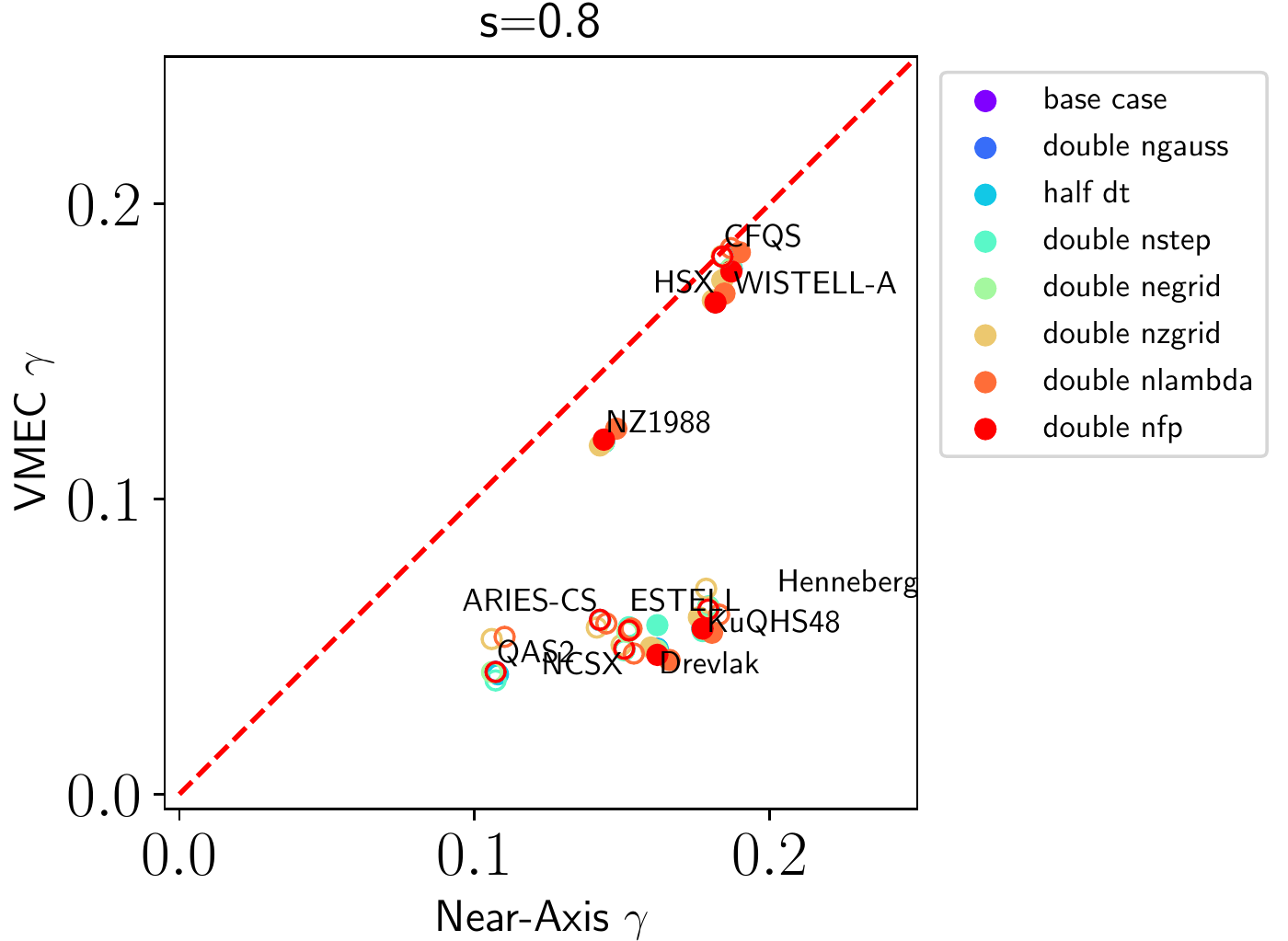}
    \caption{
    Growth rate convergence test of the base case scenario for the VMEC and near-axis geometries using the gyrokinetic GS2 code at different radii $s=\psi/\psi_a$ and $k_y \rho=1$, $a/L_T=3$ and $a/L_n=1$. Quasi-axisymmetric stellarators are represented with hollow circles while quasi-helically symmetric ones are represented with filled circles.}
    \label{fig:allgammastells}
\end{figure}

We start with the convergence scan.
A comparison between the growth rate obtained with the near-axis expansion and VMEC is shown in \cref{fig:allgammastells} for several radii $s=\psi/\psi_a=(0.01, 0.05, 0.1,0.3,0.5,0.8)$ and $k_y \rho=1$, $a/L_T=3$ and $a/L_n=1$.
In order to obtain converged base case scenarios, the spatial resolution was increased to \textit{nzgrid}=150, \textit{nlambda} increased to \textit{nlambda}=28 and, due to the presence of lower values of $\gamma$ with respect to the NAQS benchmark, the simulation time was increased to \textit{nstep=250}.
Concerning the top row in \cref{fig:allgammastells} (radii with $s<0.1$), the growth rate $\gamma$ obtained with a near-axis model is found to be in very good agreement with the ones using the corresponding VMEC design.
However, it is found that for larger radii, the near-axis expansion provides an overestimation of the growth rate.
This behaviour is observed for all cases studied here.

We distinguish between the quasi-axisymmetric and quasi-helically symmetric configurations in \cref{fig:allgammastells} by using both hollow and filled circles, respectively.
It is seen that, in general, there is no systematic difference between the two for the physical parameters simulated here, except for a small tendency of quasi-axisymmetric stellarators close to the magnetic axis to exhibit higher growth rates.
However, this tendency seems to reverse further away from the axis where, at $s=0.8$, the majority of the quasi-axisymmetric configurations have considerably lower growth rates than quasi-helically symmetric ones, widening the gap between the maximum and minimum values of the computed $\gamma$.

\begin{figure}
    \centering
    \includegraphics[width=.6\textwidth]{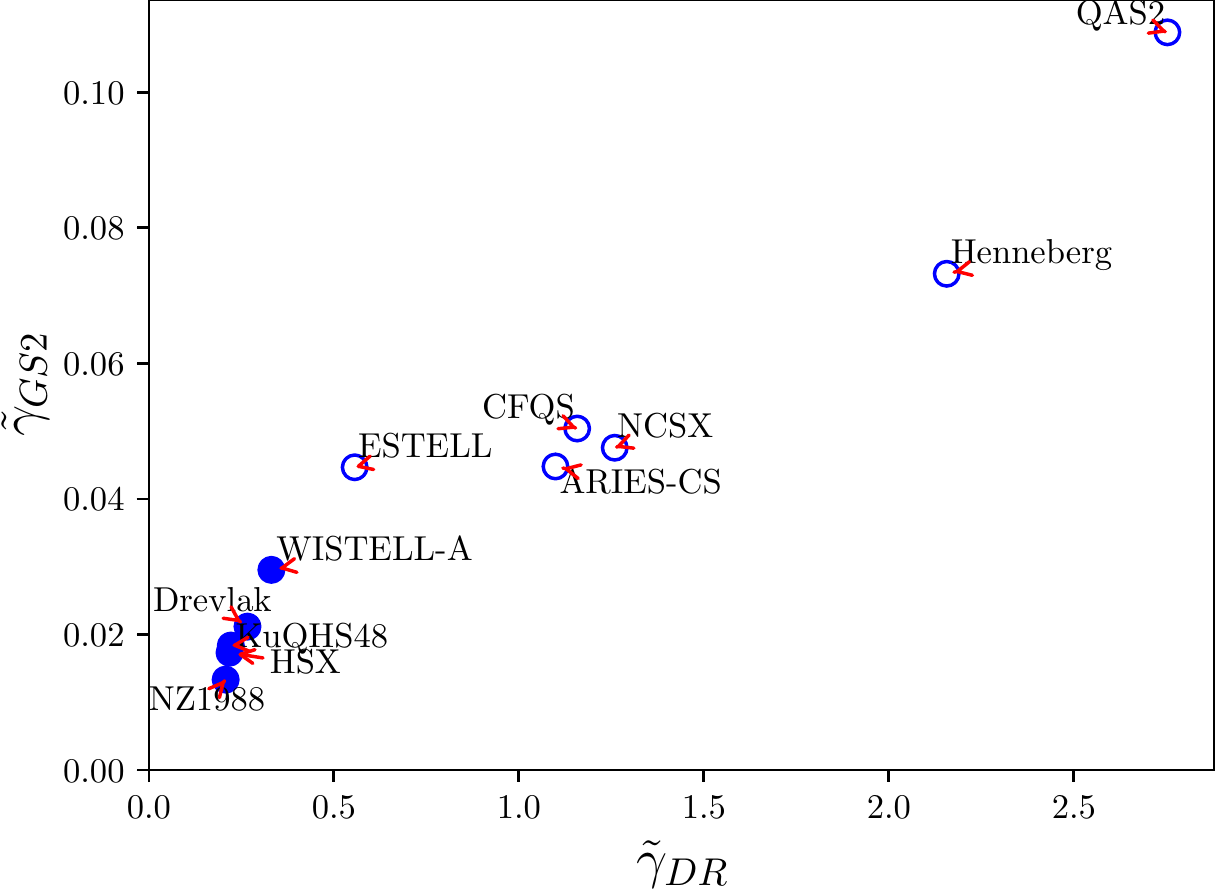}
    \caption{Horizontal axis: growth rate estimate $\gamma$ normalized to $\omega_t$ computed using the dispersion relation in \cref{eq:analyticalDR1} (denoted as $\gamma_{DR}$). Vertical axis: growth rate estimate $\gamma$ normalized to $\omega_t$ computed using the gyrokinetic GS2 code and the corresponding VMEC equilibrium files (denoted as $\gamma_{GS2}$, corresponding to the values in \cref{fig:allgammastells} multiplied by $L/2\pi a$). Quasi-axisymmetric stellarators are represented with hollow circles while quasi-helically symmetric ones are represented with filled circles and $r=0.01$.}
    \label{fig:LnvsGS2}
\end{figure}

We now comment on the accuracy of the estimate using a simplified dispersion relation in \cref{eq:analyticalDR1} when compared with the growth rates using first-principle gyrokinetic simulations in \cref{fig:allgammastells}.
This comparison is shown in \cref{fig:LnvsGS2}, where we show the growth rate estimate $\gamma$ normalized to $\omega_t$ computed using the dispersion relation in \cref{eq:analyticalDR1} (denoted as $\gamma_{DR}$) on the horizontal axis and the growth rate estimate $\gamma$ normalized to $\omega_t$ computed using the gyrokinetic GS2 code and the corresponding VMEC equilibrium files (denoted as $\gamma_{GS2}$, corresponding to the values in \cref{fig:allgammastells} multiplied by $L/2\pi a$).
It is seen that the estimate $\gamma_{DR}$ resulting from the simplified dispersion relation in \cref{eq:analyticalDR1} consistently overestimates the growth rate which, in some cases, can lead to a one order of magnitude difference between the two growth rates for quasi-axisymmetric stellarators with higher values of $\gamma$.
However, the general trend and the relative growth rates between different configurations is well predicted by the simplified dispersion relation.

\begin{figure}
    \centering
    \includegraphics[width=.9\textwidth]{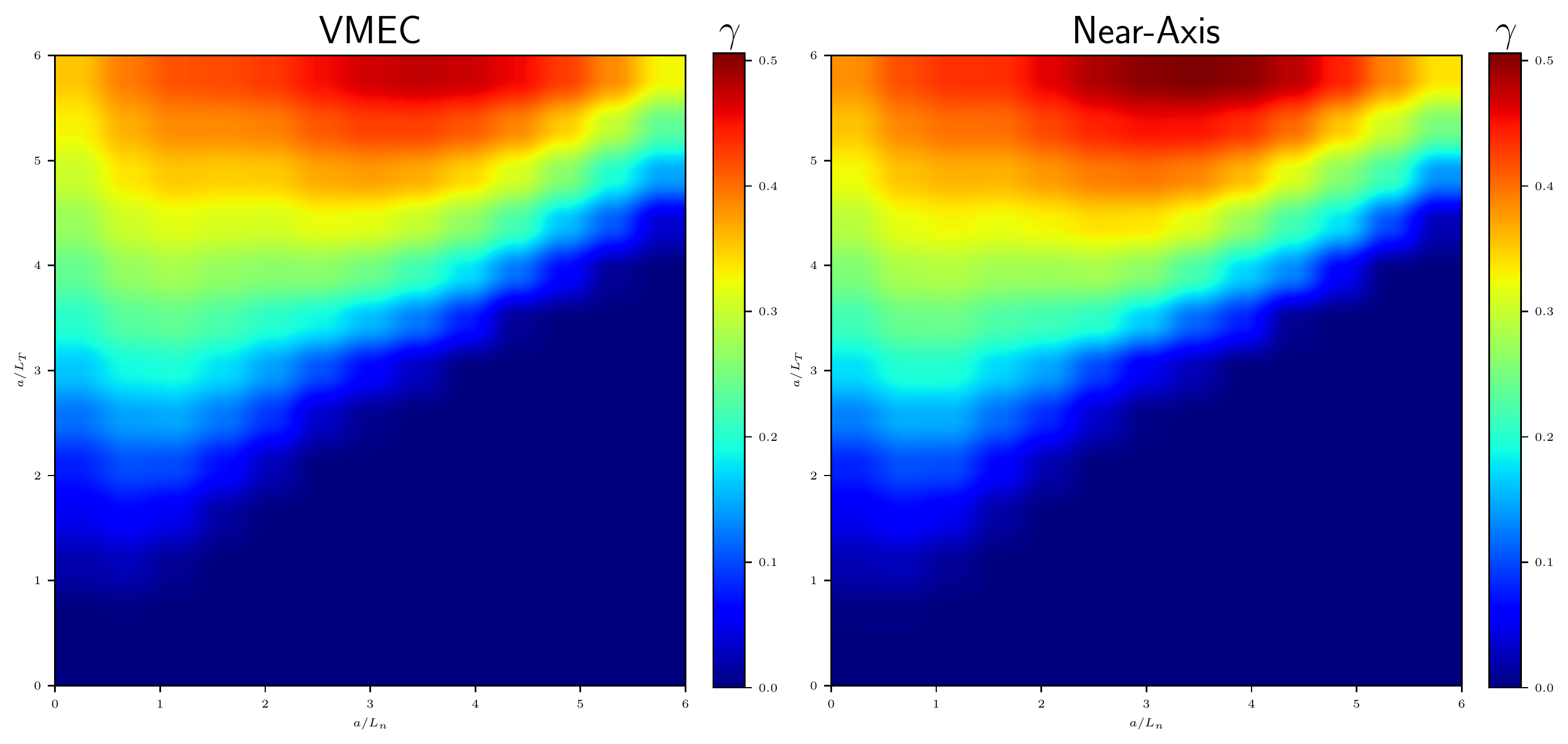}
    \includegraphics[width=.9\textwidth]{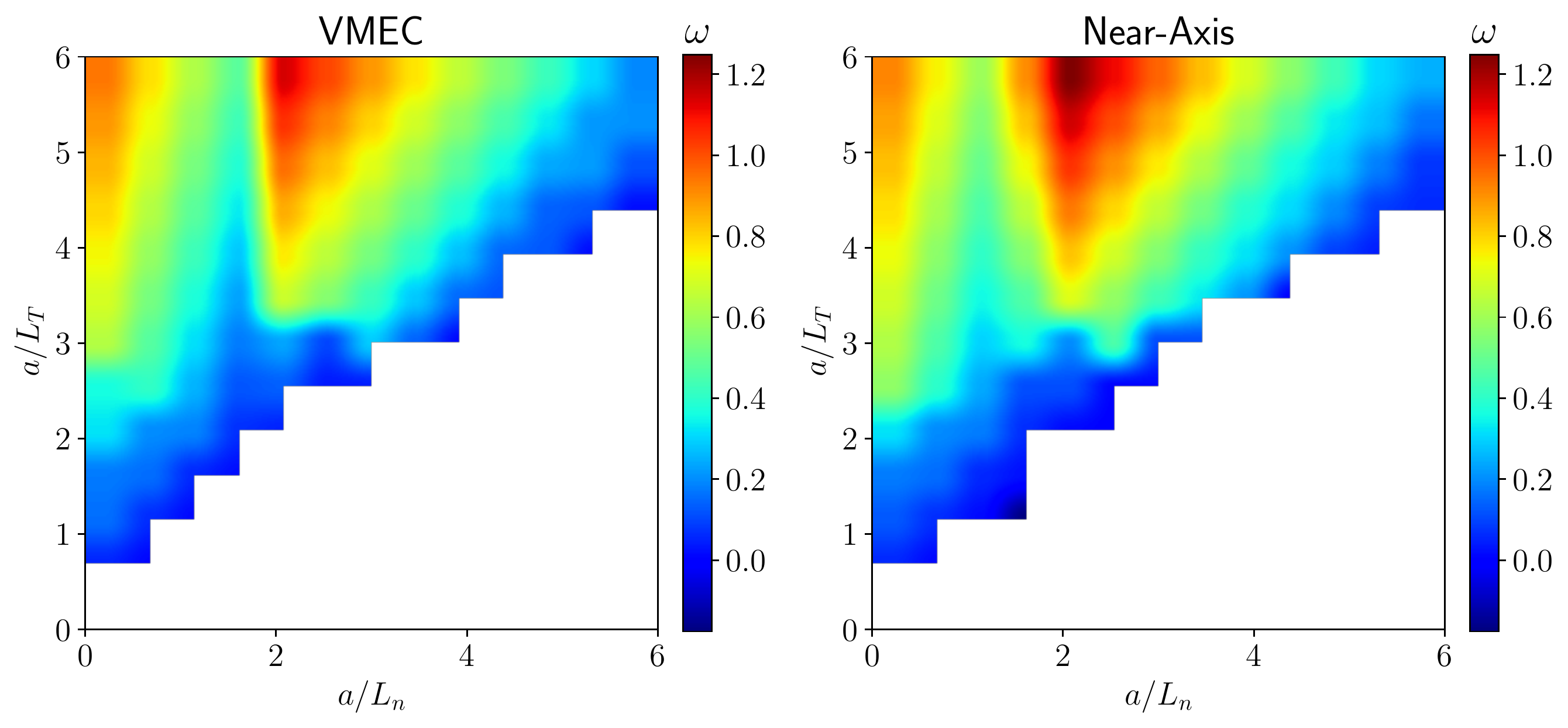}
    \includegraphics[width=.9\textwidth]{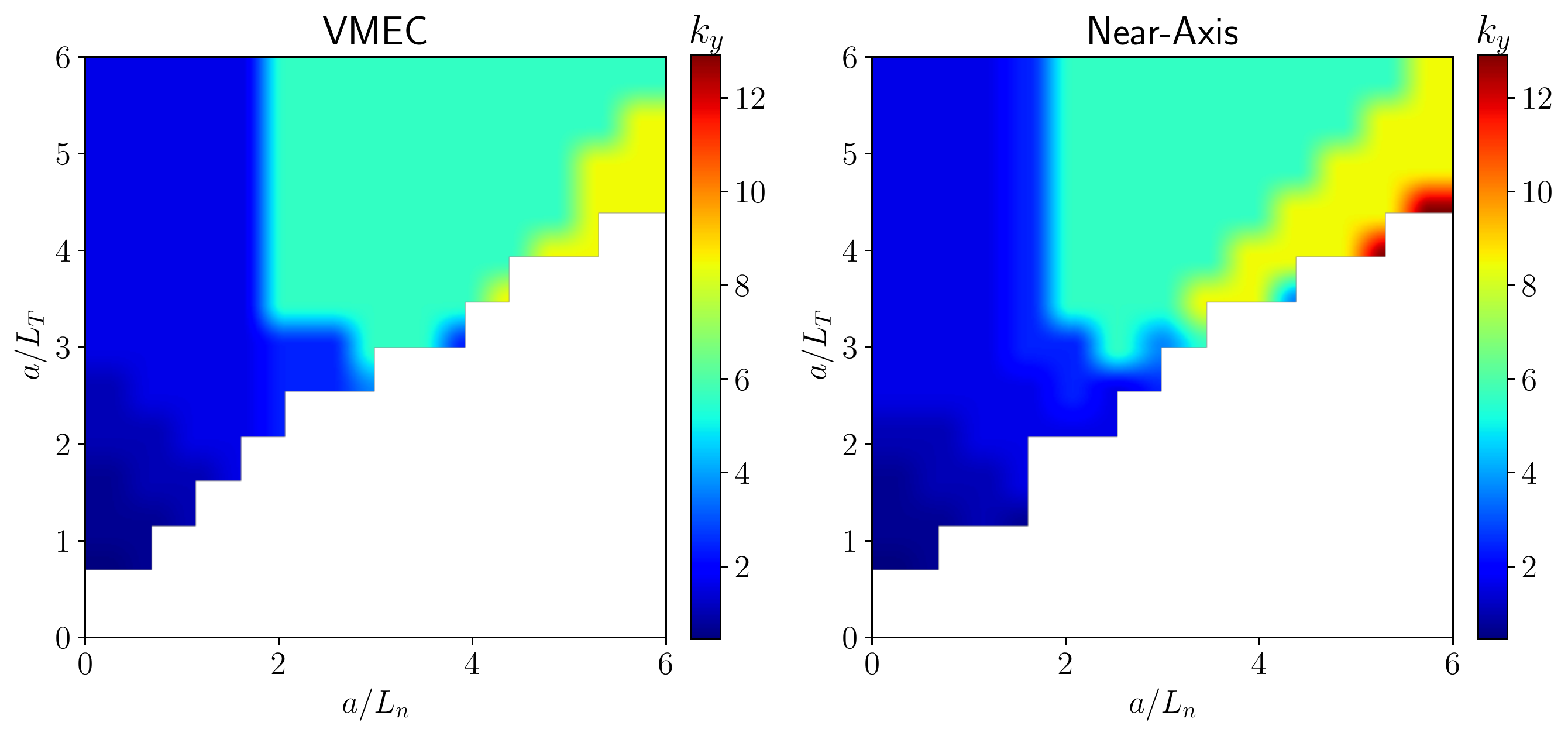}
    \caption{Comparison of the peak growth rates $\gamma$ and associated real frequencies $\omega$ and wave-numbers $k_y$  between the VMEC and near-axis geometries for the NCSX stellarator at $s=0.01$. The gradient lengths are scanned over $0\le a/L_n\le6$ and $0\le a/L_T\le6$.}
    \label{fig:NCSXmaxgammaomegaky}
\end{figure}

\begin{figure}
    \centering
    \includegraphics[width=.9\textwidth]{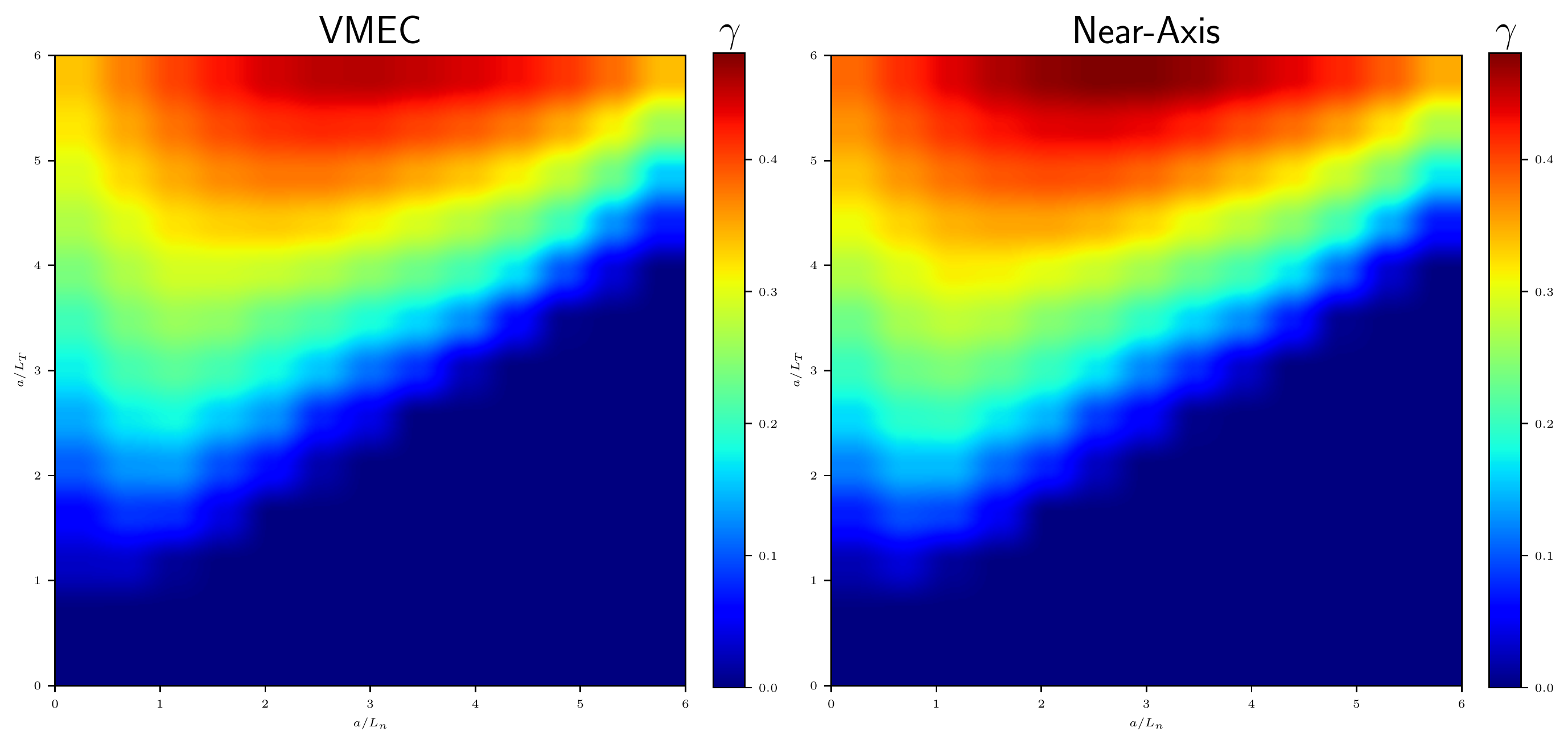}
    \includegraphics[width=.9\textwidth]{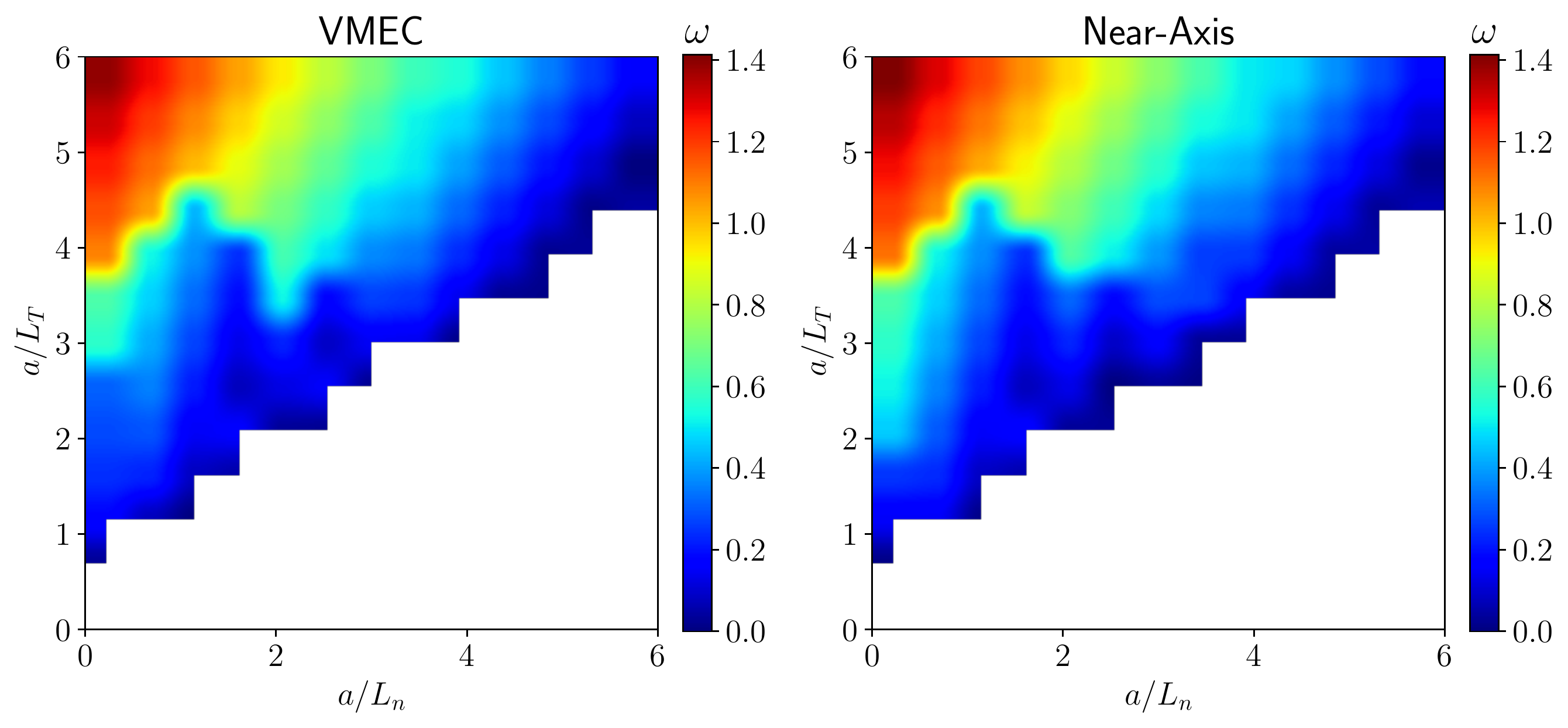}
    \includegraphics[width=.9\textwidth]{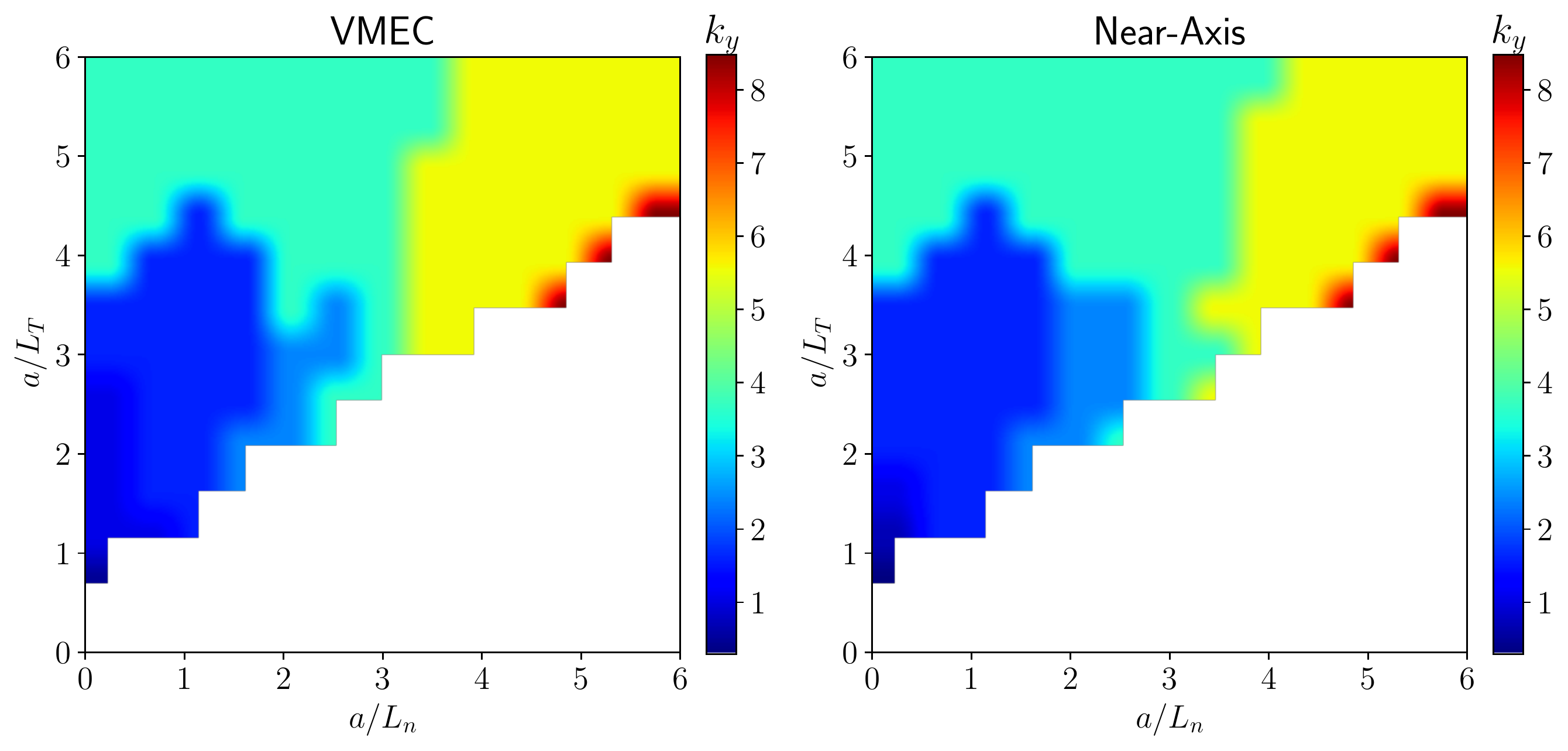}
    \caption{Comparison of the peak growth rates $\gamma$ and associated real frequencies $\omega$ and wave-numbers $k_y$  between the VMEC and near-axis geometries for the HSX stellarator at $s=0.01$. The gradient lengths are scanned over $0\le a/L_n\le6$ and $0\le a/L_T\le6$.}
    \label{fig:HSXmaxgammaomegaky}
\end{figure}

Finally, we assess the accuracy of the near-axis expansion at predicting the peak growth rates when performing a scan over $a/L_n$ and $a/L_T$.
Two examples are shown in \cref{fig:NCSXmaxgammaomegaky} and \cref{fig:HSXmaxgammaomegaky} for the NCSX and HSX stellarators, respectively.
These confirm the general trend observed in \cref{fig:NAQSmaxgammaomegaky} and the accuracy of the near-axis expansion for values of $s<0.3$.
In general, the linear critical gradient for the ITG instability near the axis of quasisymmetric stellarators lies at values of $\eta \sim 1$.
We also find that the peak growth rate increases with increasing $a/L_T$ and is maximum for values of $2<a/L_n<4$.
The values of the real frequency $\omega$ at the peak growth rate increase with increasing $a/L_T$ while peaking at either $a/L_n=0$ or $a/L_n\simeq 3$.
The wave-vector $k_y$ at the peak growth rate is observed to increase mainly with increasing $a/L_n$.

\section{Conclusions}
\label{sec:conclusions}

In this work, the ion-temperature-gradient stability of quasisymmetric stellarators is assessed near the magnetic axis by employing the gyrokinetic approximation together with a near-axis expansion framework.
It is found that the near-axis expansion is not only able to simplify analytical estimates of ITG growth rates and turbulent fluxes, but also able to reproduce the properties of eleven quasisymmetric configurations designed using numerical optimization methods.
In general, as the radius is increased beyond the inner core, it is found that the near-axis model overestimates the growth rates when compared with the corresponding optimized VMEC design.
Such overestimation could arise due to the absence of global magnetic shear $\hat s = -(r/\iota)\iota'(r)$ in the near-axis expansion formalism at the order considered here which could contribute as a stabilizing factor at larger radii where higher order geometry effects are more relevant.
Global magnetic shear may also provide additional finite-Larmor radius stabilization to ITG modes by contributing with additional terms to the perpendicular wave-vector $\mathbf k_\perp$.
Furthermore, the presence of inhomogeneities in the magnetic field of the VMEC configurations, which could lead to departures from quasisymmetry, may allow the presence of a larger fraction of trapped particles at the good curvature side and, therefore, decrease $\gamma$.

While the analysis performed here pertained the ITG instability only, a similar analysis can be carried out to other instabilities, such as the trapped-electron mode, the electron-temperature-gradient mode and the micro-tearing mode.
Furthermore, as future work, we intend to generalize the numerical methods employed here for quasisymmetric designs to maximum-J devices \cite{Rosenbluth1968} and have a more complete understanding of the supression of electrostatic microinstabilities recently found in such configurations \cite{Alcuson2020}.

\section{Acknowledgments}
\label{sec:acknowledgments}

We wish to thank I. G. Abel, W. Dorland, G. T. Roberg-Clark and G. G. Plunk for many fruitful discussions regarding the content of the manuscript.
This work was supported by a grant from the Simons Foundation (560651, ML) and from the U.S. Department of Energy, Office of Science, Office of Fusion Energy Science, under award number DE-FG02-93ER54197.

\section*{References}
\bibliographystyle{unsrt}
\bibliography{library}

\begin{thebibliography}{10}

\bibitem{Xanthopoulos2007}
P.~Xanthopoulos and F.~Jenko.
\newblock {Gyrokinetic analysis of linear microinstabilities for the
  stellarator Wendelstein 7-X}.
\newblock {\em Physics of Plasmas}, 14(4):042501, 2007.

\bibitem{Helander2015}
P.~Helander, T.~Bird, F.~Jenko, R.~Kleiber, G.~G. Plunk, J.~H.E. Proll,
  J.~Riemann, and P.~Xanthopoulos.
\newblock {Advances in stellarator gyrokinetics}.
\newblock {\em Nuclear Fusion}, 55(5):053030, 2015.

\bibitem{Nuhrenberg1988a}
J.~Nuhrenberg and R.~Zille.
\newblock {Quasi-helically symmetric toroidal stellarators}.
\newblock {\em Physics Letters A}, 129(2):113--117, 1988.

\bibitem{Boozer1995}
A.~H. Boozer.
\newblock {Quasi-helical symmetry in stellarators}.
\newblock {\em Plasma Physics and Controlled Fusion}, 37(11A):A103, 1995.

\bibitem{Garabedian1996}
P.~R. Garabedian.
\newblock {Stellarators with the magnetic symmetry of a tokamak}.
\newblock {\em Physics of Plasmas}, 3(7):2483, 1996.

\bibitem{Garren1991a}
D.~A. Garren and A.~H. Boozer.
\newblock {Existence of quasihelically symmetric stellarators}.
\newblock {\em Physics of Fluids B}, 3(10):2822, 1991.

\bibitem{Landreman2018}
M.~Landreman and W.~Sengupta.
\newblock {Direct construction of optimized stellarator shapes. Part 1. Theory
  in cylindrical coordinates}.
\newblock {\em Journal of Plasma Physics}, 84(6):905840616, 2018.

\bibitem{Jorge2020b}
R.~Jorge, W.~Sengupta, and M.~Landreman.
\newblock {Construction of quasisymmetric stellarators using a direct
  coordinate approach}.
\newblock {\em Nuclear Fusion}, 60(7):076021, 2020.

\bibitem{Plunk2020}
G.~G. Plunk.
\newblock {Perturbing an axisymmetric magnetic equilibrium to obtain a
  quasi-axisymmetric stellarator}.
\newblock {\em Journal of Plasma Physics}, 86(4):905860409, 2020.

\bibitem{Nelson2003}
B.~E. Nelson, L.~A. Berry, A.~B. Brooks, M.~J. Cole, J.~C. Chrzanowski, H.~M.
  Fan, P.~J. Fogarty, P.~L. Goranson, P.~J. Heitzenroeder, S.~P. Hirshman,
  G.~H. Jones, J.~F. Lyon, G.~H. Neilson, W.~T. Reiersen, D.~J. Strickler, and
  D.~E. Williamson.
\newblock {Design of the national compact stellarator experiment (NCSX)}.
\newblock In {\em Fusion Engineering and Design}, volume 66-68, page 169, 2003.

\bibitem{Canik2007}
J.~M. Canik, D.~T. Anderson, F.~S.B. Anderson, K.~M. Likin, J.~N. Talmadge, and
  K.~Zhai.
\newblock {Experimental demonstration of improved neoclassical transport with
  quasihelical symmetry}.
\newblock {\em Physical Review Letters}, 98(8):085002, 2007.

\bibitem{Drevlak2019}
M.~Drevlak, C.~D. Beidler, J.~Geiger, P.~Helander, and Y.~Turkin.
\newblock {Optimisation of stellarator equilibria with ROSE}.
\newblock {\em Nuclear Fusion}, 59(1):016010, 2019.

\bibitem{Helander2013a}
P.~Helander, J.~H.E. Proll, and G.~G. Plunk.
\newblock {Collisionless microinstabilities in stellarators. I. Analytical
  theory of trapped-particle modes}.
\newblock {\em Physics of Plasmas}, 20(12):122505, 2013.

\bibitem{Mynick2011}
H.~E. Mynick, N.~Pomphrey, and P.~Xanthopoulos.
\newblock {Reducing turbulent transport in toroidal configurations via
  shaping}.
\newblock {\em Physics of Plasmas}, 18(5):056101, 2011.

\bibitem{Rewoldt1982}
G.~Rewoldt, W.~M. Tang, and M.~S. Chance.
\newblock {Electromagnetic kinetic toroidal eigenmodes for general
  magnetohydrodynamic equilibria}.
\newblock {\em Physics of Fluids}, 25(3):480, 1982.

\bibitem{Rewoldt1987}
G.~Rewoldt, W.~M. Tang, and R.~J. Hastie.
\newblock {Collisional effects on kinetic electromagnetic modes and associated
  quasilinear transport}.
\newblock {\em Physics of Fluids}, 30(3):807, 1987.

\bibitem{Xanthopoulos2007a}
P.~Xanthopoulos, F.~Merz, T.~G{\"{o}}rler, and F.~Jenko.
\newblock {Nonlinear gyrokinetic simulations of ion-temperature-gradient
  turbulence for the optimized wendelstein 7-X stellarator}.
\newblock {\em Physical Review Letters}, 99(3):035002, 2007.

\bibitem{Mynick2010}
H.~E. Mynick, N.~Pomphrey, and P.~Xanthopoulos.
\newblock {Optimizing stellarators for turbulent transport}.
\newblock {\em Physical Review Letters}, 105(9):095004, 2010.

\bibitem{Dorland2000a}
W.~Dorland, F.~Jenko, M.~Kotschenreuther, and B.~N. Rogers.
\newblock {Electron Temperature Gradient Turbulence}.
\newblock {\em Physical Review Letters}, 85(26):5579, 2000.

\bibitem{Zocco2018}
A.~Zocco, P.~Xanthopoulos, H.~Doerk, J.~W. Connor, and P.~Helander.
\newblock {Threshold for the destabilisation of the ion-temperature-gradient
  mode in magnetically confined toroidal plasmas}.
\newblock {\em Journal of Plasma Physics}, 84(1):715840101, 2018.

\bibitem{Xanthopoulos2016}
P.~Xanthopoulos, G.~G. Plunk, A.~Zocco, and P.~Helander.
\newblock {Intrinsic turbulence stabilization in a stellarator}.
\newblock {\em Physical Review X}, 6(2):021033, 2016.

\bibitem{Hegna2018}
C.~C. Hegna, P.~W. Terry, and B.~J. Faber.
\newblock {Theory of ITG turbulent saturation in stellarators: Identifying
  mechanisms to reduce turbulent transport}.
\newblock {\em Physics of Plasmas}, 25(2):022511, 2018.

\bibitem{McKinney2019}
I.~J. McKinney, M.~J. Pueschel, B.~J. Faber, C.~C. Hegna, J.~N. Talmadge, D.~T.
  Anderson, H.~E. Mynick, and P.~Xanthopoulos.
\newblock {A comparison of turbulent transport in a quasi-helical and a
  quasi-axisymmetric stellarator}.
\newblock {\em Journal of Plasma Physics}, 85(5):905850503, 2019.

\bibitem{Wang2020}
H.~Y. Wang, I.~Holod, Z.~Lin, J.~Bao, J.~Y. Fu, P.~F. Liu, J.~H. Nicolau,
  D.~Spong, and Y.~Xiao.
\newblock {Global gyrokinetic particle simulations of microturbulence in W7-X
  and LHD stellarators}.
\newblock {\em Physics of Plasmas}, 27(8):082305, 2020.

\bibitem{Landreman2019}
M.~Landreman.
\newblock {Optimized quasisymmetric stellarators are consistent with the
  Garren–Boozer construction}.
\newblock {\em Plasma Physics and Controlled Fusion}, 61(7):075001, 2019.

\bibitem{Jorge2020c}
R.~Jorge and M.~Landreman.
\newblock {The Use of Near-Axis Magnetic Fields for Stellarator Turbulence
  Simulations}.
\newblock {\em Plasma Physics and Controlled Fusion}, 63(1):014001, 2020.

\bibitem{Faber2018}
B.~J. Faber, M.~J. Pueschel, P.~W. Terry, C.~C. Hegna, and J.~E. Roman.
\newblock {Stellarator microinstabilities and turbulence at low magnetic
  shear}.
\newblock {\em Journal of Plasma Physics}, 84(5):905840503, 2018.

\bibitem{Martin2018}
M.~F. Martin, M.~Landreman, P.~Xanthopoulos, N.~R. Mandell, and W.~Dorland.
\newblock {The parallel boundary condition for turbulence simulations in low
  magnetic shear devices}.
\newblock {\em Plasma Physics and Controlled Fusion}, 60(9):095008, 2018.

\bibitem{Baumgaertel2011}
J.~A. Baumgaertel, E.~A. Belli, W.~Dorland, W.~Guttenfelder, G.~W. Hammett,
  D.~R. Mikkelsen, G.~Rewoldt, W.~M. Tang, and P.~Xanthopoulos.
\newblock {Simulating gyrokinetic microinstabilities in stellarator geometry
  with GS2}.
\newblock {\em Physics of Plasmas}, 18(12):122301, 2011.

\bibitem{Howes2006}
G.~G. Howes, S.~C. Cowley, W.~Dorland, G.~W. Hammett, E.~Quataert, and A.~A.
  Schekochihin.
\newblock {Astrophysical Gyrokinetics: Basic Equations and Linear Theory}.
\newblock {\em The Astrophysical Journal}, 651(1):590, 2006.

\bibitem{Frieman1982a}
E.~A. Frieman and L.~Chen.
\newblock {Nonlinear gyrokinetic equations for low-frequency electromagnetic
  waves in general plasma equilibria}.
\newblock {\em Physics of Fluids}, 25(3):502, 1982.

\bibitem{Hirshman1983}
S.~P. Hirshman and J.~C. Whitson.
\newblock {Steepest-descent moment method for three-dimensional
  magnetohydrodynamic equilibria}.
\newblock {\em Physics of Fluids}, 26(12):3553, 1983.

\bibitem{Boozer1981}
A.~H. Boozer.
\newblock {Plasma equilibrium with rational magnetic surfaces}.
\newblock {\em Physics of Fluids}, 24(11):1999, 1981.

\bibitem{Garren1991}
D.~A. Garren and A.~H. Boozer.
\newblock {Magnetic field strength of toroidal plasma equilibria}.
\newblock {\em Physics of Fluids B}, 3(10):2805, 1991.

\bibitem{Mercier1964}
C.~Mercier.
\newblock {Equilibrium and stability of a toroidal magnetohydrodynamic system
  in the neighbourhood of a magnetic axis}.
\newblock {\em Nuclear Fusion}, 4(3):213, 1964.

\bibitem{Solovev1970}
L.~S. Solov'ev and V.~D. Shafranov.
\newblock {\em {Reviews of Plasma Physics 5}}.
\newblock Consultants Bureau, New York - London, 1970.

\bibitem{Jorge2020}
R.~Jorge, W.~Sengupta, and M.~Landreman.
\newblock {Near-axis expansion of stellarator equilibrium at arbitrary order in
  the distance to the axis}.
\newblock {\em Journal of Plasma Physics}, 86(1):905860106, 2020.

\bibitem{Nuhrenberg1988}
J.~Nuhrenberg and R.~Zille.
\newblock {Quasi-helically symmetric toroidal stellarators}.
\newblock {\em Physics Letters A}, 129(2):113, 1988.

\bibitem{Drevlak2017}
M.~Drevlak.
\newblock {Stellarator configuration design using ROSE}.
\newblock In {\em 21st International Stellarator-Heliotron Workshop}, page~2,
  2017.

\bibitem{Anderson1995}
F.~S.~B. Anderson, A.~F. Almagri, D.~T. Anderson, P.~G. Matthews, J.~N.
  Talmadge, and J.~L. Shohet.
\newblock {The Helically Symmetric Experiment, (HSX) Goals, Design and Status}.
\newblock {\em Fusion Technology}, 27(3T):273, 1995.

\bibitem{Ku2011}
L.~P. Ku and A.~H. Boozer.
\newblock {New classes of quasi-helically symmetric stellarators}.
\newblock {\em Nuclear Fusion}, 51(1):013004, 2011.

\bibitem{Bader2020}
Aaron Bader.
\newblock {Dataset for Wistell-A stellarator:
  zenodo.org/record/3952862{\#}.Xxm8OpNKgWo}, 2020.

\bibitem{Zarnstorff2001}
M.~C. Zarnstorff, L.~A. Berry, A.~Brooks, E.~Fredrickson, G.~Y. Fu,
  S.~Hirshman, S.~Hudson, L.~P. Ku, E.~Lazarus, D.~Mikkelsen, D.~Monticello,
  G.~H. Neilson, N.~Pomphrey, A.~Reiman, D.~Spong, D.~Strickler, A.~Boozer,
  W.~A. Cooper, R.~Goldston, R.~Hatcher, M.~Isaev, C.~Kessel, J.~Lewandowski,
  J.~F. Lyon, P.~Merkel, H.~Mynick, B.~E. Nelson, C.~Nuehrenberg, M.~Redi,
  W.~Reiersen, P.~Rutherford, R.~Sanchez, J.~Schmidt, and R.~B. White.
\newblock {Physics of the compact advanced stellarator NCSX}.
\newblock {\em Plasma Physics and Controlled Fusion}, 43(12A):A237, 2001.

\bibitem{Najmabadi2008}
F.~Najmabadi, A.~R. Raffray, S.~I. Abdel-Khalik, L.~Bromberg, L.~Crosatti,
  L.~El-Guebaly, P.~R. Garabedian, A.~A. Grossman, D.~Henderson, A.~Ibrahim,
  T.~Ihli, T.~B. Kaiser, B.~Kiedrowski, L.~P. Ku, J.~F. Lyon, R.~Maingi,
  S.~Malang, C.~Martin, T.~K. Mau, B.~Merrill, R.~L. Moore, R.~J. Peipert,
  D.~A. Petti, D.~L. Sadowski, M.~Sawan, J.~H. Schultz, R.~Slaybaugh, K.~T.
  Slattery, G.~Sviatoslavsky, A.~Turnbull, L.~M. Waganer, X.~R. Wang, J.~B.
  Weathers, P.~Wilson, J.~C. Waldrop, M.~Yoda, and M.~Zarnstorff.
\newblock {The ARIES-CS compact stellarator fusion power plant}.
\newblock {\em Fusion Science and Technology}, 54(3):655, 2008.

\bibitem{Garabedian2008}
P.~R. Garabedian.
\newblock {Three-dimensional analysis of tokamaks and stellarators}.
\newblock {\em Proceedings of the National Academy of Sciences of the United
  States of America}, 105(37):13716, 2008.

\bibitem{Drevlak2013}
M.~Drevlak, F.~Brochard, P.~Helander, J.~Kisslinger, M.~Mikhailov,
  C.~N{\"{u}}hrenberg, J.~N{\"{u}}hrenberg, and Y.~Turkin.
\newblock {ESTELL: A Quasi-Toroidally Symmetric Stellarator}.
\newblock {\em Contributions to Plasma Physics}, 53(6):459, 2013.

\bibitem{Shimizu2018a}
A.~Shimizu, H.~Liu, M.~Isobe, S.~Okamura, S.~Nishimura, C.~Suzuki, Y.~Xu,
  X.~Zhang, B.~Liu, J.~Huang, X.~Wang, H.~Liu, and C.~Tang.
\newblock {Configuration property of the Chinese first quasi-axisymmetric
  stellarator}.
\newblock {\em Plasma and Fusion Research}, 13:3403123, 2018.

\bibitem{Henneberg2019}
S.~A. Henneberg, M.~Drevlak, C.~N{\"{u}}hrenberg, C.~D. Beidler, Y.~Turkin,
  J.~Loizu, and P.~Helander.
\newblock {Properties of a new quasi-axisymmetric configuration}.
\newblock {\em Nuclear Fusion}, 59(2):026014, 2019.

\bibitem{Barnes2019}
M.~Barnes, F.~I. Parra, and M.~Landreman.
\newblock {stella: An operator-split, implicit–explicit
  {$\delta$}f-gyrokinetic code for general magnetic field configurations}.
\newblock {\em Journal of Computational Physics}, 391:365, 2019.

\bibitem{Proll2013}
J.~H.E. Proll, P.~Xanthopoulos, and P.~Helander.
\newblock {Collisionless microinstabilities in stellarators. II. Numerical
  simulations}.
\newblock {\em Physics of Plasmas}, 20(12):122506, 2013.

\bibitem{Plunk2014}
G.~G. Plunk, P.~Helander, P.~Xanthopoulos, and J.~W. Connor.
\newblock {Collisionless microinstabilities in stellarators. III. the
  ion-temperature-gradient mode}.
\newblock {\em Physics of Plasmas}, 21(3):032112, 2014.

\bibitem{Zocco2020}
A.~Zocco, G.~G. Plunk, and P.~Xanthopoulos.
\newblock {Geometric stabilization of the electrostatic
  ion-temperature-gradient driven instability. II. Non-axisymmetric systems}.
\newblock {\em Physics of Plasmas}, 27(2):022507, 2020.

\bibitem{Proll2015a}
J.~H.E. Proll, H.~E. Mynick, P.~Xanthopoulos, S.~A. Lazerson, and B.~J. Faber.
\newblock {TEM turbulence optimisation in stellarators}.
\newblock {\em Plasma Physics and Controlled Fusion}, 58(1):014006, 2015.

\bibitem{Beeke2020}
O.~Beeke, M.~Barnes, M.~Romanelli, M.~Nakata, and M.~Yoshida.
\newblock {Impact of plasma shaping on tokamak microstability}.
\newblock {\em arXiv:2012.02669}, 2020.

\bibitem{Landreman2019b}
M.~Landreman and W.~Sengupta.
\newblock {Constructing stellarators with quasisymmetry to high order}.
\newblock {\em Journal of Plasma Physics}, 85(6):815850601, 2019.

\bibitem{Landreman2019a}
M.~Landreman, W.~Sengupta, and G.~G. Plunk.
\newblock {Direct construction of optimized stellarator shapes. Part 2.
  Numerical quasisymmetric solutions}.
\newblock {\em Journal of Plasma Physics}, 85(1):905850103, 2019.

\bibitem{Romanelli1989}
F.~Romanelli.
\newblock {Ion temperature-gradient-driven modes and anomalous ion transport in
  tokamaks}.
\newblock {\em Physics of Fluids B}, 1(5):1018--1025, 1989.

\bibitem{Biglari1989}
H.~Biglari, P.~H. Diamond, and M.~N. Rosenbluth.
\newblock {Toroidal ion-pressure-gradient-driven drift instabilities and
  transport revisited}.
\newblock {\em Physics of Fluids B}, 1(1):109, 1989.

\bibitem{Roberg-Clark2020}
G.~T. Roberg-Clark, G.~G. Plunk, and P.~Xanthopoulos.
\newblock {Calculating the linear critical gradient for the
  ion-temperature-gradient mode in magnetically confined plasmas}.
\newblock {\em arXiv:2010.13441}, 2020.

\bibitem{Rosenbluth1968}
M.~N. Rosenbluth.
\newblock {Low-frequency limit of interchange instability}.
\newblock {\em Physics of Fluids}, 11(4):869, 1968.

\bibitem{Alcuson2020}
J.~A. Alcuson, P.~Xanthopoulos, G.~G. Plunk, P.~Helander, F.~Wilms, Y.~Turkin,
  A.~Von Stechow, and O.~Grulke.
\newblock {Suppression of electrostatic micro-instabilities in maximum-J
  stellarators}.
\newblock {\em Plasma Physics and Controlled Fusion}, 62(3):035005, 2020.

\end{thebibliography}

\end{document}